\documentstyle[12pt]{article}
\font\euler=eufm10
\newfam\eufmfam
\textfont\eufmfam=\euler
\def\g{\mbox{\euler g}}
\def\h{\mbox{\euler h}}
\def\liet{\mbox{\euler t}}

\def\spl{\mbox{\euler sl}\,}

\def\n{\mbox{\euler n}}
\def\b{\mbox{\euler b}}

\newcommand{\beq}{\begin{equation}}
\newcommand{\eeq}{\end{equation}}

\newcommand{\beqa}{\begin{eqnarray}}
\newcommand{\eeqa}{\end{eqnarray}}
\newcommand{\ot}{\otimes}
\newcommand{\longto}{\longrightarrow}
\newcommand{\nibun}{\frac{1}{2}}
\newcommand{\lmd}{\lambda}
\newcommand{\Lmd}{\Lambda}
\newcommand{\nipi}{\frac{1}{2\pi i}}
\newcommand{\ad}{{\rm ad}}
\newcommand{\Lie}{{\rm Lie}}
\newcommand{\lieh}{h}

\newcommand{\HC}{H_{\!\C}}
\newcommand{\Hc}{H_{\!\bf c}}
\newcommand{\semidir}{\tilde{\times}}
\newcommand{\met}{\mbox{\sl g}}
\newcommand{\smet}{\mbox{\scriptsize {\sl g}}}
\newcommand{\bartial}{\bar \partial}
\newcommand{\bartialA}{\bartial_{\!A}}
\newcommand{\rA}{\mbox{\tiny A}}

\newcommand{\PP}{{\rm P}}

\newcommand{\PPpk}{{\rm P}_{\!+}^{(k)}}
\newcommand{\PPptilk}{\PP_{\!+}^{(\tilk)}}
\newcommand{\Pv}{\PP^{\vee}}
\newcommand{\tilP}{\tilde{\PP}}
\newcommand{\QQ}{{\rm Q}}
\newcommand{\Qv}{\QQ^{\vee}}
\newcommand{\C}{{\bf C}}
\newcommand{\R}{{\bf R}}

\newcommand{\Z}{{\bf Z}}
\newcommand{\N}{\bf N}
\newcommand{\CP}{{\bf P}^1}
\newcommand{\Ch}{{\rm C}}

\newcommand{\alcv}{\widehat{{\rm C}}}

\newcommand{\V}{{\rm V}}
\newcommand{\Waff}{W_{\!{\rm aff}}}
\newcommand{\Waffh}{W'_{\!{\rm aff}}}
\newcommand{\Gmalcv}{\Gamma_{\alcv}}
\newcommand{\tr}{{\rm tr}}
\newcommand{\ttr}{\,{}^{\tr}\!}

\newcommand{\trVlmd}{\tr_{\!\!\:{}_{V_{\lmd}}}}
\newcommand{\trP}{\tr_{\!\!\:{}_P}}
\newcommand{\trG}{\tr_{\!\!\:{}_G}}
\newcommand{\trH}{\tr_{\!\!\:{}_H}}
\newcommand{\hp}{{\rm H}}
\newcommand{\Daff}{\Delta_{{\rm aff}}}
\newcommand{\Vaff}{{\rm V}_{\!{\rm aff}}}
\newcommand{\PC}{P_{\!\C}}
\newcommand{\A}{{\cal A}}
\newcommand{\Ass}{\A_{ss}}
\newcommand{\Assc}{\Ass^{\circ}}
\newcommand{\AP}{{\cal A}_P}
\newcommand{\APc}{\AP^{\circ}}

\newcommand{\G}{{\cal G}}
\newcommand{\GP}{\G_P}
\newcommand{\GPC}{\G_{P_{\!\bf c}}}
\newcommand{\Ph}{{\cal P}}
\newcommand{\NN}{{\cal N}}
\newcommand{\NNP}{{\cal N}_P}
\newcommand{\NNPgma}{\NN_{P\gamma}}
\newcommand{\NNPc}{\NNP^{\circ}}

\newcommand{\NNc}{\NN^{\circ}}
\newcommand{\VV}{{\cal V}}

\newcommand{\dN}{d_{\NN}}
\newcommand{\dSf}{\hat{d}_{S}}
\newcommand{\dNf}{\hat{d}_{\NN}}
\newcommand{\bnu}{\bar \nu}
\newcommand{\lnu}{\mbox{\large $\nu$}}
\newcommand{\blnu}{\mbox{\large $\bar \nu$}}
\newcommand{\un}{\mbox{\small {\sl U}}}
\newcommand{\sun}{\mbox{\scriptsize {\sl U}}}

\newcommand{\tilH}{\tilde{H}}
\newcommand{\kh}{k^{\!\bf c}}

\newcommand{\tila}{\tilde{a}}
\newcommand{\tilb}{\tilde{b}}
\newcommand{\LWZ}{{\cal L}_{\!{}_{\rm W\!Z}}}
\newcommand{\Pic}{{\rm Pic}}
\newcommand{\Jac}{{\rm Jac}}
\newcommand{\Sgmtau}{\Sigma_{\tau}}
\newcommand{\Ker}{{\rm Ker}}
\newcommand{\Vol}{{\rm vol}}
\newcommand{\vol}{{\rm vol}}

\newcommand{\tot}{{\rm tot}}
\newcommand{\Aut}{{\rm Aut}}
\newcommand{\End}{{\rm End}}

\newcommand{\vb}{\underline v}
\newcommand{\lbackslash}{\mbox{\large $\backslash$}}

\newcommand{\Lslash}{\mbox{\Large /}}

\newcommand{\bz}{\bar \zeta}
\newcommand{\z}{\zeta}

\newcommand{\pinitau}{\frac{\pi}{2\tau_2}}
\newcommand{\sgmad}{\sigma_{\!\ad}}
\newcommand{\OO}{{\cal O}}
\newcommand{\ooint}{\mbox{\footnotesize $\OO$}\!\!\!\!\!\!\!\:\int}

\newcommand{\Chi}{\mbox{\large $\chi$}}
\newcommand{\tilk}{\tilde{k}}
\newcommand{\coxh}{\lieh^{\vee}}

\newcommand{\dd}{d^2\!\!\:}

\newcommand{\sDelta}{{\sl \Delta}}

\newcommand{\NP}{Nucl. Phys.\ }
\newcommand{\PL}{Phys. Lett.\ }
\newcommand{\CMP}{Commun. Math. Phys.\ }
\newcommand{\e}{{\rm e}}
\newcommand{\Map}{{\rm Map}}
\newcommand{\vsp}{\vspace{0.2cm}}
\newcommand{\Gmhol}{\Gamma_{hol}}

\newcommand{\hvG}{{\sl g}^{\!\vee}}
\newcommand{\hvH}{h^{\!\vee}}
\newcommand{\dslash}{/\!\!/}
\newcommand{\indJ}{\hat{\cal J}}
\newcommand{\Fadgh}{{\cal F}_{\it gh}^{\ad}}
\newcommand{\tiL}{\,\,\widetilde{\!\!L\,\,\,}\!\!\!}
\newcommand{\tms}{{\times}}
\newcommand{\cd}{{\cdot}}

\newtheorem{th}{Theorem}
\newtheorem{pn}[th]{Proposition}

\begin{document}
\baselineskip=0.61cm
\normalsize

\hfill UT-685

\vspace{-0.1cm}
\hfill UT-Komaba/94-1

\vspace{-0.05cm}
\hfill August, 1994

\renewcommand{\thefootnote}{\fnsymbol{footnote}}
\vspace{1.8cm}

\begin{center}
{\Large \bf Global Aspects Of\\
\vspace{0.2cm}
 Gauged Wess-Zumino-Witten Models}

\vspace{1cm}
{\sc Kentaro Hori}\footnote[2]{e-mail address:
hori@danjuro.phys.s.u-tokyo.ac.jp}

\vspace{0.5cm}
{\it Department Of Physics, University Of Tokyo,
Bunkyo-ku, Tokyo 113, Japan}

\vspace{0.5cm}

\begin{minipage}[t]{15.5cm}
{\small
A study of the gauged Wess-Zumino-Witten models is given focusing
on the effect of topologically non-trivial configurations of gauge fields.
A correlation function is expressed as an integral over a
moduli space of holomorphic bundles with quasi-parabolic structure.
Two actions of the fundamental group of the gauge group is defined: One
on the space of gauge invariant local fields and the other on
the moduli spaces. Applying these in the integral expression, we obtain
a certain identity which relates correlation functions for configurations
of different topologies. It gives an important information on the
topological sum for the partition and correlation functions.
}
\end{minipage}

\end{center}

\vspace{0.5cm}
\renewcommand{\theequation}{1.\arabic{equation}}\setcounter{equation}{0}
\renewcommand{\thefootnote}{\arabic{footnote}}
\begin{center}
{\bf 1. Introduction}
\end{center}

The gauged Wess-Zumino-Witten model in two dimensions has two different
aspects of interest.
On the one hand, it is an exactly soluble quantum gauge theory and
is interesting from the point of view of geometry of gauge fields.
On the other hand,
it is a conformally invariant quantum field theory (CFT): There
are observations \cite{GawKup,KaSch} that
a wide class of solved CFTs such as
unitary minimal models (bosonic \cite{BPZ} or supersymmetric \cite{FQS}),
parafermionic models \cite{Parafermi}, etc
are realized by gauged WZW models as lagrange field theories.
Hence, the model provides a powerful method
for the model building and classification
of CFTs, important problems for the study of two dimensional statistical
systems and string theory.

In this paper, we focus on the former, the geometric aspects of the theory
and propose a method to take into account the topologically non-trivial
configurations of gauge fields. Then, we see that incorporation of
non-trivial topology has simple consequences which are of vital importance
in the model building of CFTs.

A gauged WZW model is a WZW sigma model with target $G$ a group manifold,
coupled to gauge field for a group $H$.
We concentrate on the case in which
$G$ is compact, connected and simply connected and $H$ is a
connected, closed subgroup of the adjoint group $G/Z_G$ where $Z_G$ is the
center of $G$. The classical action of the system (the WZW action) at level
$k\in \N$ for a closed Riemann surface $\Sigma$, a map $g:\Sigma\to G$ and
an $\h=\Lie(H)$-valued one form $A$ is given by
\beqa
kI_{\Sigma}(A,g)&=&
\frac{ik}{4\pi}\int_{\Sigma}\tr(\partial g^{-1}\bartial g)
-\frac{ik}{12\pi}\int_{B_{\Sigma}}\tr(\tilde{g}^{-1}d\tilde{g})^3
\label{actiontriv}\\
&&+\frac{ik}{2\pi}\int_{\Sigma}\tr\!\left(g\partial g^{-1}A^{01}
+A^{10}g^{-1}\bartial g+gA^{10}g^{-1}A^{01}-A^{10}A^{01}\right),\nonumber
\eeqa
where ``$\tr$'' is a trace in a representation of $G$(\footnote{
If $G$ is
simple, it is normalized by $\tr_{\g}(\ad X\ad Y)=2\hvG\tr (XY)$ for $X,
Y\in\g=\Lie(G)$ where $\hvG$ is the dual Coxeter number of $\g$.
Generalization to the non-simple case is obvious.})
and $A^{10}$ (resp.
$\!A^{01}$) is the $(1,0)$-form (resp. $\!(0,1)$-form) component of $A$.
In the second term, $B_{\Sigma}$ is a compact
three manifold having $\Sigma$ as its boundary and $\tilde{g}:B_{\Sigma}\to
G$ is an extension of $g$. The value $\e^{-kI_{\Sigma}(A,g)}$ which we call
the {\it WZW weight} is independent on the choice of $B_{\Sigma}$ and
$\tilde{g}$ and hence
may be used as the weight for the path integration over $A$ and $g$. This
weight is invariant under the gauge transformation $A\to A^h=h^{-1}Ah
+h^{-1}dh$, $g\to h^{-1}gh$ and the resulting system is a quantum gauge
theory.

A natural generalization is to consider the topologically non-trivial
configurations of $A$ and $g$. Thus, let $\{U_0,U_{\infty}\}$ be an open
covering of $\Sigma$ such that $U_0$ contains a disc $D_0$ and $U_0\cap
U_{\infty}$ is an annular neighborhood of the boundary circle $\partial D_0$.
General configuration is determined by gauge fields
$\{A_0,A_{\infty}\}$ and maps $\{g_0,g_{\infty}\}$, both
defined on $\{U_0,U_{\infty}\}$ and satisfying the relation
\beq
A_0=h_{\infty 0}^{-1}A_{\infty}h_{\infty 0}+h_{\infty 0}^{-1}dh_{\infty 0}
\quad\mbox{and} \quad g_0=h_{\infty 0}^{-1}g_{\infty}h_{\infty 0}
\label{gauge transf}
\eeq
on $U_0\cap U_{\infty}$ where $h_{\infty 0}$ is a map to $H\subset G/Z_G$.
In geometric
terms, this map $h_{\infty 0}$, called the transition function, determines a
principal $H$-bumdle $P$ over $\Sigma$, $\{A_0,A_{\infty}\}$ determines a
connection $A$ of $P$ and $\{g_0,g_{\infty}\}$ determines a section $g$ of
the associated $G$-bundle $P\tms_H G$.
The homotopy type of the loop
$\gamma_{\infty 0}=h_{\infty 0}|_{\partial D_0}$ determines
the topological type of the $H$-bundle $P$ and hence the fundamental group
$\pi_1(H)$ of $H$ classifies the topological types of
configurations.(\footnote{
Any principal $H$-bundle over $\Sigma$ admits trivialization over
$\Sigma_{\infty}=\Sigma-D_0$ as well as over $D_{0}$: Take a pants
decomposition of $\overline{\Sigma_{\infty}}$. Since $H$ is connected,
we can choose a gauge over any circle. Now it is enough to observe
that given gauges over two boundary circles of a pants can be extended
over the whole pants and determines (up to homotopy) a gauge
over the third boundary.})

One purpose of the paper is to give a method to calculate the correlation
function of gauge invariant fields $O_1\cdots O_s$ restricted to
configurations of the topological type determined by $P$:
\beq
Z_{\Sigma,P}(O_1\cdots O_s)=\frac{1}{\Vol \,\G_P}\int {\cal D}\!A
{\cal D}\!\!\,g\,\e^{-kI_{\Sigma,P}(A,g)}O_1\cdots O_s\,,
\eeq
where $\G_P$ is the group of gauge transformations and $kI_{\Sigma,P}(A,g)$
is
the WZW action defined in \S 2 for a general principal $H$-bundle $P$.

Another and the main purpose is to prove certain exact relationships of
correlators for configurations of different topologies.
Namely, we will see that the
group $\pi_1(H)$, which acts on the set of principal $H$-bundles $\gamma :P
\mapsto P\gamma$ by multiplication on the transition functions
$\gamma_{\infty 0}
\mapsto \gamma_{\infty 0}\gamma$, acts on the space of gauge invariant local
fields $\gamma :O\mapsto \gamma O$ in such a way that the following holds:
\beq
Z_{\Sigma,P}(O_1\cdots O_s \gamma O)=Z_{\Sigma,P\gamma}(O_1\cdots O_s O)\,.
\label{FI}
\eeq
We call this the {\it topological identity}.
The proof is reduced to the solution of
a problem in the geometry of moduli spaces of holomorphic $\HC$-bundles
with quasi-parabolic structure.
In addition to the case with abelian $H$ in which the problem is trivial,
it is solved for the cases
$\Sigma=$ sphere with $H$ general and $\Sigma=$ torus with $H=SO(3)$.

The significance of (\ref{FI})
can be seen if we take the sum
$\sum_P$ over topologies; the fields $O$ and $\gamma O$ are then
indistinguishable. For instance, consider the case with
$G=SU(2)\times SU(2)$ and $H=SO(3)$ diagonally embedded into $G/Z_G=SO(3)
\times SO(3)$. The gauge invariant local fields can be classified by
the rectangular grid whose squares are labeled by
$\{0,\frac{1}{2},\cdots,\frac{k}{2}\}\tms
\{0,\frac{1}{2},\cdots,\frac{k+1}{2}\}$.
The space of fields in
the square $(j_1,j)$ is identified with the degenerate representation of
the Virasoro algebra of central charge $1-\frac{6}{(k+2)(k+3)}$ and dimension
$\frac{((k+3)j_1-(k+2)j+1)^2-1}{4(k+2)(k+3)}$ as is also the case for the
corner $(\frac{k}{2}-j_1,\frac{k+1}{2}-j)$. As we shall see in \S4, this
transformation $(j_1,j)\leftrightarrow (\frac{k}{2}-j_1,\frac{k+1}{2}-j)$
corresponds precisely to the transformation $O\leftrightarrow \gamma O$ where
$\gamma$ is the non-trivial element of $\pi_1(SO(3))=\Z_2$.
Hence, only after the sum over topologies, the set of
distinguishable fields coinsides with that of the $k$-th unitary minimal
model \cite{BPZ}.
The situation is the same for general $G$ and $H$. The space of local gauge
invariant fields, acted on by the infinite conformal symmetry, is identified
\cite{KaSch} with the direct sum of Virasoro modules by coset construction
\cite{GKO}.
For each element $\gamma\in \pi_1(H)$, there is an isomorphism of coset
Virasoro modules, known as the ``field identification"
\cite{Gep,Moore-Seiberg,LVW}, that
corresponds to our transformation $O\mapsto \gamma O$. Hence,
this identification of Virasoro mudules leads via the sum over topologies to
a genuin identification of quantum fields.

\vsp
The rest of the paper consists of five sections and four appendices.
Sections 2 and 3 are the preparatory parts which follow to some extent
the route
explioted by Gaw\c edzki and others \cite{F-G-K,GawKup}. The main part is
section 4 in which a novel expression of the correlator is proposed
(see (\ref{newintexpr})) and the topological identity (\ref{FI}) is proved
at least for the cases mentioned above. An application of (\ref{FI}) is made
in section 5.
The last section includes a remark on alternative choices of the
classical action.

\renewcommand{\theequation}{2.\arabic{equation}}\setcounter{equation}{0}

\vsp
\begin{center}
{\bf 2. Wess-Zumino-Witten Model}
\end{center}

We start with the study of the WZW model in a general background gauge
field with the group $H=G/Z_G$.
The first material is a construction of the WZW action for topologically
non-trivial configurations. It is designed to satisfy the following property
of factorization.
For a Riemann surface $\Sigma$, we
choose a disc $D_0$ in $\Sigma$ and an open covering $\{U_0,U_{\infty}\}$
of $\Sigma$ as in \S 1 and put $S=\partial D_0$. Let $P$ be the principal
$H$-bundle with the transition function $h_{\infty 0}:U_0\cap
U_{\infty}\to H$.
For fields $A=\{A_0,A_{\infty}\}$ and $g=\{g_0,g_{\infty}\}$ satisfying
(\ref{gauge transf}), the WZW weight on the whole surface $\Sigma$ is
expressed as the product of the weight on $D_0$ and the weight on
$\Sigma_{\infty}=\overline{\Sigma-D_0}$:
\beq
\e^{-kI_{\Sigma,P}(A,g)}=\langle \e^{-kI_{\Sigma_{\infty}}(A_{\infty},
g_{\infty})},\ad \gamma_{\infty 0}\,\e^{-kI_{D_0}(A_0,g_0)}\rangle,
\label{WZWweight}
\eeq
Here, the weight on $D_0$ is not valued in the ordinary number field $\C$
but in a complex line $\LWZ^k|_{\gamma_0}$ associated to the loop
$\gamma_0=g_0|_S$, and the weight on $\Sigma_{\infty}$ is in
a line $\LWZ^{*k}|_{\gamma_{\infty}}$
associated to $\gamma_{\infty}=g_{\infty}|_S$.
The product is defined through a gauge transformation
$\ad\gamma_{\infty 0}:\LWZ^k|_{\gamma_0}\to \LWZ^k|_{\gamma_{\infty}}$
associated to the transition function $\gamma_{\infty 0}=h_{\infty 0}|_S$.

This factorization property goes over to the quantum theory:
A correlation function on the surface $\Sigma$ is expressed
as the pairing of two wave functions at $S=D_0\cap \Sigma_{\infty}$, one
coming from $D_0$ and the other from $\Sigma_{\infty}$.
Using the infinite dimensional symmetry of the gauge field $A_0$,
we can explicitly determine the wave function coming from the disc $D_0$
with
field insertion at one point, and thus obtain the correspondence
of fields and states.
If we change the gauge (=reference section) over the boundary $\partial D_0$,
the correspondence effectively changes
and we have transformations of states and of fields.
For a certain gauge transformation of
non-trivial homotopy, the corresponding transformation of states
(or of fields) takes a simple form that is known as the {\it spectral flow}
\cite{Moore-Seiberg}.
Consequently, we obtain a relation of correlators of the WZW model
that may be considered as a prototype of the equation (\ref{FI}) of the
gauged WZW model.

\vspace{0.5cm}
\noindent{\sc 2.1 The Line Bundle $\LWZ^k$ And The Adjoint Action Of $LH$}

\vspace{0.3cm}

We begin with defining the WZW weight on the disc
$D_0=\{\,z\!\in\! \C\,;\,|z|\leq 1\}$ with the parametrized boundary
$\theta\mapsto \e^{i\theta}\in\partial D_0$, following the line of argument
in \cite{F-G-K}.
In order to deal with the chiral gauge symmetry, we consider maps to the
complexified group $G_{\!\C}$.
For $A\in \Omega^1(D_0,\h)$ and $g\in\Map(D_0,G_{\!\C})$, by choosing
a smooth extension of $g$ to a map $\hat{g}$ defined over the Riemann
sphere $\CP=\C\cup\{\infty\}$, we can define the WZW action
$I_{\CP}(\hat{A},\hat{g})$ by (\ref{actiontriv}) where $\hat{A}=A$ on $D_0$
and $\hat{A}=0$ on $D_{\infty}=\overline{\CP-D_0}$.
Since it depends on the choice of $\hat{g}$, we consider the set
of all extensions and put a suitable equivalence relation
on $\Map(D_{\infty},G_{\!\C})\times \C$ so that the class
\beq
\e^{-kI_{D_0}(A,g)}=\{(\hat{g}|_{D_{\infty}},\e^{-kI_{\CP}(\hat{A},\hat{g})})
\}\,,
\label{wzweightdisc}
\eeq
is independent of the choice. This defines the line bundle
$\LWZ^k$ over the group $LG_{\!\C}$
of loops in $G_{\!\C}$ so that the WZW weight (\ref{wzweightdisc}) is
an element of the line $\LWZ^k|_{\gamma}$ over the boundary loop
$\gamma(\theta)=g(\e^{i\theta})$.

\vsp
The group structure of $LG_{\!\C}$ lifts to
a semigroup structure of $\LWZ^k$ by
\beq
\{(g_1,c_1)\}\{(g_2,c_2)\}=\{(g_1g_2,c_1c_2
\e^{-k\Gamma_{D_{\infty}}(g_1,g_2)})\},
\eeq
where $\Gamma_{\Sigma}(g_1,g_2)=\frac{i}{2\pi}\int_{\Sigma}\tr(g_2\partial
g_2^{-1}g_1^{-1}\bartial g_1)$. The group $(\LWZ)^{\times}$
of invertible elements for $k=1$ is isomorphic to
the basic central extension $\tiL G_{\!\C}$ \cite{P-S} of the loop group
$LG_{\!\C}$ and acts on $\LWZ^k$ on the left and on the right through the
homomorphism $\{(g,c)\}\in (\LWZ)^{\times}\mapsto \{(g,c^k)\}\in
(\LWZ^k)^{\times}$.
The Polyakov-Wiegmann (PW) identity exhibits the response
of the WZW weight to the {\it  chiral gauge transformation} $A\mapsto A^h$,
$g\mapsto g^h$ by $h\in \Map(D_0,G_{\!\C})$:
\beqa
&&(A^h)^{01}=h^{-1}A^{01}h
+h^{-1}\bartial h\,,\quad (A^h)^{10}=h^*A^{10}h^{*-1}+h^*\partial h^{*-1}
\label{def:chiA}\\
\hspace{-1cm}\mbox{and}&&g^h=h^{-1}gh^{*-1}\label{def:chig}
\eeqa
in which  $h\mapsto h^*$ is the Cartan
involution that corresponds to hermitian conjugation in a unitary
representation of $G$. It states that
\beq
\e^{-kI_{D_{0}}(A,g)}=\e^{-kI_{D_{0}}(A,h)}\e^{-kI_{D_{0}}(A^{h},g^h)}
\e^{-kI_{D_{0}}(A,h^{*})}\e^{-k\Gamma_{D_{0}}(A,h,h^{*})}\, \, ,\label{locPW}
\eeq
where $\Gamma_{D_0}$ is given by
\beq
\Gamma_{D_{0}}(A,h,h^*)=\frac{i}{2\pi}\int_{D_{0}}\tr\!\left( h^*
\partial_{\!A}h^{*-1}h^{-1}\bartial_{\!A}h\right)\,,\label{defGamma}
\eeq
in which $h^{-1}\bartial_{\!A}h=h^{-1}\bartial h+h^{-1}A^{01}h-A^{01}$ and
similarly for $h^{*}\partial_{\!A}h^{*-1}$.

If $h$ is $G$-valued, the above identity can be written as
\beq
\e^{-kI_{D_{0}}(A,g)} =\gamma \e^{-kI_{D_{0}}(A^{h},g^h)} \gamma^{-1}\,\,,
\label{cov}
\eeq
where $\gamma\in LG$ is the boundary loop of $h$. In this sense, we can say
that the WZW weight on $D_0$ is gauge invariant.
In (\ref{cov}), we have used
the fact that the adjoint action $\gamma_1\mapsto \gamma\gamma_1\gamma^{-1}$
of $\gamma\in LG$ on $LG_{\!\C}$ lifts to an action on $\LWZ^k$ by choosing
any element in $(\LWZ)^{\times}|_{\gamma}$.
In fact, the adjoint action of $LH$ on
$LG_{\!\C}$, which is apparently well-defined,
lifts to an automorphic action on $\LWZ^k$ so that
the gauge invariance (\ref{cov}) holds when $h^{-1}gh$ is
defined on $D_0$: The action of a loop $\gamma^{-1}\in LH$ on the element
$\{(\check{g},c)\}\in \LWZ^k$ for $\check{g}\in \Map (D_{\infty},G_{\!\C})$
with $\check{g}(\infty)=1$ is defined by
\beq
\ad\gamma^{-1}\{(\check{g}, c)\}=\{(\check{h}^{-1}\check{g}\check{h},
c\e^{-kC_{D_{\infty}}(\check{h},\check{g})})\},\label{liftad}
\eeq
in which $\check{h}\in\Map(D_{\infty}-\{\infty\},H)$ is any extension of
$\gamma$ and $C_{D_{\infty}}$ is given by
\beq
C_{D_{\infty}}\!(\check{h},\!\check{g})\!=\!
K_{\!D_{\!\infty}}(\check{h}^{-1}
\!\check{g}\check{h})-\!K_{\!D_{\!\infty}}(\check{g})-\frac{i}{4\pi}\!\!
\int_{D_{\!\infty}}\!\!\!\! \tr\!\left\{ \!(d\check{g}\check{g}^{-1}\!\!
+\check{g}^{-1}\!d\check{g})\check{h}d\check{h}^{\!-1}\!\!
+\check{h}d\check{h}^{\!-1}\!\check{g}\check{h}d\check{h}^{\!-1}\!
\check{g}^{-1}\right\}\!,
\eeq
where $K_{D_{\!\infty}}(\check{g})=\frac{i}{4\pi}\int_{D_{\!\infty}}
\tr(\partial \check{g}^{-1}\bartial \check{g})$. In \cite{P-S},
the adjoint
action of $LH$ on $\tiL G_{\!\C}$ is defined and is shown to be unique.
As it should be, it coincides with the action (\ref{liftad}) for $k=1$
coincides.

\vsp
Next, we construct the WZW weight on $\Sigma_{\infty}=\overline{\Sigma-D_0}$
where $D_0$ is the unit disc in an open subset of $\Sigma$
with coordinate $z$.
As in the above argument, we put an equivalence relation on
$\Map(D_0,G_{\!\C})\times \C$ defining a line bundle
$\LWZ^{*k}$ over $LG_{\!\C}$
so that the WZW weight $\e^{-kI_{\Sigma_{\infty}}(A,g)}$ for $A\in
\Omega^1(\Sigma_{\infty},\h)$ and $g\in \Map(\Sigma_{\infty},G_{\!\C})$ is
given as the class
$\{(\hat{g}|_{D_0},\e^{-kI_{\Sigma}(\hat{A},\hat{g})})\}$
in the line $\LWZ^{*k}|_{\gamma}$ over the loop
$\gamma(\theta)=g(\e^{i\theta})$.
This bundle has a semigroup structure so that
the PW identity holds:
\beq
\e^{-kI_{\Sigma_{\!\infty}}\!(A,g)}=\e^{-kI_{\Sigma_{\!\infty}}\!(A,h)}
\e^{-kI_{\Sigma_{\!\infty}}\!(A^{h},g^h)}
\e^{-kI_{\Sigma_{\!\infty}}\!(A,h^{*})}
\e^{-k\Gamma_{\Sigma_{\!\infty}}\!(A,h,h^{*})} \, .
\label{loc*PW}
\eeq

\vsp
The line bundles $\LWZ^{*k}$ and $\LWZ^k$ are dual to each other under the
product
\beq
\langle \{(g|_{D_0},c_0)\},\{(g|_{D_{\infty}},c_{\infty})\}\rangle
=c_0c_{\infty}\e^{kI_{\CP}(g)}\,,
\label{L*Lprod}
\eeq
where $g\in \Map(\CP,G_{\!\C})$. The WZW weight for general topology
is now
defined by (\ref{WZWweight}) where $\ad \gamma_{\infty 0}$ is the
adjoint action (\ref{liftad}) of the loop $\theta\mapsto
\gamma_{\infty 0}(\e^{i\theta})$.
For the trivial topology, we may take $\gamma_{\infty 0}\equiv 1$ and
(\ref{WZWweight})
reproduces the action (\ref{actiontriv}).

Since the product
(\ref{L*Lprod})
satisfies $\langle\tilde{\gamma}'_{1}\tilde{\gamma}'_{2},\tilde{\gamma}_{1}
\tilde{\gamma}_{2}\rangle=\langle\tilde{\gamma}'_{1},\tilde{\gamma}_{1}
\rangle
\langle \tilde{\gamma}'_{2},\tilde{\gamma}_{2}\rangle$ for
$\tilde{\gamma}'_{i}\in \LWZ^{*k}|_{\!\gamma_{i}}$ and $\tilde{\gamma}_{i}
\in \LWZ^{k}|_{\!\gamma_{i}}$ ($i=1,2$), the
PW identities (\ref{locPW}) and (\ref{loc*PW}) lead to the global version of
the PW identity:
\beq
I_{\Sigma,P}(A^h,g^h)=I_{\Sigma,P}(A,g)-I_{\Sigma,P}(A, hh^{\ast})\,.
\label{PW}
\eeq
In this expression, $h$ is a section of the {\it adjoint $\HC$-bundle},
namely the bundle
$P\tms_H H_{\!\C}$ associated to $P$ via the adjoint action of
$H$ on $\HC$.
The transformation $A\mapsto A^h$, $g\mapsto g^h$
(the {\it chiral gauge transformation}) is locally defined by
(\ref{def:chiA}) and (\ref{def:chig}).
If $h$ is $H$-valued, or precisely
if $h$ takes values in the adjoint $H$-bundle $P\tms_H H$, we have $hh^*=1$
and (\ref{PW}) is the statement of gauge invariance.

For a section $\epsilon$ of the {\it adjoint bundle} $\ad P=P\times_H \h$,
the action satisfies
\beq
\left(\frac{d}{dt}\right)_{\!0}\!I_{\Sigma,P}(A\, ,e^{t \epsilon}\, )
= \nipi\int_{\Sigma}\trP(\epsilon\, F_{\!A}) \, \, ,
\label{infchanom}
\eeq
where $F_A$ is the curvature of $A$ represented in $\ad P$ and $\trP$ is the
trace of the adjoint bundle normalized by $\tr_{\ad P}(\ad X\ad Y)=2\hvG
\trP(XY)$ when $G$ is simple. The properties (\ref{PW}) and (\ref{infchanom})
are just what we would expect for the chiral anomaly in the massless free
fermionic systems. Indeed, the WZW model was first introduced as the
non-abelian bosonization of spin-half fermions \cite{Witten}.

{\it Remark}. There is another way to construct the WZW action for
configurations of general topology. It is to make use of the equivariant
version of the Cheeger-Simons differential character
(see \cite{DijkWitt,Axl}).
A discussion on this is given in \S 6 and
in the future publication \cite{Hori}.

\vspace{0.5cm}
\noindent{\sc 2.2 Space Of States}

\vspace{0.3cm}
We proceed next to the quantization of the WZW model. The correlation
function
of the local fields $O_1\cdots O_s$ is given by the path-integral
\beq
Z_{\Sigma,P}(\, \met,A \,; O_1\cdots O_s\, )
=\int_{\Gamma(P\tms_H G)}\!\!\!
{\cal D}_{\!\smet}g\,\,  \e^{-kI_{\Sigma,P}(A,g)}\, O_1(g)\cdots O_s(g)\, ,
\label{Feynmanwzw}
\eeq
where $\met$ is a metric on $\Sigma$ and ${\cal D}_{\smet}g$ is the
left-right
invariant measure on the configuration space $\Gamma(P\tms_H G)$
equipped with the
metric induced by $\met$. In the following, the sign ``$\met$" will not
usually be mentioned for simplicity of notation. Suppose that the fields
$O_1\cdots O_n$ are inserted in $\Sigma_{\infty}$ whereas the fields
$O_{n+1}\cdots O_s$ are in $D_0$.
Having in mind the order of integration such that
the last is the integration over
configurations on the circle $S=D_0\cap \Sigma_{\infty}$, we see that
the correlation function is expressed as the pairing
\beq
Z_{\Sigma,P}(A;O_1\cdots O_s)=\langle Z_{\Sigma_{\infty}}(A_{\infty};O_1
\cdots O_n),\gamma_{\infty 0}.Z_{D_0}(A_0;O_{n+1}\cdots O_s)\rangle
\label{pair}
\eeq
of wave functions
\beqa
Z_{\Sigma_{\infty}}(A_{\infty};O_1\cdots O_n)\!&:&\!\gamma\mapsto
\int_{\gamma=g|_S}\!\!\!{\cal D}g \, \e^{-kI_{\Sigma_{\!\infty}}\!
(A_{\!\infty},g)}\, O_1(g)\cdots O_n(g)\, ,
\label{wf*}\\
Z_{D_0}(A_0;O_{n+1}\cdots O_s)\!&:&\!\gamma\mapsto \int_{\gamma=g|_S}\!\!
{\cal
D}g \, \e^{-kI_{D_{0}}(A_{0},g)}\, O_{n+1}(g)\cdots O_s(g)\,,\label{wf}
\eeqa
through the gauge transformation $\gamma_{\infty 0}.$ acting on the wave
functions by
\beq
(\gamma_{\infty 0}.\Phi)(\gamma)=
\gamma_{\infty 0}\Phi(\gamma_{\infty 0}^{-1}
\gamma \gamma_{\infty 0})\gamma_{\infty 0}^{-1}\,.
\label{gauge transformation}
\eeq

\vsp
\noindent The wave functions (\ref{wf}) and (\ref{wf*}) are sections of the
line bundles $\LWZ^k|_{LG}$ and $\LWZ^{*k}|_{LG}$ over $LG$ respectively,
and can be extended to the holomorphic sections over $LG_{\!\C}$. This
observation motivate us to consider the spaces $\Gmhol(\LWZ^k)$ and
$\Gmhol(\LWZ^{*k})$ of holomorphic sections of $\LWZ^k$ and $\LWZ^{*k}$.

\vsp
The group $\tiL G_{\!\C}$ acts on the space
$\Gmhol(\LWZ^k)$ by the left ($J$) and the right ($\bar J$) representations:
\beq
J(\tilde{\gamma}_1)\bar J(\tilde{\gamma}_2)\Phi(\gamma)
=\tilde{\gamma}_1\Phi(\gamma_1^{-1} \gamma\gamma_2^{*-1})\tilde{\gamma}_2^*\,
,
\eeq
where $\{(g,c)\}^*=\{(g^*,c^*)\}$. For any
smooth map $h:D_0\to G_{\!\C}$, the PW identity (\ref{locPW}) together with
the left-right invariance of the measure leads to
\beqa
J(\tilde{\gamma}){\bar J}(\tilde{\gamma})Z_{D_{0}}(A_{0};\,O_a O_b\cdots )
\!\!&=&\!\!Z_{D_{0}}(A_{0}^{h};\,h^{-1}O_a h^{-1}O_b\cdots )\label{bbbb}\\
\noalign{\vskip0.2cm}
;\hspace{2.7cm}&&\hspace{-3.5cm}\tilde{\gamma}^{-1}\,=\,\e^{-I_{D_0}(A_0,h)
-\frac{1}{2}\Gamma_{D_0}(A_0,h,h^*)},
\eeqa
where $h^{-1}O$ is defined by $(h^{-1}O)(g)=O(hgh^*)$. Hence, the
infinitesimal
generators of the representations $J$, $\bar J$ can be identified with
components of the current that are defined as the responses to infinitesimal
variations of the gauge field. The responses to the variations of the metric
under infinitesimal conformal transformations can be identified with
the Fourier components $\{L_n^{G,k}\}$ and $\{\bar L_n^{G,k}\}$
of the Sugawara energy-momentum tensor which is given in (\ref{Sug})
\cite{Witten,KZ}.
These are two copies of representations
of Virasoro algebra with central charge $c_{G,k}=\frac{k\dim G}{k+\hvG}$.

\vsp
We now determine the wave function $\Phi_O=Z_{D_0}(0;O)$
for a field insertion $O$ at $z=0$ in the unit disc $D_0$
with a fixed metric and a gauge field $A_0=0$.
To describe it explicitly, we choose
maximal tori $T_G$ of $G$ and $T=T_G/Z_G$ of $H$
and also a chambre $\Ch$ in $i\liet$
(see Appendix A for notations and basics on the root
system and Weyl groups). These choices determine, for a unitary irreducible
representation $V$ of $G$,
the weight space decomposition and the highest weight $\Lambda$.
We shall describe
the state $\Phi_{\Lambda}=\Phi_{O_{\Lmd}}$ corresponding to the matrix
element $O_{\Lmd}(g)=(v_{\Lmd},g(0)^{-1}v_{\Lmd})$ for the highest weight
vector $v_{\Lmd}\in V$. Let $g_1$ and $g_2$ be holomorphic maps of $D_0$ to
$G_{\!\C}$ such that the value $g_1(0)$ (resp. $\!g_2(0)$) at $z=0$ belongs
to
the Borel subgroup $B$ of $G_{\!\C}$ (resp. the maximal unipotent subgroup
$\!N$ of $B$) that is generated by the Cartan subalgebra $\liet_{\bf c}$ and
the positive root vectors (resp. by only
the positive root vectors). Since these preserve the gauge field $A_0=0$, the
property (\ref{bbbb}) leads to
\beq
J(\e^{-I_{D_0}(g_1)})\bar J(\e^{-I_{D_0}(g_2)})\Phi_{\Lmd}=\e^{\Lmd}(g_1(0))
\Phi_{\Lmd}\,,
\eeq
where $\e^{\Lmd}$ is a character of $B$ for the one dimensional
representation
$\C v_{\Lmd}$. It follows that the value of $\Phi_{\Lmd}$ at the loop
$\gamma_1\gamma_2^*$ ($\gamma_i=g_i|_S$) is given by
\beq
\Phi_{\Lmd}(\gamma_1\gamma_2^*)=\Phi_{\Lmd}(1)\e^{-\Lmd}(g_1(0))
\e^{-kI_{D_0}(g_1g_2^*)}\,,
\eeq
where $\Phi_{\Lmd}(1)\in \LWZ^k|_1$ is a constant that may be put $1$
by a renormalization.
Though any loop in $G_{\!\C}$ is not of the form $\gamma_1\gamma_2^*$
as above, the set $B^+(N^+)^*$ of such loops
is open and dense in $LG_{\!\C}$ \cite{P-S};
by definition, $B^+$ (resp. $\!N^+$) is the subgroup of $LG_{\bf c}$
consisting
of boundary loops of holomorphic maps $D_0\to G_{\bf c}$ such that the values
at $z=0$ are in $B$ (resp. $\!N$).
It is shown in \cite{F-G-K} that $\Phi_{\Lmd}$ extends all over $LG_{\!\C}$
if and only if $\Lmd$ is integrable at level $k$, namely,
\begin{center}
$0\leq (\Lmd,\alpha)\leq k$ for any positive root $\alpha$.
\end{center}
Hereafter, the set of weights integrable at level $k$ is
denoted by $\PPpk$.

\vsp
The state $\Phi_{\Lmd}\in \Gmhol(\LWZ^k)$ generates
an irreducible $\tiL G_{\!\C}\times\tiL G_{\!\C}$ module
${\cal H}^{G,k}_{\Lmd}\subset \Gmhol(\LWZ^k)$ which
is isomorphic to $L_{\Lmd}^{G,k}\ot\overline{L_{\Lmd}^{G,k}}$
where $L_{\Lmd}^{G,k}$
(resp. $\!\overline{L_{\Lmd}^{G,k}}$) is the holomorphic (resp.
anti-holomorphic)
irreducible representation of $\tiL G_{\!\C}$ with highest weight
$(\Lmd,k)$. The subspace
\beq
{\cal H}^{G,k}=\bigoplus_{\Lmd\in \PPpk}{\cal H}^{G,k}_{\Lmd}\,,
\eeq
of $\Gmhol(\LWZ^k)$ is in one to one correspondence under
$\Phi_O\leftrightarrow O$ with the current descendants of
the primary fields $\{O_{\Lmd};\Lmd\in \PPpk\}$.
Though it is not known whether ${\cal H}^{G,k}$
is dense in $\Gmhol(\LWZ^k)$ with respect to some natural topology,
we restrict our attention to this subspace in the rest of the paper.

\vsp
An advantage of this restriction is that the pairing
(\ref{gauge transformation})
can be given a rigorous definition. It is known \cite{P-S} that
$L_{\Lmd}^{G,k}$ is a unitary representation of the subgroup
$\tiL G=\{\tilde{\gamma};
\tilde{\gamma}\tilde{\gamma}^*=1\}\subset \tiL G_{\!\C}$
(the basic central extension of the loop group $LG$),
or equivalently,
there is a hermitian inner product on the space ${\cal H}^{G,k}_{\Lmd}$ such
that
\beq
(J(\tilde{\gamma}_1)\bar J(\tilde{\gamma}_2)\Phi_1,\Phi_2)=(\Phi_1,
J(\tilde{\gamma}_1^*)\bar J(\tilde{\gamma}_2^*)\Phi_2)
\eeq
In addition, an anti-linear map
$\Gmhol(\LWZ^{*k})\to \Gmhol(\LWZ^k)$; $\Psi\mapsto \overline{\Psi}$
is defined by
$\overline{\Psi}(\gamma)=\natural\Psi(\gamma^{*-1})$ where
$\natural:\LWZ^{*k}\to \LWZ^k$ is
the map covering $\gamma\mapsto \gamma^{*-1}$ defined by
\beq
\natural\{(g|_{D_0},c)\}=\{(g^{*-1}|_{D_{\infty}},
c^*\e^{-kI_{{\bf P}^1}(g^{*-1})+2kK_{D_0}(g^{*-1}|_{D_0})})\}\,.
\eeq
With similar
restriction $\check{\cal H}^{G,k}\subset\Gmhol(\LWZ^{*k})$, the pairing
$\langle \Psi,\Phi\rangle
=\int_{LG}\!\!{\cal D}\gamma\langle \Psi(\gamma),\Phi(\gamma)\rangle$
of $\Psi\in \check{\cal H}^{G,k}$ and $\Phi\in {\cal H}^{G,k}$ is
now defined by
\beq
\langle \Psi,\Phi\rangle =(\overline{\Psi},\Phi )\,.
\eeq
This satisfies the property that implies the left-right
invariance of the measure ${\cal D}\gamma$.

\vspace{0.5cm}
{\sc 2.3 The Spectral Flow}

\vspace{0.3cm}
Instead of the flat gauge field $A_0=0$, we next consider the following
configuration. We choose first a real valued smooth function
$\varrho:[0,1+\epsilon]\to[0,1]$ such that $\varrho(r)=0$ for
$0\leq r\leq\epsilon$ and $\varrho(r)=1$ for
$1-\epsilon\leq r\leq 1+\epsilon$ where $\epsilon$ is some number in
$[0,\frac{1}{2})$.
We also choose an element $a$ of $i\liet$ and put
\beq
A_{\varrho,a}
=\varrho(|z|)\frac{a}{2}\left(\frac{d\bar z}{\bar z}-\frac{dz}{z}
\right)=-\varrho(r)iad\theta\,,
\label{config}
\eeq
where $z=r\e^{i\theta}$. If $a$
is in the lattice $\Pv=\frac{1}{2\pi i}\Ker\{\exp:\liet \to T_H\}$,
$A_{\varrho,a}$ has trivial holonomy $\e^{2\pi i a}=1$ along the boundary
circle $S=\partial D_0$ and one can choose a horizontal gauge $s$ over $S$.
It is related to the old standard gauge $s_0$ as $s_0|_S=s\gamma$
by the loop
$\gamma(\theta)=g\e^{-ia\theta}$ in which $g\in H$ is a constant.

With respect to this horizontal gauge $s$,
the state $Z_{D_0}^{(s)}(A_{\varrho,a};O)$ coming
from the disc with field insertion at $z=0$ looks as the gauge transform
(\ref{gauge transformation}) by $\gamma$ of the state
$Z_{D_0}(A_{\varrho,a};O)$ associated to the standard gauge $s_0|_S$.
Let $h_{\varrho,a}:D_0\to H_{\!\C}$ be the solution of
$A_{\varrho,a}^{01}=h_{\varrho,a}\bartial h_{\varrho,a}^{-1}$ such that
$h_{\varrho,a}(0)=1$ and $h_{\varrho,a}(z)=c_{\varrho}^{-a}|z|^{-a}$ around
$S$ with $c_{\varrho}$ a real number.
Making use of (\ref{bbbb}), one can write
$Z_{D_0}(A_{\varrho,a};O)$ as the transform of $\Phi_O$
by a certain element $\tilde{c}_{\varrho,a}\in \tiL G_{\!\C}$ over the
constant loop $c_{\varrho}^{-a}\in T_{\C}$ and it follows that
\beq
Z_{D_0}^{(s)}(A_{\varrho,a};O)=\gamma.J(\tilde{c}_{\varrho,a})
\bar J(\tilde{c}_{\varrho,a})\Phi_O\,.
\eeq
As we see below, the transformation
$\gamma.J(\tilde{c}_{\varrho,a})\bar J(\tilde{c}_{\varrho,a})$
preserves the space ${\cal H}^{G,k}$ and permutes the irreducible components
$\{{\cal H}_{\Lmd}^{G,k}\}_{\Lmd\in\PPpk}$. This is the so-called spectral
flow. This line of argument was first suggested in ref. \cite{Moore-Seiberg}.

\vspace{0.4cm}
{\it Calculation Of $\gamma.\Phi_{\Lmd}$}

\vsp
When $O$ is the primary field $O_{\Lmd}$ with $\Lmd\in \PPpk$,
the corresponding state $\Phi_{\Lmd}$
has a definite weight and the new state is given by
\beq
Z_{D_0}^{(s)}(A_{\varrho,a};O_{\Lmd})=\mbox{const}\,\,\gamma.\Phi_{\Lmd}\,,
\label{locpreFI}
\eeq
where the constant is of the form $\e^{-k\tr( a^2) b_{\varrho}}
c_{\varrho}^{-2\Lmd(a)}$ in which $b_{\varrho}$ depends only on ${\varrho}$.

\vsp
We calculate $\gamma.\Phi_{\Lmd}$ when $\gamma\in LH$ represents an element
of
the group $\Gmalcv$ (see Appendix A) in which case $\ad \gamma$
presevers
the subgroups $B^+$ and $N^+$ of $LG_{\!\C}$ and the calculation becomes
particularly simple. Then, the loop $\gamma$ can be rewritten as
\beq
\gamma(\theta)=\e^{-i\mu \theta}\,n_w\,,
\eeq
where $n_w=g$ represents an element $w$ of the Weyl group and
$\mu=wa$ has value $1$ or $0$ for every positive root.
We denote by $h_{\gamma}(z)$ the holomorphic extension $z^{-\mu}n_w$
of $\gamma$. Note that each connected
component of $LH$ contains loops representing a unique element of $\Gmalcv$.
In \S 4,
we shall make use of such a loop to define a topology changing action
of the fundamental group $\pi_1(H)$ on the set of isomorphism classes
of holomorphic $\HC$-bundles with parabolic structure.

\vsp
It suffices to look at the behavior of $\gamma.\Phi_{\Lmd}$ over the open
dense subset $B^+(N^+)^*$.
Let $\gamma_1\in B^+$ and $\gamma_2\in N^+$ be the boundary loops
of holomorphic maps of $D_0$ to $G_{\!\C}$,
$g_1$ and $g_2$ respectively with
$g_1(0)\in B$ and $g_2(0)\in N$. Since $\ad \gamma$ preserves the subgroups
$B^+$ and $N^+$, holomorphic functions $h_{\gamma}^{-1}g_1h_{\gamma}$ and
$h_{\gamma}^{-1}g_2 h_{\gamma}$ are defined on $D_0$ and satisfy
$(h_{\gamma}^{-1}g_1h_{\gamma})(0)\in B$ and $(h_{\gamma}^{-1}g_2 h_{\gamma})
(0)\in N$. Hence we have
\beq
\gamma.\Phi_{\Lmd}(\gamma_1\gamma_2^*)=\e^{\!-\Lmd}\left((h_{\gamma}^{-1}g_1
h_{\gamma})(0)\right)\ad \gamma\!\left(\e^{-kI_{D_0}\left((h_{\gamma}^{-1}g_1
h_{\gamma})(h_{\gamma}^{-1}g_2h_{\gamma})^*\right)}\right)\,.
\eeq
If we put $g_1(0)\equiv \e^{t_0}\in T$ mod $N$, we find that
$(h_{\gamma}^{-1}
g_1 h_{\gamma})(0)\equiv \e^{w^{-1}t_0}$ and hence
\beq
\e^{\!-\Lmd}\left((h_{\gamma}^{-1}g_1 h_{\gamma})(0)\right)
=\e^{\!-w\Lmd(t_0)}
\,.
\eeq
Applying the transformation rule (\ref{liftad}) of $\ad \gamma$,
we find that
\beqa
\ad\gamma\,(\e^{-kI_{D_0}(h_{\gamma}^{-1}g_1h_{\gamma})})\!&=&\!
\e^{-k\,\tr(\mu t_0)}\e^{-kI_{D_0}(g_1)}\,,\\
\ad\gamma\,(\e^{-kI_{D_0}(h_{\gamma}^{-1}g_2h_{\gamma})})\!&=&\!
\e^{-kI_{D_0}(g_2)}\,.
\eeqa
Combining these results, we obtain the expression
\beq
\gamma.\Phi_{\Lmd} (\gamma_1\gamma_2^*)=\e^{-w\Lmd(t_0)-k\,\tr(\mu t_0)}
\e^{-kI_{D_0}(g_1g_2^*)}\,.
\eeq
Thus, the result is $\gamma.\Phi_{\Lmd}=\Phi_{\gamma\Lmd}$,
the vector of highest weight
\beq
\gamma \Lmd=w\Lmd+k\ttr\mu\,,
\eeq
in which $\ttr\mu$ denotes the weight $\ttr\mu(v)=\tr(\mu v)$. Indeed,
if $\gamma$ represents an element of $\Gmalcv$,
the transformation $\Lmd\mapsto \gamma \Lmd$ preserves
the set $\PPpk$ of integrable weights.

\vsp
This transformation of $\PPpk$ looks simple with respect to the
fundamental affine weights $\hat{\Lmd}_0,\cdots,\hat{\Lmd}_l\in\hat{\V}^*$
related to the simple affine roots by $2(\hat{\Lmd}_i,\hat{\alpha}_j)/||
\hat{\alpha}_j||^2=\delta_{i,j}$. Since $\Gmalcv$ is an automorphism group of
the extended Dynkin diagram, or an orthogonal group of permutations of
simple affine roots; $\gamma\hat{\alpha}_i=\hat{\alpha}_{\gamma i}$, we find
that $\gamma\in \Gmalcv$ permutes the fundamental affine weights modulo
$\R\times 0\times 0$. Hence, denoting the highest weight $(\sDelta_{\Lmd},
\Lmd,k)$ by $\hat{\Lmd}$, the transformation is written as
\beq
\hat{\Lmd}=\sum_{i=0}^ln_i\hat{\Lmd}_i\mapsto \hat{\gamma\Lmd}
=\sum_{i=0}^ln_i\hat{\Lmd}_{\gamma i}\qquad \mbox{mod}\,\,\R\times 0\times 0
\,.
\eeq

\vsp
{\it Remark}. This gauge transformation $\gamma.:{\cal H}^{G,k}\to
{\cal H}^{G,k}$ induces the external automorphism of
the Virasoro-Kac-Moody algebra. In fact, the spectral flow may be
considered as the consequence of such an algebra automorphism.

\vspace{0.3cm}
{\it Non-Abelian Insertion Theorem}

\vsp
Let $P$ be the principal $H$-bundle over $\Sigma$ with a connection $A$ which
is flat on the unite disc $D_0\subset \Sigma$ in a coordinatized subset.
We choose a horizontal gauge
$\sigma_0$ over $D_0$.
Gluing $(D_0\times H,A_{\varrho,a})$ to
$(P|_{\Sigma_{\infty}},A|_{\Sigma_{\infty}})$
at the boundaries by the identification
$(\e^{i\theta},1)\equiv \sigma_0(\e^{i\theta})\gamma(\theta)$, we obtain
another $H$-bundle $P\gamma$ with a connection $A^{\gamma}$.

Applying the pairing formula (\ref{pair}) to
$Z_{\Sigma_{\infty}}(A_{\infty};O_1\cdots O_s)$
and $\gamma.Z_{D_0}(A_{\varrho,a};O_{\Lmd})$
and using the above result $\gamma.\Phi_{\Lmd}=\Phi_{\gamma\Lmd}$,
we see that
\beq
Z_{\Sigma,P\gamma}(A^{\gamma};O_1\cdots O_s O_{\Lmd})
={\rm const}\cdot Z_{\Sigma,P}(A;O_1\cdots O_s O_{\gamma\Lmd})\,,
\label{preFI}
\eeq
where the constant is the same as the one in (\ref{locpreFI}).
This may be considered as the prototype of (\ref{FI}).
Equation of the same kind is already known
in the free fermionic (bosonic) system as the insertion theorem \cite{ABMNV}.

\renewcommand{\theequation}{3.\arabic{equation}}\setcounter{equation}{0}

\vsp
\begin{center}
{\bf 3. Integration Over Gauge Fields}
\end{center}

Let $H$ be a connected, closed subgroup of $G'=G/Z_G$.
We denote by $\AP$ the set of connections of a
principal $H$-bundle $P$ over a Riemann surface $\Sigma$ and by $\GP$ the set
of sections of the adjoint $H$-bundle $P\tms_H H$
which acts on $\AP$ as the {\it gauge
transformation group}. In this section, we turn to the quantization of the
gauged WZW model with target group $G$ and gauge group $H$. We
develop a method to perform the integration
\beq
Z_{\Sigma,P}(O_1\cdots O_s)=\frac{1}{\vol \GP}
\int_{\AP}\!\!{\cal D}\!A\,Z_{\Sigma,P}^{G,k}
(A\,;O_1\cdots O_s)\,,
\label{intgauge}
\eeq
of the WZW correlator (\ref{Feynmanwzw}) of the gauge invariant fields
$O_1\cdots O_s$.(\footnote{
The superscript ``$G,k$'' is introduced for specification since we shall
consider several different groups $H$, $\HC$, $G$ etc. at the same time.
The $H$-bundle $P$ and its connection $A$ under the superscript are
prescribed to mean the extension to $G'$-bundle and $G'$-connection.
})

The method takes advantage of the chiral gauge symmetry
\beq
Z_{\Sigma,P}^{G,k}(A^h;O_1\cdots O_s)=\e^{kI_{\Sigma,P}^G(A,hh^*)}
Z_{\Sigma,P}^{G,k}(A;hO_1\cdots hO_s)\,,
\label{chiral sym}
\eeq
for a section $h$ of the adjoint $\HC$-bundle $P\tms_H \HC$,
which is a consequence of the PW identity (\ref{PW}). We integrate first over
each orbit of the group $\GPC=\Gamma(P\tms_H \HC)$ of
chiral gauge transformations, and then over the orbits.
One can see that $\AP$ contains
a submanifold $\A_{ss}$ with the complement of codimension $\geq 1$
such that the orbit space $\A_{ss}/\GPC$ is approximately a finite
dimensional compact space $\NNP$
with a preferable structure.
Change to the parametrization of $\AP$ in terms of $\GPC$ and $\NNP$
induces the Jacobian factor that can be represented by the spin $(1,0)$
ghost system with values in the adjoint $\h_{\bf C}$-bundle $\ad P_{\!\C}$.
The integration over $\GPC$ mod $\GP$ leads to a
sigma model with the target space $H_{\!\C}/H$. Consequently, the
correlation function (\ref{intgauge}) is expressed as the integration over
$\NNP$ of a correlation function of the three systems coupled to common
representative gauge field --- the WZW model with the target $G$, the sigma
model with the target $H_{\!\C}/H$ and the ghost system valued in
the adjoint bundle.

\vspace{0.5cm}
{\sc 3.1 The Space Of Gauge Fields}

\vspace{0.3cm}
We give a description of the structure of $\GPC$-orbits in $\AP$ and argue
that we can neglect some orbits in the integration (\ref{intgauge}).
To start with, we note that a connection
$A$ of $P$ determines a unique holomorphic structure $\bartialA$ of the
complexified $\HC$-bundle $P_{\!\C}$: A local section $\sigma$ of $P_{\!\C}$
is holomorphic if $\bartialA\sigma=0$ when $\sigma$ is represented as a local
frame of the vector bundle associated to $\PC$ with a holomorphic
representation of $\HC$. Conversely, any holomorphic structure of $\PC$ is
obtained in this way. Since two connections $A$ and $A^h$ related by a chiral
gauge transformation $h\in \GPC$ correspond to the isomorphic holomorphic
structures $\bartialA$ and $h^{-1}\bartialA h$, we can identify the orbit set
$\AP/\GPC$ with the set of isomorphism classes of holomorphic structures of
$\PC$. Such an identification makes easy the explicit description of the
orbits for genus zero and makes possible for genus $\geq 1$
to use the well-known techniques in analytic and algebraic geometry
such as the Riemann-Roch theorem, the Atiyah-Bott stratification
and especially the Narasimhan-Seshadri theorem.

It should be noticed that the space $\AP$ is given a complex structure
$J_{\Sigma}$ so that the $\GPC$-action is holomorphic: On each tangent space
$\Omega^1(\Sigma,\ad P)$ which is the set of one forms on $\Sigma$ valued in
the adjoint bundle, $J_{\Sigma}$ acts as the Hodge $*$-operator; $*Xd\bar
z=iXd\bar z$, $*\bar X dz=-i\bar X dz$.

\vspace{0.4cm}

{\it On The Sphere}

\vsp
We begin with the case in which $\Sigma$ is the Riemann sphere $\CP$. It is
covered by the $z$-plane $U_0$ and $w$-plane $U_{\infty}$ where $z$ and $w$
are related by $zw=1$. We denote by $D_0$ (resp. $\!D_{\infty}$) the unit
disc in the $z$-plane (resp. $\!w$-plane).

\vsp
{\it For $H=U(1)$}:  A holomorphic $\HC=\C^*$-bundle
admits local sections $\sigma_0$ and $\sigma_{\infty}$ over
open neighborhoods of $D_0$ and $D_{\infty}$ respectively. If they are
related by the holomorphic transition function $h_{\infty 0}$
on a neighborhood of $S=D_0\cap D_{\infty}$
\beq
\sigma_0(z)=\sigma_{\infty}(z)h_{\infty 0}(z)\,,
\eeq
the winding number $a=\frac{i}{2\pi}
\oint_S h_{\infty 0}^{-1}dh_{\infty 0}\in \Z$ determines
the topological type.
Taking the Laurent expansion of the function $\log \{h_{\infty 0}(z)z^a\}$,
we find that
\beq
h_{\infty 0}(z)=h_{\infty}(z)z^{-a}h_0(z)^{-1}\,,
\label{FCT}
\eeq
where $h_0$ and $h_{\infty}$ are $\C^*$-valued holomorphic functions
on neighborhoods of $D_0$ and $D_{\infty}$ respectively. Hence, we
can always take the transition function of the form $z^{-a}$.
In other words, for a $U(1)$-bundle $P$,
\begin{center}
$\AP$ is itself a single $\GPC$-orbit.
\end{center}

\vsp
{\it For $H=SU(n)/\Z_n$}:  We next consider the group
$SU(n)/\Z_n$ where $\Z_n$ is the center of $SU(n)$ consisting of identity
matrices multiplied by $n$-th roots of unity. The property (\ref{FCT}) holds
also in this case provided that $a$ is an
element of $\Pv$, that is, $a$ is of the form
\beq
a=\left(\begin{array}{cccc}
a_1&   &   \\
       &\ddots&   \\
     &     & a_n
\end{array}\right)    \,\,\,\mbox{with}\,\,\, a_i+\frac{j}{n}\in{\Z}\,\,(i=1,
\cdots,n)\,\,\,\mbox{and}\,\, \sum_{i=1}^n a_i=0\,,
\label{adiag}
\eeq
for some $j\in \indJ=\{0,1,\cdots,n-1\}$. This is due to the Birkhoff
factorization theorem \cite{Grothendieck,P-S}
which also states that such $a$ is
unique up to permutations of $a_1,\cdots,a_n$. Hence, holomorphic
$\HC$-bundles over $\CP$ are classified by the countable set $\Pv/W$
in which
$W$ the Weyl group of $H$ acts on diagonal matrices as permutations of
diagonal entries. Note that the loop $\e^{i\theta}\mapsto \e^{-ia\theta}$
extends to a map on $D_0$ with values in $H$ if and only if all $a_i$ are
integers. Thus the topological type of the bundle is determined by
the number $j\in\indJ$.
Stated in another way, for each $j\in \indJ$, there is an $H$-bundle
$P^{(j)}$ and its complexification admits countably many
holomorphic structures classified by $\Pv_j/W$ in which $\Pv_j$ is
the set of matrices in $\Pv$ whose diagonal entries differ from
$-\frac{j}{n}$ by integers. Since the set $\overline{\Ch}$ of
diagonal matrices $t$ with $t^1_1\geq\cdots \geq t^n_n$ is
a fundamental domain of $W$, we see that
\beq
\A_{P^{(j)}}=\bigcup_{a\in \Pv_j\cap \overline{\Ch}}\A_a\,,
\label{stratP1}
\eeq
where $\A_a$ is the $\G_{P^{(j)}_{\bf c}}$-orbit corresponding to the
holomorphic $\HC$-bundle $\Ph_{[a]}$ with the transition function
$h_{\infty 0}(z)=z^{-a}$.

Though each $\A_a$ is infinite dimensional, one can compare
the dimensions of these orbits {\it relative to} $\G_{P^{(j)}_{\bf c}}$.
That is to consider the the group $\Aut\,\Ph_{[a]}$ of
holomorphic automorphisms of $\Ph_{[a]}$ that corresponds to
the isotropy group of $\G_{P^{(j)}_{\bf c}}$ at a point of $\A_a$.
An element $f$ of $\Aut\,\Ph_{[a]}$ is given by $\HC$-valued
holomorphic functions $f_0$ and $f_{\infty}$ on $U_0$ and $U_{\infty}$
respectively such that $f_0(z)=z^a f_{\infty}(z)z^{-a}$ on
$U_0\cap U_{\infty}$. We find that $(f_0(z))^i_j$ is a span of
$1,z,\cdots,z^{a_i-a_j}$ if $a_i\geq a_j$ and is zero if $a_i<a_j$.
The dimension of $\Aut\,\Ph_{[a]}$ is thus given by
$n-1+\sum_{i<j}(\delta_{a_i,a_j}+1+|a_i-a_j|)$ and is minimized
in $\Pv_j\cap \overline{\Ch}$ by $a=\mu_j$ where
$(\mu_j)_i=1-\frac{j}{n}$ for $i=1,\cdots,j$ and $(\mu_j)_i=-\frac{j}{n}$
for $i=j+1,\cdots,n$. Hence $\A_{P^{(j)}}$ contains an orbit $\A_{\mu_j}$ of
maximal dimension and another orbit $\A_a$ has codimension $d_a>0$ given by
\beq
d_a=\sum_{i<j}(\delta_{a_i,a_j}-1+|a_i-a_j|)=\sum_{a_i>a_j}(a_i-a_j-1)\,.
\eeq
Therefore, in the integration (\ref{intgauge}) for $P=P^{(j)}$, we have only
to take into account the single orbit $\A_{\mu_j}$.

\vsp
{\it For general $H$}: We follow the preceding argument
using the notation of Appendix A. For each $j\in\indJ$,
there is an $H$-bundle $P^{(j)}$ with the transition function
$\e^{-i\mu_j\theta}$ and any $H$-bundle is isomorphic to $P^{(j)}$
for some $j\in\indJ$. The set of connections of $P^{(j)}$ is decomposed
as the disjoint union of the form (\ref{stratP1}) in which $\Pv_j=\mu_j+\Qv$,
$\overline{\Ch}$ is the closure of a chambre $\Ch$ in $i\liet$ and
$\A_a$ is
the $\G_{P^{(j)}_{\bf c}}$-orbit corresponding to a holomorphic bundle
with the transition function $z^{-a}$. The orbit $\A_{\mu_j}$ is of maximal
dimension and the codimension of $\A_a$ is
\beq
d_a=\sum_{\alpha(a)>0}(\alpha(a)-1)
\eeq
where $\alpha$ in the sum runs over roots of $H$. Since $d_a\geq 1$ for
$a\ne \mu_j$, we may replace $\A_{P^{(j)}}$ by $\A_{\mu_j}$
in the integration (\ref{intgauge}) for $P=P^{(j)}$.

\vspace{0.4cm}
{\it On A Surface Of Genus $\geq 1$}

\vsp
For a Riemann surface $\Sigma$ of genus $\geq 1$, the set of orbits
$\AP/\GPC$ is not in general countable. This can be seen by looking at
the index $\dim H(1-g)$ of the operator $\bartialA :\Omega^0(\Sigma,\ad \PC)
\to \Omega^{0,1}(\Sigma,\ad \PC)$ which counts the dimension of
the symmetry group of $\bartialA$ minus the codimension of
the $\GPC$-orbit through $A$.

\vsp
{\it For $H=U(1)$}: Let ${\cal O}$
(resp. $\!{\cal O}^{\times}$) be the sheaf of germs of holomorphic functions
valued in $\C$ (resp. $\!\C^*$). The set of isomorphism classes of
holomorphic principal $\C^*$-bundles is identified with
the sheaf cohomology group $H^1(\Sigma,{\cal O}^{\times})$,
the Picard group $\Pic(\Sigma)$.
The long exact sequence induced by
the homomorphism ${\cal O}\to {\cal O}^{\times}$; $f\mapsto \e^{2\pi i f}$
with kernel $\Z$ gives the following description of $\Pic (\Sigma)$:
\beq
0\to \Jac(\Sigma) \to \Pic(\Sigma) \stackrel{c_1}{\to}{\Z}\to 0\,,
\label{descript.pic}
\eeq
where the projection $c_1$ counts the first chern class
and $\Jac( \Sigma)$ is
the Jacobian variety $H^1(\Sigma,{\cal O})/H^1(\Sigma, \Z)$ which is a
complex torus of dimension $g$.

 For each topological type $a\in \Z$, a choice $\Ph\in \Pic( \Sigma)$ with
$c_1(\Ph)=a$ determines an isomorphism of $\Jac (\Sigma)$ and the set
${c_1}^{-1}(\{a\})$ of holomorphic $\HC$-bundles of 1-st chern class  $a$.
Thus, for any $U(1)$-bundle $P$,
\beq
\AP/\GPC \cong \Jac (\Sigma) \,\, \,(\mbox{a complex $g$-torus})\,.
\eeq
In particular, even if $P$ and $P'$ are topologically distinct, there are
isomorphisms of $\AP/\GPC$ and $\A_{P'}/\G_{P'_{\bf c}}$. In \S 4, we shall
use a certain isomorphism to prove (\ref{FI}).

\vsp
{\it For general $H$}: If $H$ is non abelian, the situation
is a little different. For simplicity of the discussion, we assume that $H$
is simple. We make use of the following stratification (decomposition into
submanifolds) of the space of connections of a principal $H$-bundle $P$ over
$\Sigma$ which is due to Atiyah and Bott \cite{A-B}:
\beq
\AP=\bigcup_{\mu}\A_{\mu}\,.
\label{stratgen}
\eeq
This generalizes the disjoint union (\ref{stratP1}) for genus zero.
Here, $\mu$ runs over a discrete subset of $\overline{\Ch}$ and $\A_{\mu}$ is
a $\GPC$-invariant submanifold of $\AP$ of codimension
\beq
d_{\mu}=\sum_{\alpha(\mu)>0}(\alpha(\mu)+g-1)\,.
\eeq
The unique solution to $d_{\mu}=0$ is $\mu=0$ for genus $\geq 1$.
It is known \cite{A-B}
that $A\in \A_0$ if and only if the adjoint bundle $\ad \PC$ with
the holomorphic structure $\bartialA$ is {\it semi-stable}, namely, any
holomorphic subbundle has non-positive first chern class. In view of this
characterization, we hence-forth denote $\A_0$ by $\Ass$ ($ss$
means ``semi-stable''). The space $\Ass$ contains a $\GPC$-invariant,
open and dense submanifold $\Assc$ such that
the quotient $\Assc/\GPC$ is a non-empty
complex manifold whose dimension $\dN$ is $\dim H(g-1)$ for genus $\geq 2$
and is between $0$ and ${\rm rank} H$ for genus one. Hence, we may replace
$\AP$ by $\Assc$ in the integration (\ref{intgauge}).
A compactification of $\Assc/\GPC$ is given by
the quotient $\NNP=\Ass\dslash \GPC$ of $\Ass$ under
a certain equivalence relation.
The theorem of Narasimhan and Seshadri
\cite{N-S,Donaldson,Ramanathan} essentially states that
the set $\A_F$ of flat connections is included in $\Ass$ and the inclusion
map induces the identification of the moduli space $\A_F/\GP$ of flat
connections and the moduli space $\NNP=\Ass\dslash \GPC$ of semi-stable
$\HC$-bundles.

\vspace{0.4cm}

{\it Example --- Flat $SO(3)$-Connections over the Torus}

\vsp
We explicitly describe the moduli spaces of flat connections of the trivial
and the non-trivial $H=SO(3)$-bundles on the torus
$\Sigma_{\tau}=\C/({\Z}+\tau{\Z})$ of period $1$ and $\tau$ where
$\tau_2={\rm Im}\tau >0$.
We denote by $\z$ the coordinate of this plane $\C$.
The homology base $A$, $B : [0,1]\to \Sgmtau$ defined by $\z(A_t)=t$
and $\z(B_t)=t\tau$ provides a set of generators of the
fundamental group $\pi_1\Sigma={\Z}^2$. A flat connection of
an $H$-bundle $P$ defines (up to conjugation) a holonomy representation
$\rho : \pi_1\Sigma \to H$.
It is determined by the commuting elements $a=\rho(A)$ and
$b=\rho(B)$ of $H$.

If $P$ is trivial, $a$ and $b$ are represented by commuting elements
$\tila$ and $\tilb$ of $\tilH=SU(2)$.
By conjugation, we can bring them to diagonal matrices
\beq
\tila=\pmatrix{
e^{2\pi i \phi} & 0 \cr
0 & e^{-2\pi i\phi}\cr
}\,,\qquad
\tilb=\pmatrix{
e^{2\pi i\psi} & 0 \cr
0 & e^{-2\pi i\psi} \cr
}\,.
\label{holonomtriv}
\eeq
Such holonomy is provided by the gauge field of the following form:
\beq
A_u=\Bigl(\frac{\pi}{\tau_2}u\,d\bar \zeta-\frac{\pi}{\tau_2}\bar u\,
d\zeta\Bigr)\pmatrix{
1 & 0 \cr
0 & -1 \cr
}\,,
\label{flatconnSO(3)triv}
\eeq
where $u=\psi-\tau \phi$.
This $u$ can be considered as a holomorphic parameter.
$A_{u'}$ is gauge equivalent to $A_u$ if and only if
$u'=\pm u -\frac{m}{2}+\tau \frac{n}{2}$ for some $n$, $m\in {\Z}$.
Hence, the moduli space is given by
\beq
\NN_{{\rm triv}}=\C\Lslash \!\!\left\{\bigl(\mbox{$\nibun{\Z}+
\frac{\tau}{2}{\Z}$}\bigr)\semidir \{\pm1\}\right\}\,.
\eeq
It is an orbifold with four singularities
$u\equiv 0,\frac{1}{4},\frac{\tau}{4},\frac{\tau+1}{4}$ of order $2$.
The manifold $\Assc/\GPC$ in this case is
$\NN_{\rm triv}$ with these singular points deleted.

If $P$ is non-trivial, $a$ and $b$ are represented by elements $\tila$,
$\tilb$ of $\tilH=SU(2)$ that do not commute but satisfy
\beq
\tila \tilb\tila^{-1}\tilb^{-1}=\pmatrix{
-1 & 0 \cr
0 & -1 \cr
}\,.
\eeq
There is only one such pair $(\tila,\tilb)$ modulo conjugation:
\beq
\tila=\pmatrix{
i & 0 \cr
0 & -i \cr
}\,,\qquad
\tilb=\pmatrix{
0& -1 \cr
1 & 0\cr
}\,.
\label{holonomnontriv}
\eeq
Hence, on the torus,
\beq
\NN_{{\rm non-triv}}=\{\mbox{one point}\}\,.
\eeq

In contrast with the abelian case, $\NN_{{\rm triv}}$
is not isomorphic to $\NN_{{\rm non-triv}}$ and even the dimensions are
different. For a general semi-simple group $H$,
the moduli space of topologically trivial
semi-stable $\HC$-bundles over the torus $\Sgmtau$ is
\beq
\NN_{{\rm triv}}=\liet_{\bf c}\Lslash(\Pv+\tau\Pv)\semidir W\,,
\eeq
and hence of dimension ${\rm rank}H$. But for each $j\in {\cal J}$ in the
terminology of Appendix A, we have a non-trivial $H$-bundle $P^{(j)}$ and
we can see that $\dim\NN_{P^{(j)}}=\dim\Ker(w_jw_0-1)$
which is strictly less than the rank of $H$.

\vspace{0.5cm}
{\sc 3.2 The Path Integration}

\vspace{0.3cm}
To define the measure for the integration (\ref{intgauge}), we introduce
metrics on the spaces $\AP$ and $\GPC$. We identify the tangent spaces
at $A\in \AP$ and $h\in \GPC$ as
\beq
T_{\!A}^{1,0}\!\!\AP\cong\Omega^{0,1}(\Sigma,\ad P_{\!\C})\,,
\qquad T_h^{1,0}\!\GPC\cong\Omega^0(\Sigma,\ad P_{\!\C})\,,
\eeq
where $a\in \Omega^{0,1}(\Sigma,\ad P_{\!\C})$ is tangent to the curve
$\bartial_{\!A_t}=\bartial_{\!A}+ta$ at $t=0$ and $\epsilon\in
\Omega^0(\Sigma,\ad P_{\!\C})$ is tangent to the curve
$h_t=he^{t\epsilon+\bar t \epsilon^*}$ at $t=0$. We define inner products
on those spaces by $(a_1,a_2)=\frac{i}{2\pi}\int_{\Sigma}\trP(a_1^*a_2)$
and by $(\epsilon_1,\epsilon_2)
=\frac{1}{2\pi}\int_{\Sigma}*\trP(\epsilon_1^*\epsilon_2)$. Then,
$\AP$ becomes a $\GP$ invariant K\"ahler manifold and $\GPC$ becomes
a Hermitian manifold invariant under the left action of $\GPC$ and
the right action of $\GP$.

\vspace{0.3cm}
{\it Local Parametrization of Gauge Fields}

\vsp
As is noticed above, we may replace $\AP$ in the integration (\ref{intgauge})
by a submanifold $\APc$ whose $\GPC$-quotient $\NNPc$ is a complex manifold
of dimension $\dN$. For $H=U(1)$, $\APc$ is $\A_P$ itself and $\NNPc$ is
a complex $\dN=g$-torus. In general, we put

$(g=0)$\, $\APc=\A_{\mu_j}$ for $P=P^{(j)}$, \,\, $\NNc_P=$ one point,\,\,
$\dN=0$,

$(g=1)$\, $\APc=\Assc$\,, \,$\NNPc=\Assc/\GPC\subset \NN_P$\,, \,
$0\leq \dN \leq {\rm rank}H$,

$(g\geq 2)$\, $\APc=\Assc$\,,\, $\NNPc=\Assc/\GPC\subset \NN_P$\,, \,
$\dN={\rm dim}H(g-1)$.

\noindent For every point $u_0\in \NNPc$, we can take a neighborhood $U$ of
$u_0$ in $\NNc_P$  with a holomorphic family $\{A_u\}_{u\in U}$ of
representatives, that is, a holomorphic map $U\to \APc$;  $u\mapsto A_u$
such that the $\GPC$-orbit through $A_u$ is $u$. We denote by $\A_U$
the inverse image of $U$ by the quotient map $\APc\to \NNPc$ and define
a surjective map
\beq
f:U\times \GPC\longto \A_U\,\quad\mbox{by\, $f(u,h)=A_u^h$.}
\eeq
This is not injective if the symmetry group $S_u=\Aut\,\bartial_{\!A_u}$ is
non-trivial, in particular if its dimension $d_S=\dN+\dim H(1-g)$ is
non zero which is the generic situation for $g=0, 1$.

Let $(u^1,\cdots, u^{\dN})$ be a complex coordinate system on $U$.
The differentials
\beq
\lnu_{\!a}(u)=(\partial/\partial u^a)A_u^{01}\,\,\quad a=1,\cdots,\dN\,,
\eeq
provide a base of the tangent space
$T^{1,0}_u U=H^{0,1}_{\bartial_{\!A_u}}\!(\Sigma,\ad P_{\!\C})$.
We also choose a base $\{a^b(u)\}_{b=1}^{\dN}$ of the cotangent space
$(T^{1,0}_u U)^*=H_{\bartial_{\!A_u}}^{1,0}\!(\Sigma,\ad P_{\!\C})$
and a base $\{\epsilon_i(u)\}_{i=1}^{d_S}$ of the space
${\rm Lie}(S_u)=H_{\bartial_{\!A_u}}^0\!\!(\Sigma, \ad P_{\!\C})$ of
infinitesimal symmetries of $\bartial_{\!A_u}$.

At the point $f(u,h)=A_u^h$, we choose an orthonormal base
$\{ a_n(u,h)\}_{n=1}^{\infty}$ of the tangent space
$\bartial_{\!A_u^h}\Omega^0 (\Sigma,\ad \PC)$ of the $\GPC$-orbit through
$A_u^h$ and an orthonormal base $\{\epsilon_n(u,h)\}_{n=1}^{\infty}$ of
the orthogonal complement of $h^{-1}{\rm Lie}(S_u) h$
in $\Omega^0(\Sigma,\ad P_{\!\C})$. Putting
$a_{b-\dN}(u,h)=h^*a^b(u)^*h^{*-1}$ and
$\epsilon_{i-d_S}(u,h)=h^{-1}\epsilon_i(u)h$, we have a base
$\{\,a_n(u,h)\,\}_{n=1-\dN}^{\infty}$ of the tangent space of $\AP$ at
$A_u^h$ and a base $\{\,\epsilon_n(u,h)\,\}_{n=1-d_S}^{\infty}$ of the
tangent space of $\GPC$ at $h$.

Let $x=(x^{1-\dN},\cdot\cdot,x^0,x^1,\cdots)$ and
$t=(t^{1-d_S},\cdot\cdot,t^0,t^1,\cdots)$ be the complex coordinate systems
on neighborhoods of $A_u^h$ in $\A_U$ and of $h$ in $\GPC$ defined by
\beqa
A(x)^{01}&=&
\bigl(A_u^h\,\bigr)^{01}+\sum_{n=1-\dN}^{\infty}x^n a_n(u,h)\,,\\
\mbox{and}\quad\qquad h(t)&=&
{\rm exp}\biggl\{\sum_{i=1}^{d_S}t^{i-d_S}\epsilon_i(u)\biggr\}h\,
{\rm exp}\biggl\{\sum_{n=1}^{\infty}t^n\epsilon_n(u,h)\biggr\}\,.
\eeqa
Then, the pull backs of differentials $dx^n$ by the map $f$ are expressed as
\beqa
f^*dx^{a-\dN}&=&\sum_{b,c=1}^{\dN}M^{a\bar b}(u,h)\!
\left\langle a^b(u),\lnu_{\!c}(u)\right\rangle du^c\,,\label{fdta}\\
f^*dx^n&=&\sum_{m=1}^{\infty}\left(a_n(u,h),\bartial_{\!A_u^h}
\epsilon_m(u,h)\right)dt^m
+\sum_{c=1}^{\dN}\left(a_n(u,h),h^{-1}\lnu_{\!c}(u)h\right)du^c\,,
\hspace{1cm}
\label{fdtn}
\eeqa
where $M^{a\bar b}(u,h)$ is the inverse matrix of
$M_{\bar ab}(u,h)=\Bigl(a_{a-\dN}\!(u,h)\!,a_{b-\dN}\!(u,h)\Bigr)$ and
$\langle\,\,\,,\,\,\,\rangle$ is the natural pairing given by
$\langle a,\lnu\rangle=\nipi\int_{\Sigma}\tr(a\lnu)$.

\vspace{0.4cm}
{\it The Measure ${\cal D}A$}

\vsp
The pull back by $f$ of the volume element ${\cal D}A=
\det M_{\bar a b}(u,h)\prod_{n=1-\dN}^{\infty}d^2\!x^n$
of $\AP$ (where $d^2\!x^n=idx^n\wedge d\bar x^n$) is then given by
\beqa
f^*{\cal D}A_{(u,h)}&=&\frac{\left\vert \,
\det\!\left\langle a^b(u),\lnu_{\!c}(u)\right\rangle
\right\vert^2}{\det M_{\bar a b}(u,h)}\prod_{c=1}^{\dN}\dd u^c
\,\,{\det}'\Bigl(\,\bartial_{\!A_u^h}^{\dag}\bartial_{\!A_u^h}\,\Bigr)
\prod_{n=1}^{\infty}\dd t^n  \nonumber \\
&=&\frac{\left\vert \,\det\!\left\langle a^b(u),\lnu_{\!c}(u)\right\rangle
\right\vert^2}{\det\Bigl(a^a(u),a^b(u)\Bigr)}
\prod_{c=1}^{\dN}\dd u^c
\,\,{\det}'\Bigl(\,\bartial_{\!A_u}^{\dag}\bartial_{\!A_u}\,\Bigr)
\frac{1}{\det\Bigl(\epsilon_i(u),\epsilon_j(u)\Bigr)}\nonumber\\
&&\times \,\det S(u,h) \prod_{n=1}^{\infty}\dd t^n \,\,
\e^{I_{\Sigma,P}^{\ad}(A_u,hh^*)}
\label{pullback}
\eeqa
where $S_{\bar i j}(u,h)=(\epsilon_{i-d_S}(u,h),\epsilon_{j-d_S}(u,h))$ and
${\det}'(D^{\dag}\!D)$ denotes the regularized determinant of $D^{\dag}\!D$
restricted to its positive eigen-spaces. The factor
$\e^{I_{\Sigma,P}^{\ad}(A_u,hh^*)}$ is the chiral anomaly of the
infinite determinant which shall be written down shortly.

\vsp
If the dimension $d_S\ne 0$,
$f^*{\cal D}\!A$ has `lower' degree compared to the volume element
$\prod\dd u^c{\cal D}h
=\prod \dd u^c\det S(u,h)\prod_{n=1-d_S}^{\infty}\dd t^n$
of $U\times\GPC$. In order to deal with such a case, we assume
that there is a function $F_u:\GPC\to \C^{d_S}$ with the
following property: On each $S_u$-orbit in $\GPC$, $F_u$ takes the value
zero at one and only one point and that, at each zero point $h$, the
differential $F_{u,h}:{\rm Lie}(S_u)\to {\C}^{d_S}$ defined by
$F_{u,h}(\epsilon)=\left.\frac{d}{dt}\right|_{0}F_u(e^{t\epsilon}h)$ is a
linear isomorphism. Then, the last factor $\det S(u,h)\prod\cdots$
in (\ref{pullback}) can be identified with the volume form
\beq
\int_{S_u}{\cal D}h\,\, \delta^{(2d_S)}\Bigl(F_u(h)\Bigr)\!\left\vert \,
\det\!\left(F^i_{u,h}\bigl(\epsilon_j(u)\bigr)\right)\!\right\vert^2
\e^{I_{\Sigma, P}^{\ad}(A_u,hh^*)}
\label{resGF}
\eeq
on $S_u\backslash \GPC$. Since the role of $F_u$ is to fix the
`gauge degrees of freedom' $S_u$, we call it the {\it residual gauge fixing
function}.

{\it Remark}. The assumption of the existence of the residual gauge fixing
function on the whole space $\GPC$ may fail. However, we can find at least
a function $F_u$ defined on a neighborhood of a (local)
section of the fibre bundle $\GPC \to S_u\backslash \GPC$ and the factor
$\delta^{(2d_S)}\Bigl(F_u(h)\Bigr)\!\left\vert \,\det\!\left(F^i_{u,h}
\bigl(\epsilon_j(u)\bigr)\right)\!\right\vert^2$ makes sense by localizing
the integration (\ref{resGF}) in that neighborhood.

\vsp
The infinite determinant and some other factors can neatly be
expressed in terms of the system of free fermions called the adjoint
ghosts --- spin $(1,0)$ (resp. $\!(0,0)$) anti-commuting field $b$
(resp. $\!c$) valued in $(\ad \PC)^*$ (resp. $\!\ad \PC$)
and its anti-holomorphic partner $\bar b$
(resp. $\!\bar c$) --- with the classical action
\beq
I_{\Sigma,P}^{gh}(A,b,c,\bar b,\bar c)=\frac{i}{2\pi}\int_{\Sigma}b\bartialA
c+\bar b\partial_{\!A} \bar c\,.
\eeq
That is, we have
\beqa
\lefteqn{{\det}'\Bigl(\,\bartial_{\!A_u}^{\dag}\bartial_{\!A_u}\,\Bigr)
\frac{\det\!\left(F^i_{u,h}\bigl(\epsilon_j(u)\bigr)\right)}{\det\Bigl(
\epsilon_i(u),\epsilon_j(u)\Bigr)}\frac{\left\vert \,\det\!\left\langle
a^b(u),\lnu_{\!c}(u)\right\rangle \right\vert^2}{\det\Bigl(a^a(u),a^b(u)
\Bigr)}}\label{getgh}\label{detgh}\\
&&=Z_{\Sigma,P}^{\,gh}\Bigl( A_u \,;  \prod_{i=1}^{d_S}F^i_{u,h}(c)
\bar F^i_{u,h}(\bar c) \prod_{a=1}^{\dN}\Bigl\langle b,\lnu_{\!a}(u)
\Bigr\rangle\! \Bigl\langle \bar b, \blnu_{\!a}(u)\Bigr\rangle\,\Bigr)\,,
\nonumber
\eeqa
where $Z_{\Sigma,P}^{\,gh}(A\,;\cdots )$ is the correlation function of the
adjoint ghost system.

\vsp
These results (\ref{pullback}), (\ref{resGF}) and (\ref{detgh}) together
with the chiral gauge symmetry (\ref{chiral sym}) lead to the expression
\beq
Z_{\Sigma,P}(O_1\cdots O_s)
=\int_{\NNPc}\Omega_{\Sigma,P}^{\tot}(O_1\cdots O_s)\,,
\label{oldintexpr}
\eeq
where the integrand is as follows: On $U\subset \NNPc$ with representatives
$\{A_u\}_{u\in U}$,
\beqa
\lefteqn{\Omega_{\Sigma,P}^{\tot}(O_1\cdots O_s)}\\&=&
\prod_{a=1}^{\dN}\dd u^a Z_{\Sigma,P}^{\tot}
\Bigl(\,A_u\,;\delta\Bigl(F_u(h)\Bigr)\!\!
\prod_{i=1}^{d_S}F^i_{u,h}(c)\bar F^i_{u,h}(\bar c)
\prod_{a=1}^{\dN}\Bigl\langle b,\lnu_{\!a}(u)\Bigr\rangle\!
\Bigl\langle \bar b, \blnu_{\!a}(u)\Bigr\rangle hO_1\cdots hO_s\,\Bigr),
\nonumber
\eeqa
in which $Z^{\tot}_{\Sigma,P}(A;\cdots\,\cdots)$ is given by
\beq
\int_{\GPC/\G_P}\!\!{\cal D}h\,\,
\e^{kI_{\Sigma,P}^G(A,hh^*)+I_{\Sigma,P}^{\ad}(A,hh^*)}
Z_{\Sigma,P}^{G,k}(A;\cdots)Z_{\Sigma,P}^{\, gh}(A;\cdots)\,.
\label{corrtot}
\eeq
Note that the factor $\frac{1}{\vol \GP}$ in (\ref{intgauge}) corresponds to
the $\G_P$-quotient of $\GPC$ here, though we need some care  if $d_S>0$
since the residual gauge
fixing term may be gauge variant. In such a case, ``integration over
$\G_P$ with the factor $\frac{1}{\vol\,\G_P}$'' leads to a new expression of
the residual gauge fixing term that depends only on the $\G_P$-invariant
combination $hh^*$. The issue is illustrated in Appendix C for
$\Sigma =\CP$.

\vspace{0.4cm}
{\sc 3.3 A Few Remarks On The Induced Systems}

\vspace{0.3cm}
Two new systems of quamtum fields are introduced above.
One is the adjoint ghost system and
the other is the system of field $[h]\in \GPC/\GP$ with
the classical action
$I_{\it chg}=-kI_{\Sigma,P}(A,hh^*)-I_{\Sigma,P}^{\rm ad}(A,hh^*)$.
The standard calculation of anomaly \cite{Fujikawa} shows (for simple $H$)
that $I_{\it chg}=-(\tilk+2\hvH)I_{\Sigma,P}^{\tilH}(A,hh^*)$
where $\tilH$ is the universal covering of $H$ and $\tilk$ is the induced
level such that $k\trG(XY)=\tilk\trH(XY)$ for $X,Y\in \h\subset \g$.
Since $[h]\in\GPC/\GP$ is a section of $P{\times}_H(\HC/H)$, we call
this system the WZW model with the target $\HC/H$. Both systems are
conformally invariant up to the anomalies $c_{\it gh}=-2\dim H$ and
$c_{\it chg}=\frac{\tilk+2\hvH}{\tilk+\hvH}\dim H$ (if $H$ is simple)
respectively. Below, we shall give a brief description of the spaces of
states.

\vspace{0.3cm}
{\it Fermion Fock Space}

\vsp
Theory of free fermions on a Riemann surface
has been studied by many authors.
So,
we only give a minimal account on the particular system of adjoint ghosts,
referring for detail and generality to the references
\cite{Quillen,S-W,S,P-S,IMO,AGMV}.

The ghost Fock space $\Fadgh$ is the Hilbert space spanned by states at
the parametrized boundary $S$ of a unit disc $D_0$ with several ghost
insertions. The disc with flat gauge field and no insertion corresponds
with respect to a horizontal gauge to the vacuum state $|0\rangle\in\Fadgh$.
Ghost fields at $S$ act on $\Fadgh$ in the standard way
satisfying the anti-commutation relations:
\beqa
\{c_n^{\rm a},c_m^{\rm b}\}\!\!&=&\!\!\{b_{n{\rm a}},b_{m{\rm b}}\}=0\,,\\
\{c_n^{\rm a},b_{m{\rm b}}\}\!\!&=&\!\!\delta_{n+m,0}\delta_{\rm b}^{\rm a}
\,,
\eeqa
and similarly for $\bar b_{n{\rm a}}$ and $\bar c_n^{\rm a}$. Here,
$b_{n{\rm a}}$ and $c_n^{\rm a}$ are Fourier components of
$b_z(s_{\rm a}(z))$ and $s^{\rm a}(c(z))$ respectively where $\{s_{\rm a}\}$
is a horizontal frame of the adjoint bundle over $S$ and $\{s^{\rm a}\}$
is its dual.
Hilbert space structure of $\Fadgh$ are stated by
$b_{m{\rm a}}^{\dag}=\eta_{\rm ab}c^{\rm a}_{-m}$ where
$\eta_{\rm ab}=\trP(s_{\rm a} s_{\rm b})$.
A central extension $\tiL\HC^{\ad}$ of $LH_{\C}$ acts
on $\Fadgh$ by the holomorphic ($J$) and the anti-holomorphic ($\bar J$)
representations whose infinitesimal version could be read
by looking at the expression (\ref{Jgh})
of the currents $J$ and $\bar J$ in terms of the ghost fields.
The Cartan involution $\gamma\mapsto\gamma^*$ of $L\HC$ lifts to
$\tiL\HC^{\ad}$ so that $J(\tilde{\gamma}^*)=J(\tilde{\gamma})^{\dag}$ and
$\bar J(\tilde{\gamma}^*)=\bar J(\tilde{\gamma})^{\dag}$.
The gauge transformation $\gamma.$ by the loop $\gamma\in LH$ is provided by
the diagonal action $J(\tilde{\gamma})\bar J(\tilde{\gamma})$ of any element
$\tilde{\gamma}\in \tiL \HC^{\ad}$ over $\gamma$ with
$\tilde{\gamma}\tilde{\gamma}^*=1$.
The energy momentum tensor provides Virasoro generators $\{L_n^{\it gh}\}$
and $\{\bar L_n^{\it gh}\}$ acting on $\Fadgh$ with central charge
$c_{\it gh}$ (see (\ref{Tgh})).

\newcommand{\Lad}{{\cal L}^{\rm ad}}
\newcommand{\Lg}{{\cal L}_{\it chg}}
\newcommand{\bt}{\Box\!\!\!\!\!\!\times}

\vspace{0.3cm}
{\it WZW model with the target $\HC/H$}

\vsp
Let $\Lad$ denote the line bundle $\tiL \HC^{\rm ad}\tms_{\C^*}\C$ over
$L\HC$ on which the group $\tiL \HC^{\rm ad}$ acts on the left
and on the right. We pull back the line bundle $\LWZ^{-k}\ot {\Lad}^{-1}$
over $LG_{\!\C}\tms L\HC$ by the map $L(\HC/H)\to LG_{\!\C}\tms L\HC$;
$[\gamma]\mapsto (\gamma\gamma^*,\gamma\gamma^*)$ where $\gamma\gamma^*$
in the first factor is considered as a loop in $\bar \HC\subset G_{\!\C}$,
the covering group of $\HC\subset G_{\!\C}/Z_G$.
Then, we obtain a line bundle denoted by $\Lg$
over the loop space $L(\HC/H)$.
The state space of the system is the space $\Gamma(\Lg)$ of sections.

The left action of $L\bar \HC$ on $L(\HC/H)$ induces
a projective representation
${\cal J}$ of $L\bar \HC$ on $\Gamma(\Lg)$.
Unlike in the case of $\Gmhol(\LWZ^k)$,
the group $L\bar \HC$ has only one projective representation
on $\Gamma(\Lg)$. However, the infinitesimal version
splits into two copies of a representation of the Kac-Moody algebra for $H$.
This is obtained by decomposing the complexification of the Lie algebra of
$L\HC$ into holomorphic and anti-holomorphic subspaces.
These infinitesimal generators are identified with the currents.
The Sugawara forms $\{L_n^{\it chg}\}$ and
$\{\bar L_n^{\it chg}\}$ constructed by these act on
(a subspace of) $\Gamma(\Lg)$ as Virasoro generators with central charge
$c_{\it chg}$.

\vsp
To make things explicit, we consider a simple group $H$
with the universal covering $\tilH$. Then, $\Lg$ is isomorphic to
the pull back of the line bundle $\LWZ^{-\tilk-2\hvH}$ over $L\tilH_{\!\C}$
by $[\gamma]\mapsto \gamma\gamma^*$.
The representation ${\cal J}$ of
$\tiL\tilH_{\!\C}$ on $\Gamma(\Lg)$ is defined by
\beq
{\cal J}(\tilde{\gamma})\Phi([\gamma_0])
=\tilde{\gamma}\Phi([\gamma^{-1}\gamma_0])\tilde{\gamma}^*\,.
\label{defcalJ}
\eeq
For use in the next section, choosing a maximal torus $T$ and a chambre
$\Ch$, we look at the state coming from the disc $D_0$ with an insertion
of the field of the form
\beq
\left|\e^{\lmd+2\rho}(b(h))\right|^2\qquad\lmd\in \PP^{(\tilk)}_+\,,
\label{HCfield}
\eeq
where $\rho$ is half the sum of positive roots of $H$ and $b(h)$ is
the `Borel part' of the Iwasawa decomposition
$h=b(h)\un(h)$; $b(h)\in B$, $\un(h)\in H$.
(The Iwasawa decomposition is associated to the decomposition
$\h_{\bf c}=\n\oplus i\liet\oplus \h$ of the Lie algebra
where $\n$ is spanned by positive root vectors.
In this paper, we call the factor corresponding to $\n\oplus i\liet$
the `Borel-part'.)
The state at the boundary $S=\partial D_0$ is given by
\beq
\Phi_{-\lmd-2\rho}([\gamma])
=\e^{(\tilk+2\hvH)I_{D_0}(bb^*)}\left|\e^{\lmd+2\rho}(b(0))\right|^2\,,
\label{HChw}
\eeq
where $\gamma$ is bounded by a holomorphic function $b$ on $D_0$ with
$b(0)\in B$.
A calculation as in \S 2 shows that
$\gamma.\Phi_{-\lmd-2\rho}=\Phi_{-\gamma\lmd -2\rho}$ for $\gamma\in\Gmalcv$
where $\gamma\lmd=w\lmd+\tilk \ttr\mu$.
The argument is easily generalized to the case in which $H$ is not simple.

\vspace{0.3cm}
{\it The Total System}

\vsp
Space of states of the total system
--- the combined system of the WZW model and the two new systems is given by
\beq
{\cal H}^{\tot}={\cal H}^{G,k}\ot \Gamma(\Lg)\ot \Fadgh\,.
\eeq
The left and the right representations of the Kac-Moody algebra
$\Lie(\tiL \HC)$ on the three spaces determine the representations
$J^{\tot}$ and $\bar J^{\tot}$ of the loop algebra
$\Lie(L\HC)$ on ${\cal H}^{\tot}$.
In the similar way,
the representations $\{L_n^{\tot}\}$ and $\{\bar L_n^{\tot}\}$
of the Virasoro algebra is defined
with central charge $c_{\tot}=c_{G,k}+c_{\it chg}+c_{\it gh}$.

Another ingredient is the BRST operator which is the zero mode of the
meromorphic and gauge invariant fermionic current
$J_{\rm WZW}c+J_{\it chg}c+\frac{1}{2}J_{\it gh}c$ where
$J_{\rm WZW}, J_{chg}$ and $J_{gh}$ are the $H$-currents of
the three sectors.
It is nilpotent and may be used to specify the physical states or fields
by determining the cohomology group under suitable equivariant condition.
In this paper, however, we do not use it but argue in the following way.

\newcommand{\Hinv}{{\cal H}_{\rm inv}}
\newcommand{\Hhw}{{\cal H}_{\rm hw}}

\vspace{0.4cm}
{\sc 3.4 Gauge Invariant Local Fields}

\vspace{0.3cm}
We specify the set of $\GP$-invariant fields $O$ in the WZW model and
describe the dressed fields $hO$. Recall that the fields and states
in ${\cal H}^{G,k}$ are in one to one correspondence under
$O\leftrightarrow \Phi_O=Z_{D_0}(0;O)$. The gauge invariance condition
on $O$ is equivalent to the following conditions on $\Phi_O$:
\beqa
&&\left( J_0(v)+\bar J_0(v)\right)\Phi_O=0\qquad \mbox{for $\,\,v\in \h$},
\label{invcond0}\\
&&J_n(v)\Phi_O=\bar J_n(v)\Phi_O=0\qquad \!\!\mbox{for $\,v\in \h_{\!\C}$
and $n=1,2,\cdots$},\label{invcond1}
\eeqa
where $J_n(v)$ and $\bar J_n(v)$ are infinitesimal generators of $J$ and
$\bar J$ corresponding to the tangent vector to the curve
$t\mapsto \exp\{-I_{D_0}(\e^{tz^n v})\}$ in $\tiL G_{\!\C}$.

\vsp
To distinguish the space $\Hinv$ of states satisfying these conditions,
we choose
maximal tori and chambres $(T_{G},\Ch_G)$ for $G$ and $(T,\Ch)$ for $H$
and consider the decomposition of the integrable representations
$L^{G,k}_{\Lmd}$ of $\tiL G$
\`a la Goddard-Kent-Olive \cite{GKO}.
The restriction
of $\tiL G$ to $L\bar H\subset LG$ is a central extension
$\tiL\bar H$ of $L\bar H$ where $\bar H\subset G$ is the covering group
of $H\subset G/Z_G$. We decompose $L_{\Lmd}^{G,k}$
into irreducible representations of the subgroup $\tiL\bar H$:
\beq
L_{\Lmd}^{G,k}=\bigoplus_{\lmd}B_{\Lmd}^{\lmd}\otimes
L_{\lmd}^{\bar H,\tilk}\,,
\label{branch}
\eeq
in which $B_{\Lmd}^{\lmd}$ is the subspace of $L_{\Lmd}^{G,k}$ consisting
of highest weight vectors of weight $(\lmd,\tilk)$ with respect to
$\tiL\bar H_{\!\C}$ where $\tilk$ is the induced level.
We denote by ${\cal H}_{\Lmd}^{\lmd}$
the subspace of ${\cal H}_{\Lmd}^{G,k}$ corresponding to
the subspace $B_{\Lmd}^{\lmd}\ot \overline{B_{\Lmd}^{\lmd}}$
of $L_{\Lmd}^{G,k}\ot\overline{L_{\Lmd}^{G,k}}$.
Each state $\Phi\in {\cal H}_{\Lmd}^{\lmd}$
generates an irreducible $J_0(\bar H)\times \bar J_0(\bar H)$-module
$E_{\lmd}(\Phi)\subset {\cal H}_{\Lmd}^{G,k}$ which is isomorphic to
the tensor product $V_{\lmd}\ot V_{\lmd}^*$ of the irreducible
$\bar H$-module $V_{\lmd}$ of highest weight $\lmd$
and its dual $V_{\lmd}^*$.
Choosing a base $\{e_m\}$ of $V_{\lmd}$
and the dual base $\{e^m\}\subset V_{\lmd}^*$,
we denote by $\Phi_{m}^{\!\!\;\bar m}\in E_{\lmd}(\Phi)$
the vector corresponding to $e_m\ot e^{\bar m}\in V_{\lmd}\ot V_{\lmd}^*$.
Then, the $\bar H$-invariant element
$\sum_m\Phi^{\!\!\;m}_{m}\in E_{\lmd}(\Phi)$ satisfies (\ref{invcond0}) and
(\ref{invcond1}).
In this way, the space $\Hinv$ can be identified with the subspace
\beq
\Hhw=\bigoplus_{\Lmd,\lmd}{\cal H}_{\Lmd}^{\lmd}
\eeq
of ${\cal H}^{G,k}$ spanned by highest weight states
of left-right equal weights with respect to
$J(\tiL\bar H_{\!\C})\times \bar J(\tiL\bar H_{\!\C})$.

\vsp
Let ${O_{\Phi}}^{\!\bar m}_m$ denote the field corresponding to the state
$\Phi^{\!\!\;\bar m}_{m}$ and we consider it as a matrix element of a field
${\bf O}_{\Phi}$ valued in $\End(V_{\lmd})$. Then, the gauge invariant field
$O_{\Phi}$ corresponding to the state $\frac{1}{\dim V_{\lmd}}\sum_m
\Phi^{\!\!\;m}_{m}$ is expressed as
\beq
O_{\Phi}=\frac{1}{\dim V_{\lmd}}\,\trVlmd\!({\bf O}_{\Phi})\,.
\eeq
Since $J_0(\bar h)\bar J_0(\bar h)\Phi^{\!\!\;\bar m}_m$ with
$\bar h\in \bar H_{\!\C}$ is expanded as
$\sum {\bar h}^{*\bar m}_{\,{\bar m}'}
\Phi^{\!\!\;{\bar m}'}_{m'}{\bar h}^{\!m'}_m$, the dressed field for
$O_{\Phi}$ is given by
\beq
hO_{\Phi}=\frac{1}{\dim V_{\lmd}}\,\trVlmd\!({\bf O}_{\Phi}hh^*)\,.
\eeq

\newcommand{\ssstyle}{\scriptscriptstyle}
\vsp
{\it Remark}. One can construct the Sugawara forms $\{L_n^{G,k}\}$ from
the current algebra $\Lie(\tiL G_{\!\C})$ and also
from the subalgebra $\Lie(\tiL \bar \HC)$: $\{L_n^{\bar H,\tilk}\}$.
The difference
$L_n^{\ssstyle {\rm GKO}}=L_n^{G,k}-L_n^{\bar H,\tilk}$ commutes with
$\Lie(\tiL \bar \HC)$ and hence acts on $B^{\lmd}_{\Lmd}$ \cite{GKO}.
Thus we have a Virasoro actions $\{L_n^{\ssstyle {\rm GKO}}\}$ and
$\{\bar L_n^{\ssstyle {\rm GKO}}\}$
on $\Hhw$. As is shown in \cite{KaSch}, these generators coincides with the
generators $\{L_n^{\tot}\}$ and $\{\bar L_n^{\tot}\}$ up to BRST-exact
terms.

\renewcommand{\theequation}{4.\arabic{equation}}\setcounter{equation}{0}

\newpage
\vsp
\begin{center}
{\bf 4. Actions Of The Fundamental Group}
\end{center}

A prototype (\ref{preFI}) of the topological identity (\ref{FI})
is obtained in \S 2.3, but we recognize two kinds of
gaps to be filled.
One is that the field $O_{\Lmd}$ in (\ref{preFI}) is not gauge invariant
(for $H$ non-abelian) but corresponds to a highest weight state.
The other is that (\ref{preFI}) holds for certain gauge fields
of special configurations over
a neighborhood of the insertion point, while (\ref{FI}) is an equation
in the {\it quantum} gauge theory.
In \S 3, we have developed a method
to integrate over gauge fields and obtained a formula (\ref{oldintexpr})
that expresses a correlation function as an integral over the moduli space
of semi-stable $\HC$-bundles. If we are to use this formula to prove
(\ref{FI}), we must find some relation of  the moduli spaces
$\NNP$ and $\NNPgma$
of semi-stable bundles of different topology.

In this section we shall fill these gaps by taking into account
the variety of choices of highest weight conditions --- the flag manifold:
We express a correlator as an integral over a certain moduli space
of holomorphic bundles with a flag at the insertion point.
This leads us to define two actions of the fundamental group $\pi_1(H)$;
one on the set of gauge invariant local fields and the other on the
moduli spaces of bundles with flags.

\vspace{0.5cm}
{\sc 4.1 The Flag Partner}

\vspace{0.3cm}
As a step to the new expression of correlators,
a dressed gauge invariant local field $hO$ is expressed
as an integral over the flag manifold of $H$. When it is inserted into
a correlation function of the total system, the integrand is given by
contour integrals of ghost currents encircling a field $\widehat{O}$ of
ghost number $|\Delta|$ (the number of roots of $H$) which is referred
to as the {\it flag partner} of $O$.

\vspace{0.3cm}
{\it Flag Manifold and the Borel-Weil Theorem}

\vsp
We recall the representation theory of compact groups due to
Borel and Weil.

Let $Fl(H)$ be the ensemble of choices of maximal tori and chambres:
\beq
Fl(H)=\left\{\, (T,\Ch)\,\mbox{\LARGE ;} \begin{array}{ll}\mbox{$T$
is a maximal torus of $H$} \\
\noalign{\vskip0.1cm}
\mbox{ and $\Ch$ is a chambre in $i\liet$}\end{array}\,\right\}\,.
\nonumber
\eeq
A choice $(T,\Ch)\in Fl(H)$ determines an identification $Fl(H)=H/T$
which makes $Fl(H)$ a compact manifold called
the {\it flag manifold} of $H$.
Furthermore, $Fl(H)$ becomes a homogeneous complex manifold since
the embedding $H\hookrightarrow\HC$ induces the isomorphism $H/T\cong\HC/B$
where $B$ is the Borel subgroup of $\HC$ determined by $(T,\Ch)$.

A weight $\lmd\in \PP$ gives a character $\e^{\lmd}:T\to U(1)$ by
$\e^{2\pi i v}\mapsto \e^{2\pi i\lmd(v)}$ and its extension
$\e^{\lmd}:B\to \C^*$ defines a homogeneous holomorphic line bundle
\beq
L_{-\lmd}=\HC\times_B \C \longto Fl(H)\,,
\label{def:L-lmd}
\eeq
by the equivalence relation $(hb,c)\sim (h,\e^{-\lmd}(b)c)$
where $h\in \HC$,
$b\in B$ and $c\in \C$. We denote by $h\cdot c\in L_{-\lmd}$
the equivalence class represented by $(h,c)\in \HC\times \C$.
The Borel-Weil theorem states that the space $H^0(Fl(H),L_{-\lmd})$
of holomorphic sections is an irreducible $\HC$-module $V_{\lmd^*}$
of highest weight $\lmd^*=-w_0\lmd$, which is non-zero
if and only if $\lmd$ takes positive values on $\Ch$
(see \cite{Serre} and also \cite{Bott,Kostant}).

The line bundle $L_{-\lmd}$ is equipped with an $H$-invariant hermitian
metric $(\,\,,\,\,)_{-\lmd}$ such that an element $h$ of $H$ determines
a unitary frame $h\cdot 1$; $(h\cdot c_1,h\cdot c_2)_{-\lmd}=\bar c_1c_2$.
There also exists an $H$-invariant volume form $\Omega$ on $Fl(H)=H/T$.
These induces the following $H$-invariant hermitian
inner product on the space $H^0(Fl(H),L_{-\lmd})$:
\beq
(\psi_1,\psi_2)_{Fl(H)}
=\frac{1}{\Vol Fl(H)}\int_{Fl(H)}(\psi_1,\psi_2)_{-\lmd}\Omega\,.
\eeq

 Let $\{ e_m\,;\,m\in \tilP_{\lmd}\}$ be an orthonormal base of
$V_{\lmd}$ consisting of weight vectors where
$\tilP_{\lmd}$ is an indexing set. We always take the weight $\lmd$ itself
as the index for the highest weight vector.
Denoting the matrix element $(e_{m_1},h e_{m_2})$ by $(h)^{\!m_1}_{m_2}$,
we put
\beq
\psi^m(hB)=h\cdot(h)^{\!m}_{\lmd}\,,
\label{ch4:holsec}
\eeq
for $m\in \tilP_{\lmd}$. Then, $\{\, \psi^m\,;\,m\in \tilP_{\lmd}\}$ forms
an orthogonal base of $H^0(Fl(H),L_{-\lmd})$:
\beqa
(\psi^{m_1},\psi^{m_2})_{Fl(H)}&=&\frac{1}{\Vol Fl(H)}
\int_{Fl(H)}(\psi^{m_1},\psi^{m_2})_{-\lmd}\Omega \nonumber\\
&=&\frac{1}{\Vol H}
\int_H\left(\psi^{m_1}(hB),\psi^{m_2}(hB)\right)_{-\lmd}dh\nonumber \\
&=&\frac{1}{\Vol H}\int_H \overline{(h)^{\!m_1}_{\lmd}}(h)^{\!m_2}_{\lmd}
dh =\frac{1}{\dim V_{\lmd}}\delta^{m_1,m_2}\,,\label{Peter-Weyl}
\eeqa
where $dh$ is the Haar measure of $H$ and the Peter-Weyl theorem is used.

\vspace{0.3cm}
{\it Integral Expression of Gauge Invariant Fields}

\vsp
Recall that (see \S 3.4) to each $\Phi\in {\cal H}_{\Lmd}^{\lmd}$
is associated a gauge invariant field
$O_{\Phi}=\frac{1}{\dim V_{\lmd}}\trVlmd\!\!\,({\bf O}_{\Phi})$
or the dressed field
$hO_{\Phi}=\frac{1}{\dim V_{\lmd}}\trVlmd\!\!\,({\bf O}_{\Phi}hh^*)$.
In the following argument, $\Phi$ is fixed all through and
will not usually be mentioned.

We now express $hO$ as an integral over the flag manifold $Fl(H)$.
We introduce a field $\Omega(hh^*)$ with values in differential forms
on $Fl(H)$ of top degree. At the point $h_1B\in Fl(H)$ represented by
$h_1\in H$, it is expressed as
\beq
\Omega(hh^*)|_{h_1B}=h_1O^{\!\lmd}_{\lmd}\,
\Bigl| \e^{\lmd+2\rho}(b(h_1^{-1}h))\Bigr|^2\,\Omega|_{h_1B}\,,
\label{flagmeasure1}
\eeq
where $b(h_1^{-1}h)\in B$ is the `Borel-part' of the Iwasawa decomposition
of $h_1^{-1}h\in \HC$.
Let $\L_{h^{-1}}:Fl(H)\to Fl(H)$ be the left translation by $h^{-1}$.
The relation
\beq
\L_{h^{-1}}^*\Omega|_{h_1B}
=\Bigl| \e^{2\rho}(b(h_1^{-1}h))\Bigr|^2\Omega|_{h_1B}\,,
\label{page:relR}
\eeq
which shall be proved shortly shows
\beq
\L_{h}^*\Omega(hh^*)|_{h^{-1}h_1B}
=h_1O^{\!\lmd}_{\lmd}\,\Bigl|\e^{\lmd}(b(h_1^{-1}h))\Bigr|^2 \,
\Omega|_{h^{-1}h_1B}\,.
\label{eq105}
\eeq
The definition of $b(h_1^{-1}h)^{-1}$ says that
there is a representative  $\un\in H$ of $h^{-1}h_1B$ such that
$h^{-1}h_1O^{\!\lmd}_{\lmd}
=\un O^{\!\lmd}_{\lmd}\,\Bigl| e^{-\lmd}(b(h_1^{-1}h))\Bigr|^2$ and hence
(\ref{eq105}) gives
\beq
\L_{h}^*\Omega_{\lmd}(hh^*)|_{\sun B}=h\un O^{\!\lmd}_{\lmd}\,
\Omega|_{\sun B}=\sum_{{\bar m},m}hO^{\!\bar m}_m
(\un^{-1})^{\!\lmd}_{\bar m}(\un)^m_{\lmd}\,\Omega|_{\sun B}\,.
\eeq
This amounts to the following identity of top differential forms:
\beq
\L_h^*\Omega(hh^*)
=\sum_{{\bar m},m}hO^{\!\bar m}_m (\psi^{\bar m},\psi^m)_{-\lmd}\Omega\,,
\eeq
where $\psi^m$ is given in (\ref{ch4:holsec}). Due to the orthogonality
(\ref{Peter-Weyl}), it follows that
\beq
\frac{1}{\Vol Fl(H)}\int_{Fl(H)}\Omega(hh^*)
=\frac{1}{\dim V_{\lmd}}\trVlmd\left({\bf O}hh^*\right)\,.
\label{intexpr}
\eeq

\vsp
\noindent{\it Proof of the relation (\ref{page:relR}).} It is enough
to prove $\L_b^*\Omega|_B=|\e^{-2\rho}(b)|^2\Omega|_B$ for $b\in B$.
Since the holomorphic tangent space of $Fl(H)$ at $B$ is
isomorphic to $\h_{\C}/\b$, we have only to show that $\e^{-2\rho}(b)$ is
the determinant of $\ad_-(b):\h_{\C}/\b\to\h_{\C}/\b$. In view of
$2\rho=\sum_{\alpha>0}\alpha$, the proof is now trivial since
we can order the base of $\h_{\C}/\b$ consisting of negative root vectors
so that $\ad_-(b)$ is represented by an upper triangular matrix.

\vspace{0.3cm}
{\it The Flag Partner}

\vsp
Suppose that $hO$ is inserted
at $x\in \Sigma$ in a correlator $Z_{\Sigma, P}^{\tot}(A\,;\cdots)$
of the total system (\ref{corrtot}),
where we assume that the background gauge field $A$ is chosen to be flat
over a disc $D_0\subset \Sigma$ centered by $x$.
Then, (\ref{intexpr}) leads to an integral expression of $hO(x)$ over
the flag manifold $Fl(P_x)=P_x/T\cong P_{\!\C x}/B$ of
the fibre $P_x$ of $P$ over $x$:
The integrand $\Omega_x(hh^*)$ is expressed at the flag $f\in Fl(P_x)$ by
\beq
\Omega_{x}(hh^*)|_f=O^{\!\!\,\lmd}_{\lmd}(f)\,
\Bigl| \e^{\lmd+2\rho}(b_f(h))\Bigr|^2\,\Omega|_f\,.
\label{flagmeasure}
\eeq
Here, $O^{\!\!\,\lmd}_{\lmd}(f)={O_{\Phi}}^{\!\!\,\lmd}_{\lmd}(f)$ is
the field corresponding to $\Phi\in{\cal H}_{\Lmd}^{\lmd}$ and
$b_f(h)$ is the `Borel-part' of the Iwasawa decomposition of $h(x)$, both
with respect to the horizontal section $s$ of $P|_{D_0}$ with $s(x)B=f$.
$\Omega$ is the invariant volume form on $Fl(P_x)$.

This measure can be rewriten using ghost fields.
Let $U_F$ be an open subset of $Fl(P_x)$ with complex coordinates
$f^1,\cdots , f^{|\Delta_+|}$ and a family $\{ \sigma_f\}_{f\in U_F}$ of
holomorphic sections of $(\PC,\bartial_{\!A})|_{D_0}$
such that $\sigma_f(x)B=f$.
Then, the symbol $(\partial \sigma_f/\partial f^{\alpha})\sigma_f^{-1}$
determines a holomorphic section $\nu_{\!\alpha}(f)$
of $(\ad\PC,\bartial_{\!A})|_{D_0}$.
Using the singular behavior (\ref{def:regbc}) of the product of ghosts,
we obtain the following expression:
\beq
\Omega_{x}(hh^*)|_f=\prod_{\beta=1}^{|\Delta_+|}d^2\!f^{\beta}
\ooint_xb\,\nu_{\!\beta}(f) \ooint_x\bar b \,\bnu_{\!\beta}(f)\,
\widehat{O}(f)\,,
\eeq
\beq
\hspace{-2.5cm}\mbox{ where}\hspace{2cm}
\widehat{O}(f)=O^{\!\!\,\lmd}_{\lmd}(f)\,
\left| \e^{\lmd+2\rho}(b_f(h))\right|^{2}
\prod_{-\alpha<0}c^{-\alpha}(x)\bar c^{-\alpha}(x)\,.
\label{flagpartner}
\eeq
Here and henceforth, we denote the normalized contour integral
$\nipi\oint$ by ${\displaystyle \ooint}$.
In the expression (\ref{flagpartner}),
$c^{-\alpha}(x)$ is the coefficient of the ghost;
$c(x)=\sum s_{\rm a}(x) c^{\rm a}(x)$ where $\{s_{\rm a}(x)\}$
is the frame associated to any $s(x)\in P_x$
representing $f$ and to the base of $\h_{\C}$ including root vectors
$\{e_{\alpha}\}_{\alpha\in \Delta}$ normalized by
$\tr(e_{\alpha}e_{\beta})=\delta_{\alpha+\beta,0}$.
We call this field $\widehat{O}(f)$
the {\it flag partner} of $O$ associated to $f\in Fl(P_x)$.

\vsp
{\it Remark}. Construction/determination of
BRST complex/cohomology is a standard method of determining
the space of physical states of a theory with gauge symmetry.
In the literature (see \cite{BMP} and references therein),
there are several constructions
which seem to come from the gauged WZW model.
The cohomology groups include as a non-trivial element,
the state of the form
\beq
|\lmd\rangle^{G,k}\ot|-\lmd-2\rho\rangle^{\Hc/H}\ot
\prod_{-\alpha<0}c_0^{-\alpha}|0\rangle^{{\rm gh}}\,,
\eeq
where $|\lmd\rangle^{G,k}$ is a state in
$B^{\lmd}_{\Lmd}\subset L_{\Lmd}^{G,k}$ for some $\Lmd$,
$|-\lmd-2\rho\rangle^{\Hc/H}$ is a highset weight state with weight
$(-\lmd-2\rho,-\kh)$ in a suitably chosen $\Lie(\tiL H_{\!\C})$-module
and $|0\rangle^{{\rm gh}}$ is the natural vacuum of the ghost Fock space.
This state seems to correspond to (the left moving part)
of our flag partner $\widehat{O}$.

\newcommand{\rgf}{{\rm rgf}}
\newcommand{\APx}{\A_{P,x}}
\newcommand{\APgmax}{\A_{P\gamma,x}}

\vspace{0.5cm}
{\sc 4.2 A New Integral Expression}

\vspace{0.3cm}
We consider the correlation function $Z_{\Sigma,P}(O_1\cdots O_s \,O(x))$
of gauge invariant fields.
As is seen in \S 3.2, it is expressed as
an integral of $\Omega_{\Sigma,P}^{\tot}(\cdots O(x))$
on an open dense subset $\NNPc$ of the moduli space of semi-stable
$\HC$-bundles. Let $U\subset\NNPc$ be provided with a holomorphic family
$\{A_u\}_{u\in U}$ of representative gauge fields.
The result of \S 4.1 shows
that the following measure on $U\times Fl(P_x)$ reproduces
$\Omega_{\Sigma,P}^{\tot}(\cdots O(x))|_U$
after the integration along each $Fl(P_x)$:
\beqa
\tilde{\Omega}_{\Sigma,P,x}^{\tot}(\cdots \widehat{O})_U\!
&=&\!\prod_{a=1}^{\dN}\dd u^a \prod_{\alpha=1}^{|\Delta_+|}d^2\!f^{\alpha}\,
Z_{\Sigma,P}^{\tot}\Bigl(A_u\,\mbox{\Large ;}\,\,|\rgf_{\!A_u}\!(c,h)|^2
\,\,h(\cdots)
\label{omegatilde}\\
&&\,\times\prod_{a=1}^{\dN}\!\Bigl\langle b,\lnu_{\!a}(u)\Bigr\rangle\!
\Bigl\langle \bar b, \blnu_{\!a}(u)\Bigr\rangle
\prod_{\alpha=1}^{|\Delta_+|}\ooint_x b\nu_{\!\alpha}(f)
\ooint_x\bar b \bnu_{\!\alpha}(f)\, \widehat{O}(f)\,\Bigr) ,\nonumber
\eeqa
where $\widehat{O}$ is the flag partner of $O$, $h(\cdots)$ is
the dressed insertion $hO_1\cdots hO_s$ and $|\rgf_{\!A_u}\!\!\,(c,h)|^2$
denote the residual gauge fixing term:
\beq
|\rgf_{\!A_u}\!(c,h)|^2
=\delta^{(2d_S)}\Bigl(F_u(h)\Bigr)\!
\prod_{i=1}^{d_S}F^i_{u,h}(c)\bar F^i_{u,h}(\bar c)\,.
\label{rgfA}
\eeq
We ask whether this form on $U\times Fl(P_x)$
extends to a globally defined form on some well-defined flag manifold bundle
over $\NNPc$.
Below, we shall see that the answer is generally no but
$\tilde{\Omega}_{\Sigma,P,x}^{\tot}(\cdots \widehat{O})_U$
determines a new form
$\Omega_{\Sigma,P,x}^{\tot}(\cdots \widehat{O})$ which is globally defined
on a geometric object $\NNc_{P,x}$ associated to $P$ and $x$.

\vspace{0.3cm}
{\it Transformation Properties of $\tilde{\Omega}_{\Sigma,P,x}^{\tot}$}

\vsp
Let $\{A_{{}_1u}\}_{u\in U}$ and $\{A_{{}_2u}\}_{u\in U}$ be families
of representatives that are related by
\beq
A_{{}_1u}=A_{{}_2u}^{h_{{}_{21}u}}\,,
\eeq
by a family $\{h_{{}_{21}u}\}_{u\in U}$ of chiral gauge transformations.
The groups $S_{{}_iu}=\Aut\bartial_{\!A_{{}_iu}}$ of symmetries are then
related by $S_{{}_1u}=h_{{}_{21}u}^{-1}S_{{}_2u}h_{{}_{21}u}$.
Hence, if $\{F_{{}_1u}\}$ is a family of residual gauge-fixing functions
for $\{S_{{}_1u}\}$, $F_{{}_2u}(h)=F_{{}_1u}(h_{{}_{21}u}^{-1}h)$
determines a family $\{F_{{}_2u}\}$ of residual gauge-fixing functions
for the symmetries $\{S_{{}_2u}\}$. Since the chiral anomally is absent
in the total system, it follows that
\beqa
\lefteqn{Z_{\Sigma,P}^{\tot}\Bigl(A_{{}_1u}\,\mbox{\Large ;}\,
|\rgf_{\!A_{{}_1u}}\!(c,h)|^2\,h(\cdots)
\prod( b; \bar b )\, \widehat{O}(f)\,\Bigr)} \\
&=&Z_{\Sigma,P}^{\tot}\Bigl(A_{{}_2u}\,\mbox{\Large ;}\,
|\rgf_{\!A_{{}_2u}}\!(c,h)|^2\,h(\cdots)
\prod( h_{{}_{\!21}u}^{-1}b; h_{{}_{\!21}u}^*\bar b )\,
h_{{}_{\!21}u}\!\widehat{O}(f)\,\Bigr),\nonumber
\eeqa
where $\prod(b;\bar b)$ is any functional of $b$-$\bar b$
and $h^{-1}b$ is the coadjoint action of $h^{-1}\in \GPC$ on the field
with values in $(\ad \PC)^*$. Making use of the Iwasawa decomposition of
$h_{{}_{\!21}u}(x)\in P_x\tms_H\HC$
with respect to the flag $f\in Fl(P_x)$, we can see that
\beq
h_{{}_{\!21}u}\widehat{O}(f)=\widehat{O}(h_{{}_{\!21}u}f)\,,
\eeq
where the action of $\GPC$ on $Fl(P_x)={\PC}_x/B$
is induced by the action on ${\PC}_x$.

\vsp
Now, it is enough to note the relation
\beq
\delta A_{{}_1u}^{01}
=h_{{}_{\!21}u}^{-1}\delta A_{{}_2u}^{01}h_{{}_{\!21}u}
+\bartial_{\!A_{{}_1u}}\!\!\left(h_{{}_{\!21}u}^{-1}
\delta h_{{}_{\!21}u}\right)\,,
\eeq
to see that the form
$\tilde{\Omega}_{\Sigma,P,x}^{\tot}(\cdots \widehat{O})$
on
$\{1\}\tms U\tms Fl(P_x)$ with the backgrounds
$\{A_{{}_1u}\}_{u\in U}$ coincides with the one on
$\{2\}\tms U\tms Fl(P_x)$ with the backgrounds
$\{A_{{}_2u}\}_{u\in U}$, under the following identification
of the two spaces:
\beq
(1,u,f)\longleftrightarrow (2,u,h_{{}_{\!21}u}f)\,.
\label{idrule}
\eeq

\vspace{0.3cm}
{\it The Space $\NNc_{P,x}$}

\vsp
Let $\{U_i\}$ be an open covering of $\NNc_P$ such that each $U_i$ is
provided with a holomorphic family $\{A_{{}_iu}\}_{u\in U_i}$ of
representatives. If $U_i$ and $U_j$ intersect,
we can choose a family $\{h_{{}_{\!ij}u}\}_{u\in U_i\cap U_j}$ of
chiral gauge transformations such that
$A_{{}_ju}=A_{{}_iu}^{h_{{}_{\!ij}u}}$.

\vsp
If the symmetry group $S_{{}_iu}=\Aut\bartial_{\!A_{{}_iu}}$ is trivial
everywhere, the families $\{h_{{}_{\!ij}u}\}_{i,j}$ necessarily satisfy
the triangle identities:
\beq
\hspace{3cm}h_{{}_{\!ij}u}h_{{}_{\!jk}u}=h_{{}_{\!ik}u}\, ,\qquad
\mbox{for}\quad u\in U_i\cap U_j\cap U_k\,.
\label{triangle}
\eeq
Then, the identification rules as (\ref{idrule}) glue the spaces
$\{i\}\tms U_i\tms Fl(P_x)$ and forms
$\tilde{\Omega}_{\Sigma,P,x}^{\tot}(\cdots\widehat{O})_{U_i}$ together
and define an $Fl(P_x)$-bundle $\NNc_{P,x}$ over $\NNPc$
and a measure $\Omega_{\Sigma,P,x}^{\tot}(\cdots \widehat{O})$ of it.

\vsp
In general, the triangle identity (\ref{triangle}) does not hold
but modulo actions of the symmetry groups. In such a situation,
it is a natural idea to consider the quotient of $Fl(P_x)$ by
the symmetry group $S_u$. Then, we expect that the integration of
$\tilde{\Omega}_{\Sigma,P,x}^{\tot}$ along the
$S_u$-orbits determine a globally defined form on some fibre bundle over
$\NNPc$ having the quotient $S_u\!\lbackslash Fl(P_x)$
as the fibre over $u$.
However, $S_u$ is generically non-compact and the quotient space $S_u\!
\lbackslash Fl(P_x)$ is not even Hausdorff. At this stage, we assume that
we can find an open dense subset $Fl_{S_u}(P_x)$ of $Fl(P_x)$
consisting of $S_u$-orbits of maximum dimension such that the family
$\cup_{u\in U}\{u\}\tms S_u\!\lbackslash Fl_{S_u}(P_x)$ of quotients
is given a good geometric structure (such as manifold or orbifold).
Under this assumption, the spaces
$\{i\}\tms\!\cup_{u\in U_i}^{}\!\{u\}\tms S_{{}_iu}\!
\lbackslash Fl_{S_{{}_iu}}(P_x)$
are glued together by the identification rules as (\ref{idrule})
and result in a space denoted by $\NNc_{P,x}$.

Note that the space $\NNc_{P,x}$ can be considered as a subset of
the quotient of $\APx=\A_P\times Fl(P_x)$ by $\GPC$ (where $\GPC$ acts
on $\APx$ by $h:(A,f)\mapsto (A^h,h^{-1}f)$). In \S 4.3, we shall identify
elements of $\APx/\GPC$ with certain holomorphic objects over the Riemann
surface and make sure in simple cases that the assumptions involved
in this argument hold true.

\vspace{0.3cm}
{\it Local Coordinatization of $\NNc_{P,x}$}

\vsp
Recall that $\dN$ and $d_S$ denote the dimensions of the moduli space
$\NNc_P$ and the symmetry group $S_u$ for $u\in \NNc_P$ respectively.
We denote by $\dSf$ the dimension of the group $S_{u,f}$ of symmetries of
$A_u$ that fix the flag $f$.
Then, the dimension $\dNf$ of the space $\NNc_{P,x}$ is given by
$\dNf=\dN+|\Delta_+|-d_S+\dSf$.

We assume without proof that the following holds:
For a generic point $v_0\in \NNc_{P,x}$, we can find a coordinatized
neighborhood $\VV$ of $v_0$ in $\NNc_{P,x}$ so that
there is a family $\{(A_v,f_v)\}_{v\in \VV}\subset \APx$ of
representatives depending holomorphically on the coordinates
$v^1,\cdots, v^{\dNf}$.
We choose families $\{\sigma_0(v)\}$, $\{\sigma_{\infty}(v)\}$
of holomorphic trivializations on $U_0$, $U_{\infty}$ such that
$\sigma_0(v)B=f_v$ where $U_0$ is a neighborhood of $x$ and $U_{\infty}$ is
a neighborhood of $\Sigma-U_0$.
We assume that the transition function $h_{\infty 0}(v)$ of $\sigma_0(v)$
and $\sigma_{\infty}(v)$ depends holomorphically on $v$.

We choose the coordinate system $v^1,\cdots, v^{\dNf}$ in such a way
that $A_v$ and $\sigma_{\infty}(v)$ depend only on the first $\dN$-tuples
$\vb=(v^1,\cdots,v^{\dN})$. The symmetry group and the residual gauge fixing
function for $A_v$ are then denoted as $S_{\vb}$ and $F_{\vb}$ respectively.
We also choose a family $\{h_{v,t}\}_{v,t}\subset \GPC$ parametrized
by $v\in\VV$ and $t=(t^1,\cdots,t^{d_S-\dSf})$ such that the family
$\{h_{v,t}\}_t$ for a fixed $v$ is in $S_{\vb}$ and is transversal to
$S_{\vb,f_v}$-orbits.
Then, the holomorphic sections
\beqa
\nu_{\!\dN+\alpha}(v)&=&
\sigma_0(v)\cdot h_{\infty0}(v)^{-1}\!
\frac{\partial}{\partial v^{\dN+\alpha}}h_{\infty 0}(v)\qquad
\alpha=1,\cdots,\dNf-\dN\,,\\
\nu_i(v,t)&=&\Bigl(\frac{\partial}{\partial t^i}h_{v,t}\Bigr)h_{v,t}^{-1}
\qquad\,\,i=1,\cdots,d_S-\dSf\,,
\eeqa
of $\ad \PC|_{U_0}$ provide a base of the tangent space of $Fl(P_x)$
at $h_{v,t}f_v$.

Let $s_0$ and $s_{\infty}$ be fixed sections
of $P|_{U_0}$ and $P|_{U_{\infty}}$ and we put
$s_0=\sigma_0(v)h_0(v)$ and $s_{\infty}=\sigma_{\infty}(v)h_{\infty}(v)$.
Since the connection $A_v$ is represented by
$\bartial_{\!A_v}s_I=s_I\cdot h_I(v)^{-1}\bartial h_I(v)$ on $U_I$
($I=0,\infty$), the variation of $A_v$ is expressed as
$\delta A_v^{01}=\bartial_{\!A_v}(s_I\cdot h_I(v)^{-1}\delta h_I(v))$.
For a holomorphic differential $b$ valued in $\ad\PC$,
we thus have
\beqa
\nipi\int_{\Sigma} b\,\delta A_v^{01}
&=&\nipi \oint_xb
\Bigl(s_0\cdot h_0(v)^{-1}\delta h_0(v)
- s_{\infty}\cdot h_{\infty}(v)^{-1}\delta h_{\infty}(v)\Bigr)\\
&=&\nipi \oint_xb\,
\sigma_0(v)\cdot h_{\infty 0}(v)^{-1}\delta h_{\infty0}(v)\,,
\eeqa
where the contour encircles the point $x$.

The measure $\tilde{\Omega}_{\Sigma,P,x}^{\tot}(\cdots \widehat{O})_U$
is then expressed as
\beqa
\tilde{\Omega}_{\Sigma,P,x}^{\tot}(\cdots \widehat{O})_U\!
=\!\prod_{\rA=1}^{\dNf}\dd v^{\rA}\hspace{-0.6cm}&&\hspace{-0.2cm}
\prod_{i=1}^{d_S-\dSf}\!\!\!d^2 t^i\,Z_{\Sigma,P}^{\tot}\Bigl(\,A_v\,
\mbox{\Large ;} |\rgf_{\!A_v}\!(c,h)|^2\,h(\cdots)\\
\times\hspace{0.2cm}
&&\hspace{-0.7cm}\prod_{\rA=1}^{\dNf}\!
\ooint_x \!b\,\nu_{\!\rA}(v)\ooint_x \!\bar b\, \bnu_{\!\rA}(v)\!\!\!
\prod_{i=1}^{d_S-\dSf}\!\!\!\ooint_x \!b\,\nu_{\!i}(v,t)
\ooint_x\!\bar b \,\bnu_{\!i}(v,t)\,
\widehat{O}(h_{v,t}\!f_v)\,\Bigr) ,\nonumber
\eeqa
where $\nu_{\!\rA}(v)$ are the holomorphic sections of
$\ad\PC|_{U_0\cap U_{\infty}}$
\beq
\nu_{\!\rA}(v)=\sigma_0(v)\cdot h_{\infty0}(v)^{-1}\!
\frac{\partial}{\partial v^{\rA}}h_{\infty 0}(v)
\qquad\rA=1,\cdots,\dNf\,. \label{nuA}
\eeq
Due to the absence of chiral anomaly, we can rewrite the above measure by
\beqa
\prod_{\rA=1}^{\dNf}\dd v^{\rA}\hspace{-0.3cm}
\prod_{i=1}^{d_S-\dSf}\!\!\!d^2 t^i\,&&\hspace{-0.7cm}
Z_{\Sigma,P}^{\tot}\Bigl(\,A_v\,\mbox{\Large ;}
|\rgf_{A_v}(h_{v,t} c,h_{v,t} h)|^2\,h(\cdots) \\
\times\hspace{0.2cm}&&\hspace{-0.7cm}
\prod_{\rA=1}^{\dNf}\!\ooint_x \!b\,\nu_{\!\rA}(v)\ooint_x \!\bar b\,
\bnu_{\!\rA}(v)\!\!\! \prod_{i=1}^{d_S-\dSf}\!\!\!
\ooint_x \!h_{v,t}b\nu_{\!i}(v,t)
\ooint_x\!h_{v,t}^{*-1}\bar b \,\bnu_{\!i}(v,t)\,\widehat{O}(f_v)\,\Bigr) .
\nonumber
\eeqa

\vspace{0.3cm}
{\it The New Expression}

\vsp
Now we integrate over each $S_{\vb}$-orbit in $Fl_{S_{\vb}}(P_x)$.
We exchange the order of integration; we perform $\int \dd t^i$'s before
$\int{\cal D}h$.
Then, the delta function $\delta^{(2d_S)}(F_{\vb}(h_{v,t}h))$
in the residual gauge fiexing term serves another delta function of
lower dimension $2\dSf$ multiplied by a certain determinant factor. On the
other hand, deforming the contours of the integrals
$\oint h_{v,t}b \nu_{\!i}$
so that they encircle the $2d_S$ $c\bar c$-insertions in the
residual gauge fixing term, we get another $2\dSf$ $c\bar c$-insertions
multiplied by the determinant which is reciprocal to the one from
the delta function.
Thus, the integration over $t^i$'s serves the following
residual gauge fixing term for $(A_v,f_v)$:
\beq
|\rgf_{\!A_v\!,f_v}\!(c,h)|^2=\delta^{(2\dSf)}\left(F_v(h)\right)
\prod_{i=1}^{\dSf}F^i_{v,h}(c)\bar F^i_{v,h}(\bar c)
\eeq
where $F_v:\GPC\to\C^{\dSf}$ is a gauge fixing function
for $S_{\vb,f_v}\subset \GPC$.

Finally, we have reached to the following measure on $\VV$:
\beqa
\lefteqn{\Omega_{\Sigma,P,x}^{\tot}(O_1\cdots O_s \widehat{O})}
\label{newform}\\
 &=&\prod_{\rA=1}^{\dNf}\dd v^{\rA}\,
Z_{\Sigma,P}^{\tot}\!\Bigl(\,A_v\,\mbox{\Large ;}\,
|\rgf_{\!A_v\!,f_v}\!(c,h)|^2\,hO_1\cdots hO_s
\prod_{\rA=1}^{\dNf}\ooint_x \!b\,\nu_{\!\rA}(v)\ooint_x \!\bar b\,
\bnu_{\!\rA}(v)\,\widehat{O}(f_v)\,\Bigr) .\nonumber
\eeqa
We can check that this expression is independent on the choice of
the representatives $\{(A_v,f_v)\}_{v\in \VV}$. This shows that the form
$\Omega_{\Sigma,P,x}^{\tot}(\cdots\widehat{O})$ extends to a well-defined
measure on the space $\NNc_{P,x}$. We have thus obtained
the new integral expression for the correlation function:
\beq
Z_{\Sigma,P}(O_1\cdots O_s\,O(x))
=\frac{1}{\vol Fl(H)}\int_{\NNc_{P,x}}\!\!
\Omega_{\Sigma,P,x}^{\tot}(O_1\cdots O_s\,\widehat{O})\,.
\label{newintexpr}
\eeq

\vspace{0.5cm}
{\sc 4.3 The Moduli Space Of Holomorphic Principal Bundles\\
\hspace{2cm}With Flag Structure --- Examples}

\vspace{0.3cm}
In the preceding subsection, we have introduced the space $\NNc_{P,x}$ which
can be considered as a subset of the quotient of $\A_{P,x}$ by $\GPC$.
This quotient $\A_{P,x}/\GPC$ can naturally be identified with the set of
isomorphism classes of certain holomorphic objects
--- holomorphic $\HC$-bundles with quasi-parabolic structure at $x$.
Using this fact, we give an explicit description of the space $\NNc_{P,x}$
for some simple cases.

\vspace{0.3cm}
{\it Holomorphic $\HC$-Bundles With Quasi-Parabolic Structure}

\vsp
We fix a maximal torus $T$ and a chambre $\Ch$ of $H$ and denote by $B$
the corresponding Borel subgroup of $\HC$.
For a holomorphic $\HC$-bundle $\Ph$ over $\Sigma$,
a choice of flag $f\in \Ph_x/B$ at $x\in \Sigma$ is called a
{\it quasi-parabolic structure} of $\Ph$ at $x$.\cite{Mehta-Seshadri}
In this paper, we shall simply call it a {\it flag structure} instead.
Two holomorphic $\HC$-bundles with
flag structure at $x $, $(\Ph_1,f_1)$ and $(\Ph_2,f_2)$, are said to be
{\it isomorphic} when there is an isomorphism $\Ph_1\to \Ph_2$
which sends $f_1$ to $f_2$.
Notice that the set of isomorphism classes of flag structures at $x$ of
a holomorphic $\HC$-bundle $\Ph$ is given by $\Aut\Ph\lbackslash \Ph_x /B$.
As in the case without flags, for a principal $H$-bundle $P$,
the set $\A_{P,x}/\GPC$
can naturally be identified with the set of isomorphism classes
of holomorphic $\HC$-bundles of topological type $\PC$
with flag structure at $x$.

For a holomorphic $\HC$-bundle $\Ph$ with flag structure $f$ at $x$,
we denote by $\Aut(\Ph,f)$ the group of automorphisms of $\Ph$
that preserve the flag $f$. Then, $(\Ph,f)$ represents an element of
$\NNc_{P,x}$ if and only if $\Ph$ represents an element  of $\NNc_P$
and $\dim\Aut(\Ph,f)\leq \dim\Aut(\Ph,f')$ for other choices $f'$ of flags.

\vspace{0.3cm}
{\it On the Sphere}

\vsp
We classify the holomorphic principal bundles over the Riemann sphere $\CP$
with flag structure at one point. We follow the notation of section 3.1.

We start with the case $H=SU(n)/\Z_n$.
The Borel subgroup $B$ we choose is represented by the set of
upper triangular matrices.
Let $(\Ph,f)$ be a holomorphic $\HC$-bundle with a flag at $z=0$.
We choose a section $\sigma_0$ on the $z$-plane $U_0$ with $\sigma_0(0)B=f$
and a section $\sigma_{\infty}$ on the $z^{-1}$-plane $U_{\infty}$,
and let $h_{\infty 0}:U_0\cap U_{\infty}\to \HC$ be
the holomorphic transition function.
The Birkhoff theorem states \cite{P-S} that
there is a unique element $a\in \Pv$ such that $h_{\infty 0}(z)
=h_{\infty}(z)z^{-a}h_0(z)^{-1}$ for some holomorphic maps
$h_I:U_I\to \HC$ ($I=0,\infty$)
with  $h_0(0)\in B$. Thus, the set of isomorphism classes
of holomorphic $\HC$-bundles with flag structure at $z=0$ is represented by
$\{\Ph_a;a\in \Pv\}$
where $\Ph_a=(\Ph_{[a]},f_a)$ is an $\HC$-bundle $\Ph_{[a]}$
described by the transition relation $\sigma_0=\sigma_{\infty}z^{-a}$ with
the flag $f_a=\sigma_0^{(a)}(0)B$.

The automorphism group $\Aut \Ph_a$ of $(\Ph_{[a]},f_a)$ is a subgroup
of $\Aut \Ph_{[a]}$.
Recall that an element $h$ of $\Aut \Ph_{[a]}$ is represented
with respect to $\sigma_0$ by an $SL(n,\C)$-valued function
whose $i$-$j$-th entry $(h_0)^{\! i}_j(z)$ is a span of
$1, z,\cdots,z^{a_i-a_j}$ if $a_i\geq a_j$ and zero if $a_i<a_j$.
It belongs to $\Aut\Ph_a$ if $(h_0)_j^{\!i}(0)=0$ for $i>j$.
Thus, we see
\beq
\dim \Aut \Ph_a
=\dim \Aut \Ph_{[a]}-\sum_{\stackrel{i>j}{a_i\geq a_j}}1
=n-1+\sum_{i<j}\left(\,|a_i-a_j|+\theta_{a_i,a_j}\right)\,,
\eeq
where $\theta_{x,y}=0$ if $x<y$ and $\theta_{x,y}=1$ if $x\geq y$.
An element $a\in \Pv$ minimizes this value in its permutation class
if and only if the entries satisfy $a_1\leq a_2\leq\cdots\leq a_n$.

Remember that for each $j\in \indJ=\{0,\cdots,n-1\}$ there is a smooth
$SU(n)/\Z_n$-bundle $P^{(j)}$ such that $\NNc_{P^{(j)}}=\{\Ph_{[\mu_j]}\}$.
By the above argument, the set of distinct flag structures on
$\Ph_{[\mu_j]}$ is identified with the Weyl orbit $W\mu_j$.
Let $n_{w_jw_0}$ denote the matrix
\beq
\pmatrix{
0 & {\bf 1}_j \cr
{\bf 1}_{n-j}\hspace{-0.3cm} & 0 \cr
}(-1)^{\frac{j(n-j)}{n}}\,,
\eeq
where ${\bf 1}_j$ is the $j\times j$ identity matrix.
Since $a=\ad n_{w_jw_0}^{-1}\mu_j$ satisfies
$a_1\leq \cdots \leq a_n$, it is the unique element in $W\mu_j$
that minimizes the dimension of the symmetry group. Thus,
$\NNc_{P^{(j)},x}$ consists of one point represented by
$\Ph_j=\Ph_{(w_jw_0)^{-1}\mu_j}$.

\vsp
For a general compact group $H$, the story is essentially the same.
Each $a\in \Pv$ indexes
an isomorphism class represented by $\Ph_a=(\Ph_{[a]},f_a)$,
an $\HC$-bundle $\Ph_{[a]}$ described by the transition rule
$\sigma_0^{(a)}=\sigma_{\infty}^{(a)}z^{-a}$ with the flag
$f_a=\sigma_0^{(a)}(0)B$. The group $\Aut\Ph_a$ of automorphisms
has
\beq
\dim\Aut\Ph_a=l+\sum_{\alpha>0}\left(\,|\alpha(a)|+\theta_{\alpha(a),0}\,
\right)\,,
\eeq
which is minimized by $a=(w_jw_0)^{-1}\mu_j$ ($j\in \indJ$).
Thus, $\NNc_{P^{(j)},x}$ consists of one point represented by
$\Ph_j$, an $\HC$-bundle described by the transition rule
\beq
\sigma_0(z)=\sigma_{\infty}(z)z^{-\mu_j}n_{w_jw_0}\,,
\eeq
with the flag $\sigma_0(0)B$, where $n_{w_iw_0}$ is an element of $N_T$
that represents $w_jw_0\in W$.

\vspace{0.3cm}
{\it On Torus with $H=SO(3)$}

\vsp
Next, we describe $\NNc_{P,x}$ for the trivial or the non-trivial
principal $SO(3)$-bundle over the torus $\Sigma_{\tau}=\C/\Z+\tau\Z$.
This time, we realize $\Sigma_{\tau}$ as
$\C^*/q^{\Z}$ where $q^{\Z}$ is the subgroup of $\C^*$
generated by $q=\e^{2\pi i\tau}$. We take $z(x)\equiv 1$
mod $q^{\Z}$.

Below, we list up some holomorphic $\HC=PSL(2,\C)$-bundles over $\Sgmtau$
that are relevant to our story.
Every bundle $\Ph$ is obtained by putting
the relation
\beq
\sigma(qz)=\sigma(z)h(q;z)
\eeq
on the section $\sigma(z)$ of the trivial bundle $\C^*\tms\HC$.
In the following list, bundles and transition functions are exhibited as
$\Ph:h(q;z)$ (where we put $t_u=\e^{-2\pi i u}$).
\begin{center}
[Some Holomorphic Principal $PSL(2,\C)$-Bundles]

\vsp
\begin{minipage}[t]{5cm}
\hspace{0.7cm}trivial
$$
\begin{array}{ll}
\Ph^{(0)}_u\,:&\pmatrix{
t_u& \!0 \cr
0 & t_u^{-1}\! \cr
} \\
&\!\!\!\!\!\mbox{$u\sim \pm u +\frac{m}{2}+\frac{n}{2}\tau$}\\
\noalign{\vskip0.2cm}
\Ph^{(0)}_{00}\,:&\pmatrix{
1 & 1 \cr
0 & 1 \cr
}
\end{array}
$$
\end{minipage}\quad
\begin{minipage}[t]{5cm}
\hspace{0.5cm}non-trivial
$$
\begin{array}{ll}
\Ph_F^{(1)}\,:&\pmatrix{
0 & \!\!\!\!q^{-\frac{1}{4}}z^{-\frac{1}{2}} \cr
-q^{\frac{1}{4}}z^{\frac{1}{2}} &\!\!\!\! 0 \cr
} \\
\noalign{\vskip0.2cm}
\Ph^{(1)}_u\,:&\pmatrix{
it_u^{-1}z^{-\frac{1}{2}} & \!\!\!\,\!\!\!\!0 \cr
0 & \!\!\!\!\!\!-it_u z^{\frac{1}{2}} \cr
}\\
&\!\!\!\!\!\mbox{$u\equiv u +\frac{m}{2}+\frac{n}{2}\tau$}
\end{array}
$$

\end{minipage}
\end{center}
Remark that $\Ph^{(0)}_u\cong \Ph^{(0)}_{u'}
\Leftrightarrow u\equiv \pm u'$ mod $\frac{1}{2}\Z+\frac{\tau}{2}\Z$ and
also that
$\Ph^{(1)}_u\cong \Ph^{(1)}_{u'}\Leftrightarrow u\equiv u'$ mod
$\frac{1}{2}\Z+\frac{\tau}{2}\Z$.
$\{\Ph_u^{(0)}\}$, $\Ph_{00}^{(0)}$ and $\Ph_F^{(1)}$ are all the
semi-stable $\HC$-bundles over $\Sgmtau$. This is implicitly seen in
\cite{At} but here we content ourselves by stating that
$\Ph^{(0)}_u$ and $\Ph_F^{(1)}$ come from the flat $SO(3)$
connections
whose holonomies are (\ref{holonomtriv}) and (\ref{holonomnontriv})
respectively (under $z=\e^{-2\pi i \z}$).
For use in \S 4.4,
some unstable (i.e. non semi-stable) bundles $\Ph^{(1)}_u$
are also included in the list above.

\vsp
To determine isomorphism classes of flag structures at $z\equiv 1$ of these
bundles, we list below the groups of holomorphic automorphisms. An
automorphism $\Ph\to \Ph$ is described by
\beq
h:\sigma(z)\mapsto \sigma(z)h(z)\,,
\eeq
where $h:\C^*\to PSL(2,\C)$ satisfies $h(q;z)h(zq)=h(z)h(q;z)$.
Typical elements $h(z)$ are exhibited in the list below:
\begin{center}
[Automorphism Groups]
\end{center}
\beqa
\noalign{\vskip-0.3cm}
\Aut\Ph_u^{(0)}&\cong&\left\{ \begin{array}{ll}  \C^*\qquad \pmatrix{
c & 0 \cr
0 & c^{-1} \cr
}\,\,\,\quad &\mbox{if $u\not\sim 0,\frac{1}{4},\frac{\tau}{4},
\frac{1+\tau}{4}$}\\
\noalign{\vskip0.2cm}
PSL(2,\C)\qquad h\in PSL(2,\C)\quad &\mbox{if $u\sim 0$}\\
\noalign{\vskip0.2cm}
\C^*\semidir \Z_2\quad \pmatrix{
c & 0 \cr
0 & c^{-1} \cr
}\!,\,\,\pmatrix{
0 & c \cr
-c^{-1} & 0 \cr
}\,\,\quad &\mbox{if $u\sim \frac{1}{4}$}\\
\noalign{\vskip0.2cm}
\C^*\semidir \Z_2\quad \pmatrix{
c & 0 \cr
0 & c^{-1} \cr
}\!,\,\,\pmatrix{
0 & \!\!cz^{\nibun} \cr
-c^{-1}z^{-\nibun} & \!\!0 \cr
}\,\,\, \,\, &\mbox{if $u\sim \frac{\tau}{4},\,\frac{1+\tau}{4}$}
\end{array} \right.\\
\Aut\Ph_{00}^{(0)}&\cong&\C\qquad \pmatrix{
1 & x \cr
0 & 1 \cr
} \,\,\,\,\\
\Aut\Ph_F^{(1)}&\cong&\Z_2\!\times\Z_2=\left\{\pmatrix{
1 & 0 \cr
0 & 1 \cr
}\!,\,\,\pmatrix{
i & 0 \cr
0 & -i \cr
}\!,\,\,\pmatrix{
0 & i \cr
i & 0 \cr
}\!,\,\,\pmatrix{
0 & -1 \cr
1 & 0 \cr
}\right\},\label{symflatnontriv}\\
\Aut\Ph_u^{(1)}&\cong&B_0^-\qquad \pmatrix{
c & \!\!0 \cr
x\vartheta_{\!\tau,u}(z) & \!\!c^{-1} \cr
}
\eeqa
In the above expression, $\vartheta_{\!\tau,u}$ is
given by $\vartheta_{\!\tau,u}(z)
=\vartheta(\tau, \zeta+2u+\frac{1+\tau}{2})$ where $\vartheta(\tau, \zeta)$
is the Riemann's theta function $\sum_{n\in \Z}q^{\nibun n^2}z^{-n}$;
$q=e^{2\pi i\tau}$, $z=e^{-2\pi i \zeta}$. Note that
$\vartheta_{\!\tau,u}(1)=0$ if and only if $u\equiv 0$
mod $\frac{1}{2}\Z+\frac{\tau}{2}\Z$.

\vsp
The flag manifold over $z\equiv 1$ is identified with the Riemann sphere
$\C\cup\infty$ by
\beq
y\in \C\cup\{\infty\}\mapsto \sigma(1)\pmatrix{
a & b \cr
c & d \cr
}B\in\Ph_{z=1}/B\qquad;\quad y=c/a.
\eeq
Looking at the action of $\Aut \Ph$ on $\Ph_{z=1}/B$,
we see that the flag structures over our
holomorphic bundles are classified as follows:
\begin{center}
[Some Holomorphic $PSL(2,\C)$-Bundles With Flag Structure]

\vsp
\begin{minipage}[t]{6cm}
\hspace{1.5cm}trivial
\beqa
&&\left.
\begin{array}{lc}
(\Ph_u^{(0)},1)&1\\
(\Ph_u^{(0)},0)&\C^*\\
(\Ph_u^{(0)},\infty)&\C^*
\end{array}
\right\}
\mbox{$u\not\sim 0,\frac{1}{4},\frac{\tau}{4},\frac{\tau+1}{4}$}\nonumber\\
\noalign{\vskip0.2cm}
&&\left.
\begin{array}{lc}
(\Ph_u^{(0)},1)&\Z_2\\
(\Ph_u^{(0)},0)&\C^*
\end{array}
\right\}
\mbox{$u\sim\frac{1}{4},\frac{\tau}{4},\frac{\tau+1}{4}$}\nonumber\\
\noalign{\vskip0.2cm}
&&\,\,\,(\Ph_0^{(0)},1)\quad B \nonumber\\
\noalign{\vskip0.2cm}
&&\left.
\begin{array}{lc}
(\Ph_{00}^{(0)},\infty)&1\\
(\Ph_{00}^{(0)},0)&\C
\end{array}
\right\}\nonumber
\eeqa

\end{minipage}\quad
\begin{minipage}[t]{6cm}
\hspace{1.5cm}non-trivial
\beqa
&&\left.
\begin{array}{lcl}
(\Ph_F^{(1)},y)&1&y\not\sim 0,1,i\\
(\Ph_F^{(1)},y)&\Z_2&y\sim 0,1,i
\end{array}
\right\}\nonumber\\
\noalign{\vskip0.1cm}
&&\qquad \mbox{$y\sim -y\sim y^{-1}\sim -y^{-1}$}\nonumber\\
\noalign{\vskip0.2cm}
&&\left.
\begin{array}{lc}
(\Ph_u^{(1)},0)&\C^*\\
(\Ph_u^{(1)},\infty)&B
\end{array}
\right\}
u\not\equiv 0 \nonumber\\
\noalign{\vskip0.2cm}
&&\left.
\begin{array}{lc}
(\Ph_0^{(1)},1)&\C\\
(\Ph_0^{(1)},0)&B\\
(\Ph_0^{(1)},\infty)&B
\end{array}
\right\}\nonumber
\eeqa

\end{minipage}

\end{center}
Note that $(\Ph_F^{(1)},y)\cong (\Ph_F^{(1)},y')$ if and only if
$y'=y,-y,y^{-1}$ or $-y^{-1}$. The group $\Aut(\Ph,f)$ is presented
in the list on the right of $(\Ph,f)$.

\vsp
Recall that the moduli space $\NNc_{{\rm triv}}$ for the trivial topology
is represented by the family $\{\Ph_u^{(0)}\}_{u\not\sim 0,\frac{1}{4},
\frac{\tau}{4},\frac{1+\tau}{4}}$ whereas for the non-trivial topology
$\NNc_{P,x}$ is the one point set represented by
$\Ph_F^{(1)}$. Counting the dimension of the automorphism groups,
we see that $\NNc_{{\rm triv},x}$ and $\NN_{{\rm non-triv},x}$ are
represented by the families $\{(\Ph_u^{(0)},1)\}_{u\not\sim 0,\frac{1}{4},
\frac{\tau}{4},\frac{\tau+1}{4}}$ and $\{(\Ph_F^{(1)},y)\}_{y\in \C}$
respectively. Namely,
\beqa
\NNc_{{\rm triv},x}\!&\cong&\!\!\C\mbox{\LARGE /}
\!\Bigl\{(\mbox{$\nibun\Z+\frac{\tau}{2}\Z$})\semidir \Z_2\Bigr\}
-\Bigl\{\mbox{$[0],[\frac{1}{4}],[\frac{\tau}{4}],[\frac{\tau+1}{4}]$}
\Bigr\},\\
\NNc_{{\rm non-triv}, x}\!\!&\cong&\!\! \left(\Z_2\times \Z_2\right)
\!\mbox{\LARGE $\backslash$} \CP\,.
\label{trivnontriv}
\eeqa
If $\NNc_{{\rm triv},x}$ is compactified by attaching the points
$(\Ph_u^{(0)},1)$; $u=\frac{1}{4},\frac{\tau}{4},\frac{\tau+1}{4}$ and
$(\Ph_{00}^{(0)},\infty)$, then we see that the compactified moduli space
$\overline{\NNc_{{\rm triv},x}}$ coinsides topologically with
$\NNc_{{\rm non-triv},x} \cong S^2$.
Moreover, it seems that the families of automorphism groups coincide with
each other: Generically there is no non-trivial symmetry,
but there are three points with $\Aut\cong \Z_2$.
In \S 4.4, we shall construct a bijection between
$\A_{{\rm triv},x}/\G_{\rm triv}^{\bf c}$ and
$\A_{{\rm non-triv},x}/\G_{\rm non-triv}^{\bf c}$
which induces an isomorphism
$\overline{\NNc_{{\rm triv},x}}\cong \NNc_{{\rm non-triv},x}$.
In fact, this is an essential step in the derivation of (\ref{FI}).

\vspace{0.5cm}
{\sc 4.4 Action Of $\pi_1(H)$ On The Moduli Spaces}

\vspace{0.3cm}
Let $(\Sigma, x)$ be a closed Riemann surface with a point in it.
We choose a neighborhood $U_0$ of $x$ with a coordinate $z$
such that $z(x)=0$ and that $z(U_0)$ is an open disc.
A holomorphic principal $\C^*$-bundle admits trivializations over
$U_0$ and $U_{\infty}=\Sigma-x$ that are related by
a holomorphic transition function
$h_{\infty 0}:U_0\cap U_{\infty}\to \C^*$. For each $a\in \Z$,
the transformation
\beq
h_{\infty 0}(z)\mapsto h_{\infty 0}(z)z^{-a}
\eeq
of transition functions induces the translation of the group $\Pic(\Sigma)$
by an element of first chern class $a$ (see \S 3.1). This defines an action
of $\pi_1(U(1))\cong\Z$ on $\Pic(\Sigma)$ that covers the natural action
on the set $H^2(\Sigma,\Z)\cong\Z$ of topological types of $U(1)$-bundles.
This action depends on $x$ but not on the choice of coordinate $z$.

We ask whether such an action exists for
a general compact connected group $H$:
Does the natural action of $\pi_1(H)$ on the set of topological types
of principal $H$-bundles lift to an action on the set $\Pic^{\Hc}(\Sigma)$
of isomorphism classes of holomorphic principal $\HC$-bundles?
As an answer, we shall find that, instead of on $\Pic^{\Hc}(\Sigma)$,
$\pi_1(H)$
acts on the set $\Pic^{\Hc}(\Sigma,x)$ of isomorphism classes of
holomorphic principal $\HC$-bundles with flag structure at $x$.
We conjecture that the action permutes the moduli spaces $\NNc_{P,x}$.
It is verified on the sphere for a general group
and on torus for $H=SO(3)$.

\vspace{0.3cm}
{\it Action Of $\pi_1(H)$ On $\Pic^{\Hc}(\Sigma,x)$}

\vsp
We first describe $\Pic^{\Hc}(\Sigma,x)$ in terms of the loop group $L\HC$.
For a holomorphic principal $\HC$-bundle $\Ph$ with a flag $f\in \Ph_x/B$,
we say that a section of $\Ph|_{U_0}$ is {\it admissible}
with respect to $f$ when it represents $f$ over $x$.
Let $h:(\Ph,f)\to(\Ph',f')$ be an isomorphism. Under a choice of
trivializations $\{\sigma_0,\sigma_{\infty}\}$ and
$\{\sigma'_0,\sigma'_{\infty}\}$ over $\{U_0,U_{\infty}\}$ of $\Ph$ and
$\Ph'$ respectively such that $\sigma_0$ and $\sigma'_0$ are admissible
with repect to $f$ and $f'$, $h$ is represented by $\HC$-valued
holomorphic functions $\{h_0,h_{\infty}\}$ on $\{U_0,U_{\infty}\}$ with
$h_0(x)\in B$ : $\sigma_I\mapsto\sigma'_Ih_I$ ($I=0,\infty$).
Then, the transition functions $h_{\infty 0}$ and $h'_{\infty 0}$
are subject to the relation
\beq
h'_{\infty 0}(z)=h_{\infty}(z)h_{\infty 0}(z)h_0(z)^{-1}\qquad
z\in U_0\cap U_{\infty}.
\label{auto}
\eeq
For an open Riemann surface $U$,
we denote by $L^U\!\HC$ the group of holomorphic
maps $U\to \HC$. Pulling back by inclusions $U_{\infty 0}=U_0\cap U_{\infty}
\hookrightarrow U_0,U_{\infty}$, the groups $L^{U_0}\!\HC$ and
$L^{U_{\infty}}\!\HC$ may be considered as subgroups of
$L^{U_{\!\!\,\infty 0}}\!\HC$.(\footnote{
With a choice $S^1\hookrightarrow U_{\infty 0}$ of parametrized circle,
the group $L^{U_{\infty 0}}\HC$ is identified with a dense open subgroup
of the loop group $L\HC$. This is the origin of the notation.})
We denote by $B^{U_0}$ the subgroup of $L^{U_0}\!\HC$ consisting of maps
with values at $x$ in $B$. By the above argument,
the set $\Pic^{\Hc}(\Sigma,x)$ of isomorphism classes is identified
with the set of double cosets:
\beq
\Pic^{\Hc}(\Sigma,x)\cong L^{U_{\infty}}\!\HC\lbackslash
L^{U_{\!\!\,\infty 0}}\!\HC
/B^{U_0}\,.
\label{dblcosets}
\eeq

\vsp
The fundamental group $\pi_1(H)$ is isomorphic to
the subgroup $\Gmalcv$ of the affine Weyl group $\Waffh$
consisting of elements that preserve the al\^ove $\alcv$ (see Appendix A).
For each $\gamma\in \Gmalcv$, there is a holomorphic extension
$h_{\gamma}:\C^*\to \HC$.
For example, $h_{\gamma_j}(z)=z^{-\mu_j}n_{w_jw_0}$.
Using the coordinate $z:U_{\infty 0}\to \C^*$,
we identify $h_{\gamma}$ as
an element of $L^{U_{\infty 0}}\HC$.
Since the adjoint action of $h_{\gamma}$ on
$L^{U_{\infty 0}}\HC$ preserves the subgroup
$B^{U_0}\subset L^{U_{\infty 0}}\HC$, we find in view of (\ref{dblcosets})
that the transformation
\beq
h_{\infty 0}(z)\mapsto h_{\infty 0}(z)h_{\gamma}(z)
\eeq
of transition functions induces the transformation
\beq
\gamma_x : \Pic^{\Hc}(\Sigma,x)\longto \Pic^{\Hc}(\Sigma,x)\,.
\eeq
This transformation changes the homotopy type of the transition function by
$\gamma\in\pi_1(H)$ and hence permutes the subsets
$\{\APx/\GPC\}_P$:
\beq
\gamma_x : \APx/\GPC \longto \A_{P\gamma,x}/\G_{P\gamma_{\bf c}}\,.
\eeq
Thus $\gamma\mapsto \gamma_x$ is the desired action of $\pi_1(H)$ on
$\Pic^{\Hc}(\Sigma,x)$.

\vspace{0.3cm}
{\it The Conjecture}

\vsp
One important thing to notice is that this action preserves
the automorphism groups.
Namely, if the class of $(\Ph,f)$ is mapped by $\gamma_x$ to a class
represented by $(\Ph^{\gamma},f^{\gamma})$, we have
\beq
\Aut(\Ph,f)\cong \Aut(\Ph^{\gamma},f^{\gamma})\,.
\eeq
This can be seen by multiplying $h_{\gamma}(z)$ on the right to both sides
of (\ref{auto}) in which we put $h'_{\infty 0}=h_{\infty 0}$.

Recall that an element of the moduli space
$\NNc_{P,x}\subset\Pic^{\Hc}(\Sigma,x)$ is represented by $(\Ph,f)$
that satisfies the following conditions: $\Ph$ represents an element of
$\NNc_P$ and $\dim\Aut(\Ph,f)\leq \dim\Aut(\Ph,\tilde{f})$ for every flag
$\tilde{f}$ at $x$.

\vsp
Having these in mind, we conjecture that the following holds:
{\it There is a method to compactify the moduli space $\NNc_{P,x}$
in such a way that for each $\gamma\in \Gmalcv$, $\gamma_x$ maps
the compactified moduli space $\overline{\NNc_{P,x}}$
isomorphically onto another space $\overline{\NNc_{P\gamma,x}}$.}

If, furthermore, we can compactify $\NNc_P$ so that the natural projection
$\NNc_{P,x}\to \NNc_P$ (forgetting the flags) extends to a surjective map
$\overline{\NNc_{P,x}}\to \overline{\NNc_P}$,
we have the following double fibration:
\beq
\begin{array}{ccccc}
&&\!\!\!\!\!\!\!\overline{\NNc_{P,x}}\cong\overline{\NNc_{P\gamma,x}}
\!\!\!\!\!\!\!\!\!&&\\
\noalign{\vskip0.2cm}
&\mbox{\Large $\swarrow$}&&\mbox{\Large $\searrow$}&\\
\noalign{\vskip0.2cm}
\overline{\NNc_P}\!\!\!&&&&\!\!\!\overline{\NNc_{P\gamma}}
\end{array}
\eeq
This seems to be what mathematicians call the {\it Hecke correspondence}.
\cite{Narasimhan-Ramanan}

\vspace{0.3cm}
{\it Verification On The Sphere}

\vsp
As seen in \S 4.3, the moduli space $\NNc_{P^{(i)},x}$ for $z(x)=0$
is the one point represented by $\Ph_i$ that is described
by the transition relation $\sigma_0(z)=\sigma_{\infty}(z)h_{\gamma_i}(z)$
where $\sigma_0$ is admissible.
The $\gamma_{jx}$-transform of $\Ph_i$ is then described by
$\sigma^{\gamma_j}_0(z)
=\sigma^{\gamma_j}_{\infty}(z)h_{\gamma_i}(z)h_{\gamma_j}(z)$ where
$\sigma^{\gamma_j}_0$ is now admissible.
Hence we see that
\beq
\gamma_{jx}:\NNc_{P^{(i)},x}\longto\NNc_{P^{(i\circ j)},x}\,,
\eeq
where $\gamma_{i\circ j}=\gamma_i\gamma_j$. The conjecture is thus verified
on the sphere.

\vspace{0.3cm}
{\it Verification On Torus With $H=SO(3)$}

\vsp
For $H=SO(3)$, the non-trivial element $\gamma$ of $\Gmalcv\cong\Z_2$
is represented by a path
\beq
\gamma(\theta)=\pmatrix{
0 & -\e^{-\frac{i}{2}\theta} \cr
\e^{\frac{i}{2}\theta} & 0 \cr
},\qquad 0\leq\theta\leq 2\pi
\eeq
in $SU(2)$. We apply $\gamma_x$ to the topologically trivial semi-stable
bundles with flag structure at $z(x)=1$.

A $PSL(2,\C)$-bundle $\Ph$ we consider is described by
the transition relation
$\sigma(zq)=\sigma(z)h(q;z)$ and a flag is parametrized by
$y\in \C\cup\{\infty\}$.
If we choose a matrix $h_f\in SL(2,\C)$ such that
$(h_f)^{\!2}_1/(h_f)^{\!1}_1=y$, then, $\sigma_0(z)=\sigma(z)h_f$ is an
admissible section on a small neighborhood $U_0$ of $z=1$.
Hence, the $\gamma_x$-transform of $(\Ph,y)$ is represented by
a bundle $\Ph^{\gamma}$ with an admissible section $\sigma_0^{\gamma}$
on $U_0$ and a section $\sigma'$ on $\C^*-q^{\Z}$
that are related by
\beqa
\sigma_0^{\gamma}(z)&=&\sigma'(z)h_f \,h_{\gamma}(z-1)\,,
\qquad z\in U_0-\{1\},\label{4.4rel-1}\\
\sigma'(zq)&=&\sigma'(z)h(q;z)\,,\qquad
\mbox{$z\not\equiv 1$ mod $q^{\Z}$}.\label{4.4rel-2}
\eeqa

The conservation $\Aut(\Ph,f)\cong \Aut(\Ph^{\gamma},f^{\gamma})$
of automorphism groups enables us to guess
how $\gamma_x$ transforms the bundles listed in \S 4.3.
After a calculation, we find
the following solution (see Appendix D for the proof):
\begin{center}
[The Transformation $\gamma_x$]
\end{center}
\beqa
\noalign{\vskip-0.2cm}
\hspace{1.2cm}(\Ph_u^{(0)},1)&\to&(\Ph_F^{(1)},y_u)\quad u\not\sim 0
\nonumber\\
(\Ph_{00}^{(0)},\infty)&\to&(\Ph_F^{(1)},y_0)\nonumber\\
(\Ph_u^{(0)},0)&\to&(\Ph_u^{(1)},0)\label{rulegmax}\\
(\Ph_u^{(0)},\infty)&\to&(\Ph_{-u}^{(1)},0)\nonumber\\
(\Ph_{00}^{(0)},0)&\to&(\Ph_0^{(1)},1)\nonumber\\
\noalign{\vskip0.3cm}
\hspace{-5cm}\mbox{where}\hspace{5cm}&&\hspace{-2.5cm}
y_u=iq^{\frac{1}{4}}e^{2\pi i u}
\displaystyle{\frac{\vartheta(2\tau, 2u+\tau)}{\vartheta(2\tau, 2u)}}.
\label{hecketor}
\eeqa
The most important point to notice is that the compactified moduli space
$\overline{\NNc_{{\rm triv},x}}$ represented by
$\{(\Ph_u^{(0)},1)\}_{u\not\sim 0}$ and $(\Ph_{00}^{(0)},\infty)$
is mapped bijectively to the (compact) moduli space
$\NNc_{{\rm non-triv},x}$ of flag structures on the semi-stable bundle
$\Ph_F^{(1)}$:
\beq
\gamma_x:\overline{\NNc_{{\rm triv},x}}\longto\NNc_{{\rm non-triv},x}\,.
\label{gmaxtorus}
\eeq
In terms of the coordinates $u$ and $y$, this map is given by
$u\mapsto y_u$ where $y_u$ is given in (\ref{hecketor}) and satisfies
$y_{-u}=y_{u}$, $y_{u+\frac{1}{2}}=-y_u$ and
$y_{u+\frac{\tau}{2}}=-y_u^{-1}$.
The orbifold points
$u\sim \frac{1}{4},\frac{\tau}{4},\frac{\tau+1}{4}$ are mapped to
the orbifold points $y\sim 0,i,1$ respectively,
whereas $(\Ph_{00}^{(0)},\infty)$ is mapped to
the smooth point $y_0$. If we decide to take $u^2$ as the complex
coordinate around the point $(\Ph_{00}^{(0)},\infty)$,
the bijection (\ref{gmaxtorus}) becomes
an isomorphism and the conjecture is verified also in this case.

{\it Remark}. Even though bundles $\{\Ph^{(0)}_u\}$ are semi-stable,
the points $\{(\Ph_u^{(0)},0)\}$ which do not
lie in $\overline{\NNc_{{\rm triv},x}}$ are mapped by $\gamma_x$ to
$\{(\Ph_u^{(1)},0)\}$ that are projected to unstable bundles
by the `flag forgetful map'.

\newcommand{\Htot}{{\cal H}^{\tot}}
\newcommand{\Hflg}{\,\,\widehat{\!\!\cal H\!\!}_{\,\,\rm inv}}

\vspace{0.5cm}
{\sc 4.5 The Topological Identity}

\vspace{0.3cm}
We have just seen that the holomorphic function $h_{\gamma}(z)$ for
$\gamma\in \Gmalcv\cong\pi_1(H)$ induces the bijection
$\gamma_x:\APx/\GPC\to \A_{P\gamma,x}/\G_{P\gamma_{\bf c}}$.
In \S 2.3, we have seen (for the case $H=G/Z_G$) that the gauge
transformation by the same function $h_{\gamma}(z)$ induces
the spectral flow which maps the highest weight state $\Phi_{\Lmd}$
to another $\Phi_{\gamma\Lmd}$. Applying such spectral flow to
the flag partners, we define the action of $\pi_1(H)$ on the space of
gauge invariant local fields. Under the assumption that the conjecture
is verified, we shall observe that the double role of $h_{\gamma}$
results in the identity (\ref{FI}).

\vspace{0.3cm}
{\it States Corresponding To The Flag Partners}

\vsp
{}From the state space
$\Htot={\cal H}^{G,k}\ot\Gamma(\Lg)\ot\Fadgh$
of the total system,
we shall select out the subspace $\Hflg$ corresponding to the flag partner
fields. Let $\Phi\in{\cal H}_{\Lmd}^{\lmd}$. The expression
(\ref{flagpartner}) shows that the following is the state
corresponding to the flag partner of $O_{\Phi}$
associated to the standard flag $f_0=s_0(0)B$:
\beq
Z^{\tot}_{D_0}(0;\widehat{O}_{\Phi}(f_0))=\Phi\ot \Phi_{-\lmd-2\rho}\ot
|\Omega\rangle\,.
\eeq
Here, $\Phi_{-\lmd-2\rho}\in\Gamma(\Lg)$ is the highest
weight state (\ref{HChw}) and $|\Omega\rangle\in \Fadgh$ is defined by
$\prod_{-\alpha<0}c_0^{-\alpha}\bar c_0^{-\alpha}|0\rangle$ and
satisfies
\beqa
b_n(v)|\Omega\rangle\!&=&\!0\qquad\mbox{for $n\geq 1$, $v\in \h_{\C}$
or $n=0$, $v\in \b$},\label{condrayb}\\
c_n(v^*)|\Omega\rangle\!&=&\!0\quad\mbox{for $n\geq 1$,
$v^*\in\h_{\C}^*$ and}
\quad c_0^{-\alpha}|\Omega\rangle =0\quad \alpha\in \Delta_+\,.
\label{condrayc}
\eeqa
Thus we see that
\beq
\Hflg=\bigoplus_{\Lmd,\lmd}{\cal H}^{\lmd}_{\Lmd}\ot \Phi_{-\lmd-2\rho}
\ot |\Omega\rangle\,.
\eeq

\vspace{0.3cm}
{\it Action Of $\pi_1(H)$ On Gauge Invariant Local Fields}

\vsp
Let $\gamma\in \Gmalcv\cong \pi_1(H)$ be represented by the loop
$\gamma(\theta)=\e^{-i\mu\theta}n_w$. We consider the configuration
$A_{\varrho,w^{-1}\mu}$ of gauge field which is introduced in \S 2.3.
For a gauge invariant local field $O$, we shall define another field
$\gamma O$ by
\beq
Z^{\tot}_{D_0}(0;\,\,\,\widehat{\!\!\!\gamma O\!\!\!}\,\,\,(f_0))
=\gamma.Z^{\tot}_{D_0}(A_{\varrho,w^{-1}\mu};\widehat{O}(f_0))\,,
\eeq
where $\gamma.$ is the gauge transformation on $\Htot$ induced by the loop
$\gamma\in LH$. Below, we show that the right hand side belongs to
$\Hflg$ and is independent on the choices involved, such as
the representative $n_w$ of $w$ or as the function $\varrho$ etc.

As we have seen in \S 2.3, $A_{\varrho,w^{-1}\mu}$ can be made flat by a
chiral gauge transformation and the right hand side is expressed as
$\gamma.{\cal J}(\tilde{c})Z^{\tot}_{D_0}(0;\widehat{O}(f_0))$
where ${\cal J}(\tilde{c})$ is the action of an element $\tilde{c}$ of
$\tiL \HC$ over a constant loop in $T_{\!\C}$.
Since the total central charge vanishes and the state
$Z^{\tot}_{D_0}(0;\widehat{O}(f_0))$ has weight zero,
we see that the right hand side for $O=O_{\Phi}$ is given by
$\gamma.\Phi\ot\gamma.\Phi_{\lmd-2\rho}\ot\gamma.|\Omega\rangle$.

Since $\gamma.$ preserves the highest weight condition for $\tiL H_{\C}$,
it sends ${\cal H}_{\Lmd}^{\lmd}$ to
${\cal H}_{\gamma_G\Lmd}^{\gamma\lmd}$ where $\gamma_G$ is the image of
$\gamma$ under the natural map $\pi_1(H)\to\pi_1(G/Z_G)$.
Also, it preserves the ray $\C|\Omega\rangle$ characterized
by (\ref{condrayb}) and (\ref{condrayc}), and
a calculation shows that $\gamma.|\Omega\rangle=|\Omega\rangle$.
Finally, we recall that
$\gamma.\Phi_{-\lmd-2\rho}=\Phi_{-\gamma\lmd-2\rho}$.
Combining these all, we have
\beq
Z^{\tot}_{D_0}(0;{\,\,\,\widehat{\!\!\!\gamma O\!\!\!}\,\,\,}_{\Phi}(f_0))
=\gamma.\Phi\ot\Phi_{-\gamma\lmd-2\rho}\ot |\Omega\rangle\,,
\eeq
for $\Phi\in{\cal H}_{\Lmd}^{\lmd}$. Thus, the action of $\pi_1(H)$ on
the gauge invariant local fields is defined and is identified
with the action of $\Gmalcv(H)$ on $\oplus {\cal H}_{\Lmd}^{\lmd}$.

\vspace{0.3cm}
{\it Proof Of (\ref{FI})}

\vsp
We are now in a position to prove (\ref{FI}). We make use of
the new integral expression (\ref{newintexpr}).
Let $\VV$ be an open subet of $\NNc_{P,x}$ with a holomorphic family
$\{(A_v,f_v)\}_{v\in \VV}$ of representatives.
The non-anomalous chiral gauge symmetry of the total system enables us
to take the representatives so that there is a family
$\{\sigma_0(v)\}_{v\in \VV}$ of {\it horizontal} and {\it admissible}
sections on a neighborhood $U_0$ of $x$.

We choose a complex coordinate $z$ on $U_0$ such that $z(x)=0$ and
$z(U_0)$ contains the unit disc $D_0$. Let $\gamma\in LH$ be
a representative loop of $\gamma\in\Gmalcv\cong\pi_1(H)$.
We put $\Sigma_{\infty}=\overline{\Sigma-D_0}$.
Gluing $(P|_{\Sigma_{\infty}}, A_v|_{\Sigma_{\infty}})$ and
$(D_0\times H,A_{\varrho,w^{-1}\mu})$ by the identification of
$\sigma_0(v,\e^{i\theta})\gamma(\theta)$ and $s_0(\e^{i\theta})
:=(\e^{i\theta},1)$, we obtain an $H$-bundle $P\gamma$ over $\Sigma$ with
a smooth connection $A_v^{\gamma}$.
We denote by $\sigma(v)$ and $s_0^{\gamma}(v)$ the sections of $P\gamma$
over $\Sigma_{\infty}\cap U_0$ and $D_0$ respectively
which had been $\sigma_0(v)|_{\Sigma_{\infty}\cap U_0}$ and $s_0$
before the gluing. If we put $f_v^{\gamma}=s_0^{\gamma}(v,x)B$,
we can find a holomorphic and admissible section $\sigma_0^{\gamma}(v)$
of $(A_v^{\gamma},f_v^{\gamma})$ over $U_0$ such that
$\sigma_0^{\gamma}(v,z)=\sigma(v,z)h_{\gamma}(z)$ on a neighborhood of $S$.
This shows that the transformation
\beq
(A_v,f_v)\mapsto (A_v^{\gamma},f_v^{\gamma})\,,
\eeq
represents $\gamma_x:\VV\to \gamma_x(\VV)$.
If the conjecture is verified,
$\{(A_v^{\gamma},f_v^{\gamma})\}_{v\in \VV}$ is a family of representatives
over $\gamma_x(\VV)\subset \overline{\NNc_{P\gamma,x}}$.

\vsp
By construction of the $\pi_1(H)$ action on gauge invariant local fields,
we have
\beq
Z^{\tot(\sigma_0(v))}_{D_0}
(A_v;\,\,\,\widehat{\!\!\!\gamma O\!\!\!}\,\,\,(f_v))
=Z^{\tot(\sigma(v))}_{D_0}(A_v^{\gamma};\widehat{O}(f_v^{\gamma}))\,.
\label{locFI}
\eeq
If we choose the gauge fixing function $F_v$ for $\Aut(A_v,f_v)$
so that $F_v(h)$ is independent on $h|_{U_0}$,
then it defines the gauge fixing function $F_v^{\gamma}$ for
$\Aut(A_v^{\gamma},f_v^{\gamma})$ by the identification
$P|_{\Sigma_{\infty}}\cong P\gamma|_{\Sigma_{\infty}}$. Now we see that
(\ref{locFI}) leads to the equality
\beqa
\lefteqn{Z_{\Sigma,P}^{\tot}\Bigl(\,A_v\,\mbox{\Large ;}\,
|\rgf_{\!A_v\!,f_v}\!(c,h)|^2
h(\cdots)
\prod_{\rA=1}^{\dNf}\!\ooint_x \!b\,\nu_{\!\rA}(v)
\ooint_x \!\bar b\, \bnu_{\!\rA}(v)\,
\,\,\,\widehat{\!\!\!\gamma O\!\!\!}\,\,\,(f_v)\,\Bigr)}\\
&&=
Z_{\Sigma,P\gamma}^{\tot}\Bigl(\,A_v^{\gamma}\,\mbox{\Large ;} \,
|\rgf_{\!A_v^{\gamma}\!,f_v^{\gamma}}\!(c,h)|^2
h(\cdots)
\prod_{\rA=1}^{\dNf}\!\ooint_x \!b\,\nu^{\gamma}_{\!\rA}(v)
\ooint_x \!\bar b\, \bnu^{\gamma}_{\!\rA}(v)\,
\widehat{O}(f_v^{\gamma})\,\Bigr)\,,\nonumber
\eeqa
where $\nu^{\gamma}_{\rA}(v)$ is given by
\beq
\nu^{\gamma}_{\!\rA}(v)
=\sigma(v)\cdot
h_{\infty 0}(v)^{-1}\frac{\partial}{\partial v^{\rA}}h_{\infty 0}(v)
=\sigma^{\gamma}_0(v)\cdot
h_{\infty 0}^{\gamma}(v)^{-1}
\frac{\partial}{\partial v^{\rA}}h_{\infty 0}^{\gamma}(v)\,,
\eeq
in which $h_{\infty 0}^{\gamma}(v)=h_{\infty 0}(v)h_{\gamma}$.
This amounts to
\beq
\Omega^{\tot}_{\Sigma,P,x}(O_1\cdots O_s
\,\,\,\widehat{\!\!\!\gamma O\!\!\!}\,\,\,)
=\gamma_x^*\Omega^{\tot}_{\Sigma,P\gamma,x}(O_1\cdots O_s \widehat{O})\,,
\eeq
which shows (\ref{FI}).

\renewcommand{\theequation}{5.\arabic{equation}}\setcounter{equation}{0}

\newcommand{\Heq}{\dot{\cal H}}
\newcommand{\Heqhw}{\dot{\cal H}_{\rm hw}}
\newcommand{\Ptriv}{P_{\rm triv}}
\newcommand{\gmaG}{\gamma^{}_{\!{}_{\ssstyle G}}}
\newcommand{\triv}{{\rm triv}}
\newcommand{\nt}{{\rm non-triv}}

\newpage
\vsp
\begin{center}
{\bf 5. Sum Over Topologies}
\end{center}

The full correlation function of the gauge invariant fields $O_1\cdots O_s$
is given by
\beq
Z_{\Sigma}(O_1\cdots O_s)=\sum_P Z_{\Sigma,P}(O_1\cdots O_s)\,,
\label{defull}
\eeq
where the sum is over all topological types of principal $H$-bundles
over $\Sigma$. If we use the topological identity (\ref{FI}), we have
\beq
Z_{\Sigma}(O_1\cdots O_s O)=Z_{\Sigma,P}(O_1\cdots O_s \underline{O})
\,,\label{useFI}
\eeq
in which $P$ is any principal $H$-bundle and
$\underline{O}:=\sum_{\gamma\in\pi_1(H)}\gamma O$.
We say that two gauge invariant local fields are {\it equivalent} when they
are indistinguishable in any full correlator, and we denote by $\Heq$
the set of equivalence classes of gauge invariant local fields.
The equation (\ref{useFI}) shows that $O$ and $O'$ are equivalent
if and only if $\underline{O}=\underline{O}'$. In particullar, $O$ and
$\gamma O$ are equivalent and so are $O$ and $\frac{1}{2}(O+\gamma O)$.
Thus, $\Heq$ is the quotient of the space ${\cal H}$ of gauge invariant
local fields by the kernel of the operator $\sum_{\gamma\in\pi_1(H)}\gamma$.
Since the $\pi_1(H)$-action on ${\cal H}$ is identified with
the $\Gmalcv$-action on $\Hhw$, $\Heq$ is in one to one correspondence
with the quotient $\Heqhw$ of $\Hhw$ by the kernel of
$\sum_{\gamma\in\Gmalcv}\gamma.$. By the general principle of CFT,
we expect that the torus partition funtion $Z_{\Sgmtau}(1)$
satisfies
\beq
Z_{\Sgmtau}(1)
=\tr^{}_{\Heqhw}\!(q^{L_0-\frac{c}{24}}\bar q^{\bar L_0-\frac{c}{24}})\,,
\label{principle}
\eeq
in which $c=c_{\tot}=c_{G,k}-c_{\tilH,\tilk}$
and
$L_0$ and $\bar L_0$ are dilatation generators
$L_0^{\ssstyle {\rm GKO}}$ and $\bar L_0^{\ssstyle {\rm GKO}}$
by GKO construction; the generators $L_n^{\ssstyle {\rm GKO}}$ and
$\bar L_n^{\ssstyle {\rm GKO}}$ commute with
the operators $\gamma.$ for $\gamma\in \Gmalcv$ and hence can act
on the quotient space $\Heqhw$.

In this section, we calculate the full partition funtion on the torus
$\Sgmtau$ and see whether (\ref{principle}) holds.
For simplicity of the argument, we consider only the case in which
$H$ is semi-simple. In this case, $\pi_1(H)$ is a finite group and
the quotient $\Heqhw$ is mapped by $\frac{1}{|\pi_1(H)|}\sum_{\gamma}\gamma$.
isomorphically (as Virasoro module) onto the subspace $\Hhw^{\Gmalcv}$ of
$\Gmalcv$-invariant elements.

\vspace{0.5cm}
{\sc 5.1 Torus Partition Function For The Trivial Topology}

\vspace{0.3cm}
We start with the calculation of the partition function for
the trivial bundle $\Ptriv=\Sgmtau\tms H$. We recall that the moduli space
$\NN_H=\NN_{\Ptriv}$ is parametrized by $u\in \liet_{\C}$ with
the representative family
\beq
u\mapsto A_u=\frac{\pi}{\tau_2}ud\bz-\frac{\pi}{\tau_2}\bar ud\z
\eeq
of flat gauge fields. $A_{u'}$ is gauge equivalent to $A_u$ if and only if
$u'=wu+n+\tau m$ for some $w\in W$ and $n,m\in \Pv$.

\vsp
The partition function of the WZW model with the target $G$ is given by
\beq
Z^{G,k}_{\Sgmtau,\Ptriv}(A_u;1)=e^{\pinitau k \trG(u-\bar u)^2}\!\!\!\!
\!\!\!\sum_{\Lmd\in \PPpk(G)}
\!\!\left|\Chi_{\Lmd}^{G,k}(\tau,u)\right|^2\,,
\label{torWZW}
\eeq
(see \cite{GW,B,GawKup})
in which $\Chi_{\Lmd}^{G,k}$ is the character of
the representation $L_{\Lmd}^{G,k}$ of $\tiL G$:
\beq
\Chi_{\Lmd}^{G,k}(\tau,u)
=\tr^{}_{L_{\Lmd}^{G,k}}(q^{L_0-\frac{c_{G,k}}{24}}e^{2\pi i J_0(u)})\,,
\label{defaffch}
\eeq
where $u\in\liet_{\C}$ is considered as an element of $\g_{\C}$.
As it should be, (\ref{torWZW}) is invariant under
the gauge transformation $u\mapsto wu+n+\tau m$. This can be seen
by looking at the transformation rule (\ref{affact})
in which $(x,t,y)$ is identified with $xL_0+J_0(t)+ky$
and by noting (i) since $T\subset T_{G'}$, $\Pv$ is a sublattice of
$\Pv_{G'}$ and (ii) for any $w\in W$, there is an element $w'\in W_G$
such that $w=w'$ on $\liet_{\C}$ \cite{Lie}. It is also invariant under
the modular transformations generated by
$T$ and $S$: $(\tau, u)\mapsto (\tau+1,u)$ and
$(-\frac{1}{\tau},\frac{u}{\tau})$. This is due to unitarity \cite{Kac}
of the modular transformation matrices for the characters
$\Chi_{\Lmd}^{G,k}$.

\vsp
Now we calculate
\beq
Z_{\Sgmtau,\Ptriv}(1)=\int^{}_{\NN_{\!H}}\!\prod_{i=1}^l\dd u^i
Z^{\tot}_{\Sgmtau,\Ptriv}\Bigl(A_u\,;\,|\rgf_{\!A_u}\!(c,h)|^2
\prod_{j=1}^l
\frac{i}{2\pi}\!\int_{\Sgmtau}\!\!\!b\frac{\partial A_u^{01}}{\partial u^j}
\frac{i}{2\pi}\!
\int_{\Sgmtau}\!\!\!\bar b\frac{\partial A_u^{10}}{\partial \bar u^j}\,
\Bigr)\,.
\eeq
The symmetry group of $A_u$ for generic $u$ is the group
$T_{\C}=T\tms\e^{i\liet}$ of constant gauge transformations.
Parametrizing $hh^*$ as $n_+\e^{\varphi}n_+^*$ where $n_+$ is $N$-valued
and
$\varphi$ is $i\liet$-valued,
the residual gauge fixing term can be chosen as
\beq
|\rgf_{\!A_u}\!(c,h)|^2
=\frac{\delta^{(l)}(\varphi(x_0))}{\vol(T)}
\prod_{i=1}^lc^i(x_0)\bar c^i(x_0)\,,
\eeq
where $x_0$ is any point of $\Sigma_{\tau}$. As calculated essentially
in \cite{GawKup}, we have
\beqa
&&Z^{\rm gh}_{\Sgmtau,\Ptriv}\Bigl(\,A_u\,;\,
\prod_{i=1}^lc^i(x_0)\bar c^i(x_0)\prod_{j=1}^l
\frac{i}{2\pi}\!\int_{\Sgmtau}\!\!\!
b\frac{\partial A_u^{01}}{\partial u^j}\,
\frac{i}{2\pi}\!
\int_{\Sgmtau}\!\!\!\bar b\frac{\partial A_u^{10}}{\partial \bar u^j}\,
\Bigr)\\
&&\hspace{6.9cm}=\left(\frac{\pi}{\tau_2}\right)^{\!2l}\!
{\det}'_{\ad}\Bigl(\,\bartial_{\!A_u}^{\dag}\bartial_{\!A_u}\Bigr)\,,
\hspace{2cm}\nonumber\\
\noalign{\vskip0.2cm}
&&Z^{\Hc/H,-\tilk-2\coxh}_{\Sgmtau,\Ptriv}\!\Bigl(A_u;
\frac{\delta^{(l)}(\varphi(x_0))}{\vol(T)}\,\Bigr)
=\frac{(2\tau_2(\tilk+\hvH))^{\frac{l}{2}}}{(2\pi)^l\vol(T)}
\left({\det}'_{\ad}\Bigl(\,\bartial_{\!A_u}^{\dag}\bartial_{\!A_u}\Bigr)
\right)^{-\frac{1}{2}}\!\!\!\!,
\label{torHCWZW}
\eeqa
where ${\det}'_{\ad}\Bigl(\,\bartial_{\!A_u}^{\dag}\bartial_{\!A_u}\Bigr)$
is the $\z$-regularized determinant of the Laplace operator
$\bartial_{\!A_u}^{\dag}\bartial_{\!A_u}$ acting on sections
of the adjoint bundle. Calculation of the determinant is done in
\cite{Ray-Singer} and the result is
\beq
{\det}'_{\ad}\Bigl(\,\bartial_{\!A_u}^{\dag}\bartial_{\!A_u}\Bigr)
=(2\tau_2)^{2l}\e^{\frac{\pi}{2\tau_2}2\hvH \tr(u-\bar u)^2}\!
\left|\Pi_{\tilH}(\tau, u)\right|^4\!,
\eeq
in which $\Pi_{\tilH}(\tau,u)$ is the Weyl-Kac denominator defined by
\beq
\Pi_{\tilH}(\tau, u)
=q^{\frac{\dim H}{24}}\!\!
\prod_{\alpha\in \Delta_+}(\e^{\pi i\alpha(u)}-\e^{-\pi i \alpha(u)})
\prod_{n=1}^{\infty}\Bigl\{(1-q^n)^l\!
\prod_{\alpha\in \Delta}(1-q^n\e^{-2\pi i\alpha(u)})\,\Bigr\}\,.
\label{defdenom}
\eeq
Thus, $Z_{\Sgmtau,\Ptriv}(1)$ is equal to
\beq
\left(\frac{\tilk+\hvH}{2\tau_2}\right)^{\!\frac{l}{2}}\!\!\!
\frac{(2\pi)^l}{\vol(T)}\!
\int^{}_{\NN_{\!H}}\!\prod_{i=1}^l\dd u^i
\e^{\frac{\pi}{2\tau_2}(\tilk+\hvH)\tr(u-\bar u)^2}\!
\sum_{\Lmd}\!
\left| \Chi_{\Lmd}^{G,k}(\tau,u)\Pi_{\tilH}(\tau, u)\right|^2\!.
\label{interm}
\eeq
The branching rule (\ref{branch}) leads to the expansion
\beq
\Chi^{G,k}(\tau,u)
=\sum_{\lmd}b^{\lmd}_{\Lmd}(\tau)\Chi^{\tilH,\tilk}_{\lmd}(\tau,u)\,,
\eeq
in which the branching function $b^{\lmd}_{\Lmd}$ for $\Lmd\in\PPpk(G)$
and $\lmd\in\PPptilk(\tilH)$ is defined by
$b^{\lmd}_{\Lmd}(\tau)=\tr^{}_{B^{\lmd}_{\Lmd}}(q^{L_0-\frac{c}{24}})$
where $c=c_{G,k}-c_{\tilH,\tilk}$ and $L_0=L_0^{\ssstyle {\rm GKO}}$.
Since the Virasoro generators by the GKO construction commute with the
spectral flow, we have $b_{\gmaG\Lmd}^{\gamma\lmd}=b_{\Lmd}^{\lmd}$.
This enables us to replace the integration $\int_{\NN_H}$ in
(\ref{interm}) by $\frac{1}{|\Pv/\Qv|^2}\int_{\NN_{\tilH}}$.
Using the obvious identity $\vol(T)=(2\pi)^l\vol(i\liet /\Pv)$ and
the orthogonalty
\beq
\int^{}_{\NN_{\!\tilH}}\!\prod_{i=1}^l\dd u^i
\e^{\frac{\pi}{2\tau_2}(\tilk+\hvH)\tr(u-\bar u)^2}\!\!
\Chi_{\lmd}(\tau,u)\overline{\Chi_{\lmd'}(\tau,u)}|\Pi(\tau,u)|^2
=\vol(i\liet/\Qv)\left(\frac{2\tau_2}{\tilk+\hvH}\right)^{\!\frac{l}{2}}\!\!
\delta_{\lmd,\lmd'}
\label{orthogonality}
\eeq
of characters for $\tiL\tilH$ at level $\tilk$, we finally have
\beq
Z_{\Sgmtau,\Ptriv}(1)
=\frac{1}{|\pi_1(H)|}\sum_{\Lmd,\lmd}\left|b_{\Lmd}^{\lmd}(\tau)\right|^2\,,
\label{torpar1}
\eeq
in which $(\Lmd,\lmd)$ rums over $\PPpk(G)\tms\PPptilk(\tilH)$.
Due to the invariance
$b_{\gmaG\Lmd}^{\gamma\lmd}=b_{\Lmd}^{\lmd}$, it can also be expressed as
\beq
Z_{\Sgmtau,\Ptriv}(1)=\sum_{[\Lmd,\lmd]}^{\cdot}\,
\frac{1}{|{\cal S}_{\Lmd}^{\lmd}|}\left|b_{\Lmd}^{\lmd}(\tau)\right|^2\,,
\label{torpar2}
\eeq
where the sum is over the quotient $(\PPpk(G)\tms\PPptilk(\tilH))/\Gmalcv$
and ${\cal S}_{\Lmd}^{\lmd}$ is the isotropy subgroup of $\Gmalcv$
at $(\Lmd,\lmd)$.

\vsp
If ${\cal S}_{\Lmd}^{\lmd}=1$ for every $(\Lmd,\lmd)$,
obviously we have
\beq
Z_{\Sgmtau,\Ptriv}(1)
=\tr^{}_{\Heqhw}\!(q^{L_0-\frac{c}{24}}\bar q^{\bar L_0-\frac{c}{24}})\,.
\eeq
As we shall see shortly, in this case, topologically non-trivial
bundles do not contribute to the partition function and hence
$Z_{\Sgmtau,\Ptriv}(1)$ is itself the full partition function.
Thus, (\ref{principle}) holds if $\pi_1(H)$ acts freely on
$\PPpk(G)\tms \PPptilk(\tilH)$.

\vspace{0.5cm}
{\sc 5.2 Field Identification Fixed Points}

\vspace{0.3cm}
To an element $\gamma\in\pi_1(H)$ is associated a principal $H$-bundle
$P_{\gamma}=\Ptriv \gamma$ over $\Sgmtau$. Due to the topological identity
(\ref{FI}), the partition function for $P_{\gamma}$ is the
one point function for the trivial bundle:
\beq
Z_{\Sgmtau,P_{\gamma}}(1)=Z_{\Sgmtau,\Ptriv}(\gamma(1))\,,
\eeq
where $\gamma(1)$ is associated to the state $\gamma.\Phi_0$ in
${\cal H}_{\gmaG \!0}^{\gamma 0}$. It is expressed as an
integral over $\NN_H$ whose integrand contains a factor
$Z_{\Sgmtau,\Ptriv}^{G,k}(A_u;{O_{\!\gamma.\Phi_0}}_m^{\!\!\,\bar m})$.
For this to be non-vanishing, the fusion rule \cite{GW,Verlinde,FG} requires
\beq
\sum_{\Lmd\in\PPpk(G)}N_{\gmaG\!0\,\Lmd}^{\Lmd}\ne 0\quad
\mbox{and}
\sum_{\lmd\in\PPptilk(\tilH)}N_{\gamma 0\,\lmd}^{\lmd}\ne 0\,,
\label{condnv}
\eeq
where $N_{\Lmd\,\Lmd'}^{\Lmd''}$ (resp. $\!N_{\lmd\,\lmd'}^{\lmd''}$) is the
fusion coefficient of the WZW model with target $G$ and level $k$
(resp. target $\tilH$ and level $\tilk$). From
the Gepner's observation \cite{Gep} $S_{\lmd}^{\gamma\lmd'}
=(-1)^{l(w)}\e^{-2\pi i(\lmd+\rho)(\mu)}S_{\lmd}^{\lmd'}$
for $\gamma(\theta)=\e^{-i\mu\theta}w$ on the modular transformation matrix
and from the Verlinde formula $N_{\lmd_1\,\lmd_2}^{\lmd_3}
=\sum_{\lmd}S_{\lmd_1}^{\!\!\,\lmd}S_{\lmd_2}^{\!\!\,\lmd}
S_{\lmd_3}^{\!\!\,\lmd *}/S_{0}^{\!\!\,\lmd}$ \cite{Verlinde},
it follows that
$N_{\gamma_1 \lmd\,\gamma_2\lmd'}^{\gamma_1\gamma_2\lmd''}
=N_{\lmd\,\lmd'}^{\lmd''}$.
Since $N_{0\,\lmd}^{\lmd'}=\delta_{\lmd}^{\lmd'}$,
(\ref{condnv}) is equivalent to the condition that
there exist $\Lmd\in\PPpk(G)$ and $\lmd\in\PPptilk(\tilH)$ such that
$\gmaG\Lmd=\Lmd$ and $\gamma\lmd=\lmd$.

This observation is desirable in the following sense. If there is a pair
$(\Lmd,\lmd)$ at which the isotropy
${\cal S}_{\Lmd}^{\lmd}\subset \pi_1(H)$ is not
$\{1\}$ (such a pair is called the {\it fixed point} in the literature),
the partition function for the trivial topology has fractional coefficients
in the $q,\bar q$-expansion (see (\ref{torpar2}))
and we can hardly expect that
this function is expressed as a trace of
$q^{L_0-\frac{c}{24}}\bar q^{\bar L_0-\frac{c}{24}}$ in any Virasoro module.
In algebraic treatments of coset models \cite{LVW,Schell-Yank},
this was recognized as the field identification problem in the presence
of fixed points. We expect that a natural resolution is provided by
the sum over topologies: If ${\cal S}_{\Lmd}^{\lmd}\ne \{1\}$,
the contribution $Z_{\Sgmtau,P_{\gamma}}(1)$ for
$\gamma\in {\cal S}_{\Lmd}^{\lmd}-\{1\}$ may be non-vanishing and
the integrality of the coefficients may be restored
for the {\it full} partition function.(\footnote{
In \cite{Schell-Yank}, a method for ``fixed point resolution'' is presented.
Characters of the ``fixed point CFTs'' in that reference may be related
to the partition functions for non-trivial topologies.})
In the next subsection, choosing a concrete example,
we examine whether this happens.

The partition function (\ref{torpar1}) for the trivial topology is
manifestly modular invariant, and
so is expected for any topology $P$ since $P$
and $f^*P$ are topologically isomorphic
for any diffeomorphism $f$ of $\Sgmtau$. Hence, the modular invariance may
still hold for the full partition function.
This is also examined below.

\vspace{0.5cm}
{\sc 5.3 Models With $G=SU(2)\tms SU(2)$ and $H=SO(3)$}

\vspace{0.3cm}
We consider the case in which $G=SU(2)\tms SU(2)$ and $H$ is the subgroup
$\{(g,g);g\in SO(3)\}$ of the adjoint group $G/Z_G=SO(3)\tms SO(3)$.
For the level $k=(k_1,k_2)$, the induced level is $\tilk=k_1+k_2$. Since
a highest weight representation of $SU(2)$ is convensionally labeled by
the spin$\in\! \frac{1}{2}\Z$, we identify $\PPpk=\PPpk(SU(2))$ with the set
$\{0,\frac{1}{2},1,\cdots,\frac{k}{2}\}$ of ``integrable spins''.
The non-trivial element of $\pi_1(H)=\Z_2$ induces the involution
$((j_1,j_2),j)\leftrightarrow
((\frac{k_1}{2}-j_1,\frac{k_2}{2}-j_2),\frac{\tilk}{2}-j)$
in $\PPpk(G)\tms \PPptilk$. If $k_1$ or $k_2$ is an odd integer,
there is no fixed point and the full partition function is given by
$\frac{1}{2}\sum_{j_1,j_2,j}|b_{(j_1,j_2)}^j(\tau)|^2$.
For the case $k_2=1$,
it is the diagonal modular invariant partition function
of the $k_1$-th unitary minimal model.

\vspace{0.3cm}
{\it Partition Function For The Non-trivial Topology}

\vsp
In the following, we assume that $k_1$ and $k_2$ are both even integers.
Then, there is a unique fixed point
$((\frac{k_1}{4},\frac{k_2}{4}),\frac{\tilk}{4})$ and the topologically
non-trivial configurations contribute to the partition function.
Recall that the moduli space $\NN_{\rm non-triv}$ of
semi-stable $\HC$-bundles of non-trivial topology consists of one point
represented by $\Ph_F^{(1)}$ which is obtained by the identification
\beq
\sigma(zq)=\sigma(z)\pmatrix{
0 & \!\!\!\!q^{-\frac{1}{4}}z^{-\frac{1}{2}} \cr
-q^{\frac{1}{4}}z^{\frac{1}{2}} &\!\!\!\! 0 \cr
}
\eeq
of a holomorphic section $z\mapsto \sigma(z)$ of the bundle $\C^*\tms \HC$
over $\C^*$. Denoting by $A_F$ the flat $SO(3)$-connection
corresponding to the holomorphic bundle $\Ph_F^{(1)}$, we have
\beq
Z_{\Sgmtau,\nt}(1)
=Z^{\tot}_{\Sgmtau,\nt}(A_F;\mbox{$\frac{1}{4}$})\,,
\eeq
where $\frac{1}{4}$ is the residual gauge fixing term for
$\Aut\Ph_F^{(1)}=\Z_2\tms\Z_2$. This factorises into the product
of the partition functions for the three (or four) constituents:
\beq
\frac{1}{4}\prod_{i=1}^2 Z^{SU(2),k_i}_{\Sgmtau,\nt}(A_F;1)
Z^{\Hc/H,-\tilk-4}_{\Sgmtau,\nt}(A_F;1)
Z^{\rm gh}_{\Sgmtau,\nt}(A_F;1)\,.
\eeq
We shall show that each is constant, that is, independent on $\tau$.
For this, we introduce the Green's function of the operator
$\bartial_{\!A_F}$ for the adjoint bundle.
Let $\sgmad(z):\g_{\C}\stackrel{\simeq}{\to}\ad \PC|_z$ be the frame
associated to $\sigma(z)\in\PC|_z$. The Green function is then expressed as
$G_w(z)=\sgmad(w)g(w,z)\sgmad(z)^{-1}\!\ot dz$ where
$g(w,z)\in{\rm End}(\h_{\C})$ is represented by the matrix
\beq
g(w,z)=\pmatrix{
\sum\frac{q^n}{z-q^{2n}w}&0&-\sum\frac{z^{-1}q^{n-\nibun}}{z-q^{2n-1}w}\cr
0&f(w,z)&0 \cr
-\sum\frac{wq^{n-\nibun}}{z-q^{2n-1}w}&0&\sum\frac{wz^{-1}q^n}{z-q^{2n}w} \cr
}\,,
\label{Greennontriv}
\eeq
with respect to the base $(\sigma_+,\sigma_3,\sigma_-)$ of
$\h_{\C}=\spl(2,\C)$ ($\sigma_{\pm}=(\sigma_1\pm i\sigma_2)/2$) in which
$\sigma_i$'s are Pauli matrices.
The sums $\sum_n$ in the four entries are over all integers and
$f(w,z)$ is expressed by the theta function $\vartheta$
and its derivative $\vartheta'=\frac{\partial}{\partial \z}\vartheta$ as
\beq
f(\e^{-2\pi i\xi},\e^{-2\pi i\z})=\frac{1}{2\pi i z}
\frac{\vartheta(\tau,\xi-\z+\frac{\tau}{2})}{\vartheta(\tau,
\xi-\z+\frac{\tau+1}{2})}
\frac{\vartheta'(\tau,\frac{\tau+1}{2})}{\vartheta(\tau,\frac{\tau}{2})}.
\eeq
The partition function $Z^{SU(2),k}_{\Sgmtau,\nt}(A_F;1)$ for
the $SU(2)$-WZW model
satisfies the following Ward identities for the chiral gauge symmetry:
\beqa
Z^{SU(2),k}_{\Sgmtau,\nt}(A_F;J)\!&=&\!0\\
Z^{SU(2),k}_{\Sgmtau,\nt}(A_F;J\cd \epsilon(z)\,J\cd\epsilon'(w))\!\!
&=&\!\!\!
k\trP(\partial_{\!A_F}^{(w)}G_w\epsilon(z)\epsilon'(w))
Z^{SU(2),k}_{\Sgmtau,\nt}(A_F;1)\,.\nonumber
\eeqa
Putting these into the expression (\ref{Sug}) for
the energy momentum tensor, we find
\beq
\frac{\partial}{\partial \tau}Z^{SU(2),k}_{\Sgmtau,\nt}(A_F;1)=0\,.
\eeq
This also holds for the WZW model with the target $\HC/H$.
As for the ghost system, putting the identity
\beq
Z^{\rm gh}_{\Sgmtau,\nt}(A_F;c(w)b(z))=
G_w(z)Z^{\rm gh}_{\Sgmtau,\nt}(A_F;1)\,,
\eeq
into the expression (\ref{Tgh}) of the energy momentum tensor, we find
\beq
\frac{\partial}{\partial \tau}Z^{\rm gh}_{\Sgmtau,\nt}(A_F;1)=0\,.
\eeq
Thus, the partition function is a constant:
\beq
Z_{\Sgmtau,\nt}(1)=C_{\nt}\,.
\label{cst}
\eeq

\vspace{0.3cm}
{\it The Full Partition Function}

\vsp
The partition function for topologically trivial configurations is
given by
\beq
Z_{\Sgmtau,\triv}(1)
=\sum_{[(j_1,j_2),j]}^{\cdot}\!\!\!\!\!\!{}^{\circ}
\left|b_{(j_1,j_2)}^j(\tau)\right|^2
+\frac{1}{2}
\Bigl|b_{(k_1/4,k_2/4)}^{\tilk/4}(\tau)\Bigr|^2
\,,\label{extriv}
\eeq
where the sum $\!\dot{\,\sum^{\circ}}$ is over the $\Z_2$-quotient of
$\PPpk(G)\tms\PPptilk-\{((\frac{k_1}{4},\frac{k_2}{4}),\frac{\tilk}{4})\}$.
For the non-trivial topology, we have (\ref{cst}). Since we have no way to
determine $C_{\nt}$ now, we change the question to the following form:
Can we tune $C_{\nt}$ so that (\ref{principle}) holds?

The term $\!\dot{\,\sum^{\circ}}$ in (\ref{extriv})
is the trace of $q^{L_0-\frac{c}{24}}\bar q^{\bar L_0-\frac{c}{24}}$
on the space $\Heqhw^{\circ}$ where
\beq
\Hhw^{\circ}
=\Hhw\ominus {\cal H}^{\rm f}\qquad;\,\,\,
{\cal H}^{\rm f}:=
{\cal H}_{(k_1/4,k_2/4)}^{\tilk/4}\,.
\eeq
Hence, the question is whether there is a constant $C_{\nt}$ such that
\beq
\frac{1}{2}
\Bigl|b_{(k_1/4,k_2/4)}^{\tilk/4}(\tau)\Bigr|^2
+C_{\nt}=
\tr^{}_{\dot{\cal H}^{\rm f}}\!
(q^{L_0-\frac{c}{24}}\bar q^{\bar L_0-\frac{c}{24}})\,.
\label{question}
\eeq

We answer this in the case $k_2=2$. The Virasoro modules by
the GKO construction $SU(2)\tms SU(2)/SU(2)$ at level $(k_1,2)$ are
known \cite{GKO} to be
the ones appearing in the $k_1$-th $N=1$ superconformal minimal model.
Among others,
$B^{\rm f}:=B_{(k_1/4,k_2/4)}^{\tilk/4}$ is in the Ramond sector
and contains a unique ground state with $L_0=\frac{c}{24}$.
In particular, there is a supercharge $G_0:B^{\rm f}\to B^{\rm f}$
such that $G_0^2=L_0-\frac{c}{24}$. One can show that
$\gamma.:{\cal H}^{\rm f}\to{\cal H}^{\rm f}$ induces an involution
$U_{\gamma}$ of the Virasoro module $B^{\rm f}$ such that
\beq
G_0 U_{\gamma}+U_{\gamma} G_0=0\,.
\label{G0gma}
\eeq
If $B^{\rm f}$ is decomposed as
$B^{\rm f}=\oplus_{n=0}^{\infty}B_n$ in which $B_n$ is
the $L_0$-eigen space with $G_0^2=n$,
we may put $U_{\gamma}=1$ on $B_0\cong \C$ and
the anti-commuting relation (\ref{G0gma}) shows that
\beq
B_n=B_n^{(+)}\oplus B_n^{(-)}\,,\qquad
B_n^{(+)}
\begin{array}{c}
{\scriptstyle G_0}\\
\noalign{\vskip-0.3cm}
\longto\\
\noalign{\vskip-0.35cm}
\longleftarrow\\
\noalign{\vskip-0.3cm}
{\scriptstyle G_0}
\end{array}
B_n^{(-)}\,\,\,\mbox{(isomorphic)}
\eeq
for $n\geq 1$, where $B_n^{(\pm)}$ is the subspace of $B_n$ on which
$U_{\gamma}=\pm 1$.
Thus, we have
${\cal H}^{\rm f}={\cal H}^{(+)}\oplus{\cal H}^{(-)}$ where
\beqa
{\cal H}^{(+)}&\cong&
\bigoplus_{n,m=0}^{\infty}B_n^{(+)}\ot \overline{B_m^{(+)}}\oplus
\bigoplus_{n,m=1}^{\infty}B_n^{(-)}\ot\overline{B_m^{(-)}}
\\
\hspace{-2cm}\mbox{and}\hspace{2cm}
{\cal H}^{(-)}&\cong&
\bigoplus_{n\geq 0,m\geq 1}^{\infty}
\Bigl\{
B_n^{(+)}\ot \overline{B_m^{(-)}}\oplus
B_m^{(-)}\ot\overline{B_n^{(+)}}
\Bigr\}
\eeqa
are subspaces on which $\gamma.=1$ and $\gamma.=-1$ respectively.
Since $\dot{\cal H}^{\rm f}$ is isomorphic to ${\cal H}^{(+)}$, we see that
(\ref{question}) and hence (\ref{principle}) hold
if we tune $C_{\nt}=\frac{1}{2}$.

\renewcommand{\theequation}{6.\arabic{equation}}\setcounter{equation}{0}
\newpage
\vsp
\begin{center}
{\bf 6. Concluding Remarks}
\end{center}

So far, we have been considering the gauged WZW model whose classical action
is defined by (\ref{WZWweight}).
However, we could have started with another choice of an action
generalizing (\ref{actiontriv}). One familiar way to modify
the action is to add the ``theta term''
\beq
\int_{\Sigma}\frac{i}{2\pi}\theta(F_A)\,,
\label{thetaterm}
\eeq
where ``$\theta$'' is some adjoint invariant linear form $\h\to \R$.
Then, the equivalence relation of gauge invariant local fields is modified
by a phase factor. If $H$ contains a $U(1)$-factor, we thus have
a continuous series of
quantum field theories having one common partition function.

If $H$ is semi-simple, the term (\ref{thetaterm}) vanishes. However,
there is another way to modify the action. It arises from the variety of
WZW actions constructed via the equivariant differential characters.
Construction of an action in terms of Cheeger-Simons differential character
was initiated by Dijkgraaf and Witten in Chern-Simoms gauge theory
\cite{DijkWitt}
and the method was elaborated in ref. \cite{Axl} (see also \cite{Witten3}).
It provides a way to define topological lagrangians satisfying
suitable physical conditions such as locality, unitarity, gluing property,
etc.
According to it,
WZW actions with the target $G$ and the gauge group $H$ are
classified by the
equivariant cohomology (Borel cohomology)
$H_H^3(G;\Z):=H^3(EH\tms^{}_H G;\Z)$ in which $EH$ is
the universal $H$-bundle and $H$ acts on $G$ via adjoint transformations.
Importantly, for a semi-simple group $H$,
we have
\beq
H^3_H(G;\Z)=H^3(G;\Z)\oplus {\rm Hom}(\pi_1(H),\R/\Z)\,.
\label{isom:theta}
\eeq
The levels are classified by $H^3(G;\Z)$ and presumably the torsion part
${\rm Hom}(\pi_1(H),\R/\Z)$ classifies the ``theta terms''.
In the quantum theory, such a theta term would modify
the equivalence relation of gauge invariant local fields.
In a theory with fixed points, it would modify the partition function
as well. For example, when $G=SU(2)\tms SU(2)$ and $H=SO(3)$,
the theory corresponding to $(k_1,k_2,\pm 1)\in \Z\oplus \Z\oplus\Z_2\cong
H^3_{SO(3)}(SU(2)^2;\Z)$ with even $k_1,k_2$ would have the full
partition function
\beq
Z_{\Sgmtau,{\rm triv}}(1)\pm C_{\nt}
=\frac{1}{2}\Bigl|b_{(k_1/4,k_2/4)}^{\tilk/4}(\tau)\Bigr|^2\pm C_{\nt}
+\cdots.
\eeq
For $k_2=2$, both have positive integral coefficients
in the $q,\bar q$-expansions if and only if
$C_{\nt}=\pm\frac{1}{2}$. If $C_{\nt}=\frac{1}{2}$,
(\ref{principle}) holds in each theory. Due to the relation (\ref{G0gma}),
the involution $\gamma.$ can be identified with the mod two fermion number
$(-1)^F$ and the theory for $(k_1,2,\pm 1)$ is the {\it spin model}
\cite{FQS,Cappelli}
with the projection $(-1)^F=\pm 1$ on the Ramond sector.
We expect in a general model that
adding a torsion (theta term) has such a simple and significant consequence
in physics. This will be elaborated in a future work \cite{Hori}.

\vsp
In this paper, we have been concentrated on the model whose matter theory
is the WZW model with a compact simply connected target group.
However, our argument is readily applicable to other models such as
(i) the model whose target group is compact and connected
but non simply connected,
(ii) the model whose matter theory is a free fermionic system of arbitrary
spin and
(iii) a combined system of free fermions and WZW models.
Study of models of the type (i) may be important for the classification of
rational CFTs. An interesting class of theories of the type (iii) is the
(twisted) $N=2$ coset conformal field theory (Kazama-Suzuki model)
\cite{KazSuz,LVW}.
Algebraic structure of the spectral flows of such a model
has been studied by many authors \cite{LVW,HT,Sch,Nakatsu-Sugawara}.
In the ref. \cite{Nakatsu-Sugawara}, a geometric interpretation of
field identification is attempted along the line similar to ours. However,
the argument in that reference uses the counterpart
(in the twisted $N=2$ system) of our old integral expression
(\ref{oldintexpr}) and hence is applicable only for
abelian gauge groups. Our method completes this. The fixed point resolution
in these systems (see \cite{Sch,FuSch} for algebraic approaches)
by the topological sum with theta terms will be interesting and perhaps
of some importance in superstring theory.

\renewcommand{\theequation}{A.\arabic{equation}}\setcounter{equation}{0}

\newcommand{\tilT}{\tilde{T}}
\newcommand{\MC}{{\rm M}_{\Ch}}
\newcommand{\tiLtilH}{\tiL\tilH}

\vspace{0.7cm}
\begin{center}
{\bf Appendix A}
\end{center}

We describe some basic facts on root systems and Weyl groups
\cite{Bourbaki}.
Let $H$ be a compact connected Lie group and let $\pi:\tilH\to H$ be the
universal covering with kernel $\pi^{-1}(1)\cong \pi_1(H)$.
We choose a maximal torus $T$ of $H$ and put $\tilT=\pi^{-1}(T)$.
The Lie algebras of $T$ and $\tilT$ are identified and
the imaginary part $i\liet$ of its complexification is denoted by $\V$.
We introduce lattices $\Qv\subset \Pv$ in $\V$ so that the exponential maps
induce isomorphisms $\liet/2\pi i\Pv\cong\Pv$ and
$\liet/2\pi i\Qv\cong \tilT$. Then we have $\Pv/\Qv\cong \pi_1(H)$.

\vspace{0.4cm}
{\sc A.1 For A Simple Centerless Group}

\vspace{0.25cm}
We first consider the case in which $H$ is simple and centerless. Then,
$\tilH$ is compact, $\tilT$ is its maximal torus
and $\pi_1(H)$ is the center of $\tilH$. $\Pv$ and $\Qv$ are
dual to the root lattice $\QQ$ of $\tilH$ and the weight lattice $\PP$ of
$\tilT$ respectively:
\beq
\begin{array}{ccc}
\V^*\supset \PP\hspace{-0.05cm}&\cdots\cdots&\hspace{-1.05cm}\Qv \\
\cup \hspace{-1.08cm}& &\hspace{-1.1cm}\!\!\cap\\
\QQ\hspace{-1.08cm}&\cdots\cdots&\hspace{-0.12cm}\Pv\subset \V
\end{array}
\eeq
where $A\cdots\cdots B$ means that $A$ is the dual of $B$.

\vspace{0.25cm}
{\it Weyl Group}

\vsp
The {\it Weyl group} $W$ of $(H,T)$ is defined by $W=N_{T}/T$ where
$N_{T}$ is the normalizer of $T$ in $H$. The adjoint action of $W$ on $T$
induces its linear actions on $\V^*$ and $\V$ leaving invariant
the four lattices and the set $\Delta$ of roots.
For each root $\alpha \in \Delta$, we introduce a hyperplane
$\hp_{\alpha}=\{ x\in V; \alpha(x)=0\}$ and we denote by $s_{\alpha}\in W$
the reflection with respect to $\hp_{\alpha}$.
Since $W$ preserves $\Delta$,
the family $\{\hp_{\alpha}\}_{\alpha\in\Delta}$
of hyperplanes is invariant by $W$.
A $chambre$ of $\Delta$ is, by definition, a connected component of
$\V-\cup_{\alpha\in \Delta}\hp_{\alpha}$. We now have the
\begin{th}
(1) $W$ acts simply transitively on the set of chambres.\\
(2) If \(\hp_{1}, \cdots ,\hp_{l}\) are walls of a chambre $\Ch$,
for each $i$ there exist a unique root $\alpha_{i}$ such that
$\hp_{\alpha_{i}}=\hp_{i}$ and that
$\alpha_{i}$ takes positive values on $\Ch$.\\
(3) The set ${\rm B}(\Ch)=\{ \alpha_{1},\cdots ,\alpha_{l}\}$ forms
a base of the free abelian group $\QQ$.\\
(4) The set $S(\Ch)=\{ s_{\alpha_{1}},\cdots , s_{\alpha_{l}}\}$
generates $W$.\\
(5) Any root $\alpha\in \Delta$ is expressed as
$\alpha=\sum_{i=1}^{l}n_{i}\alpha_{i}$ where $n_{i}$ are all non-negative
integers or all non-positive integers.
\label{page:theoremWeyl}
\end{th}

We fix a chambre $\Ch$. By {\it (3)} and {\it (5)}, we can choose a base
$\{ \mu_1,\cdots, \mu_l\}$ of $\Pv$ such that
$\alpha_{i}(\mu_j)=\delta_{i,j}$.
{\it (5)} of the theorem shows that $\Delta$ is decomposed as
a disjoint union of  the set $\Delta_{+}$ of {\it positive roots}
and the set $\Delta_{-}=-\Delta_{+}$ of {\it negative roots}
where a root is {\rm positive} if it takes positive values on $\Ch$.
We see from {\it (1)} that there exists a unique element $w_{0}\in W$
such that $w_{0}\Delta_{+}=\Delta_{-}$. It is the longest element
where the length $l(w)$ of $w\in W$ is the minimun length $n$ of
such sequence $s_{i_{1}},\cdots, s_{i_{n}}$ in $S(\Ch)$ that
$w=s_{i_{1}}\cdots s_{i_{n}}$.
The highest weight of the adjoint representation is called {\it highest
root} and is denoted by $\tilde{\alpha}$. It can be shown that, for any
$\alpha \in \Delta$, $\tilde{\alpha} -\alpha$ is a span of
$\alpha_{1}, \cdots, \alpha_{l}$ with non-negative integral coefficients.
In particular, $\tilde{\alpha}$ is expressed as
$\tilde{\alpha}=\sum_{i=1}^{l}n_{i}\alpha_{i}$ for $n_{i}\geq 1$.
We define ${\cal J}\subset\{ 1, \cdots , l\}$ by
$j\in {\cal J}\Leftrightarrow n_{j}=1$.
We put $\mu_0=0$, $\indJ=\{0\}\cup {\cal J}$ and
$\MC=\{\,\mu_j;j\in \indJ\}$.
Then, we have the following
\begin{pn}
 $a\in \Pv$ satisfies $\alpha(a)=0$ or $1$ for any $\alpha\in \Delta_+$
if and only if $a\in \MC$.
Moreover, any $\Qv$ orbit in $\Pv$ contains one and only one element
of $\MC$.
\end{pn}
The latter part can be understood after we introduce the group
$\Gamma_{\alcv}$.

\vspace{0.3cm}

{\it Affine Weyl Groups}

\vsp
The {\it affine Weyl groups} of $\tilH$ and $H$ are defined
by $\Waff={\rm Hom}(U(1),\tilT)\semidir W\cong \Qv\semidir W$ and
$\Waffh={\rm Hom}(U(1),T)\semidir W\cong \Pv\semidir W$
respectively.
Since $\Qv\subset \Pv$, $\Waff$ is considered as a subgroup of $\Waffh$.
$\Waff$ can also be defined as the Weyl group of
$U(1)\semidir \tiLtilH$ with respect to the torus
$U(1){\times}\tiLtilH|_{\tilT}$ where $U(1)$ acts on $\tiLtilH$
covering the rotation action $\e^{ix}:\gamma(\theta)\to \gamma(\theta-x)$
on $L\tilH$. Hence comes the linear action of $\Waff$ (and also of $\Waffh$)
on $\Lie(U(1){\times}\tiLtilH|_{\tilT})=i\hat{\V}$ where
$\hat{\V}=\R_r\oplus\V\oplus\R_c$
\beqa
e^{-ia\theta}w: ( x, t,y )\in \hat{\V} \!\!\!&\mapsto&\!\!\!
\mbox{$(x, wt-xa,y-\tr(awt)+\frac{x}{2} \tr(a^2))$}
\in \hat{\V},\label{affact}\\
 e^{-ia\theta}w : ( n, \lmd,k)\in \hat{\V}^*\! \!\!&\mapsto& \!\!\!
\mbox{$(n+w\lmd(a)+\frac{k}{2}\tr(a^2),w\lmd+k\ttr a, k)$}\in \hat{\V}^*\,.
\label{affact*}
\eeqa
where ``$\tr$'' is a normalized trace in $\h$ that induce the inner
product on $\V^*$ with $(\alpha,\alpha)=2$ for a long root $\alpha$.
The dual action (\ref{affact*}) preserves the following inner product on
$\hat{\V}^*$:
\beq
\Bigl((\,n_1\,,\lmd_1\,,k_1\,),(\,n_2\,,\lmd_2\,,k_2\,)\Bigr)
=(\,\lmd_1\,,\lmd_2\,)-n_1k_2-k_1n_2\,.\label{scalarproduct}
\eeq

An {\it affine root} of $LG$ is, by definition, a weight of the adjoint
action of $U(1)_r\times \tiLtilH|_{\tilH}$ on $\Lie(L\HC)$.
The set $\Daff\subset \hat{\V}^*$\label{a.2:Daff} of non-zero affine roots
is invariant under the action of $\Waffh$ and is given by
$\Daff= {\Z}_{\neq 0}\!\times \!\{0\}\!\times\!\{0\}
\cup {\Z}\!\times\! \Delta\!\times\!\{0\} $.

$\Waffh$ acts linearly on the quotient $\Vaff=\hat{\V}/\R_c$ and affinely
on each hyperplane $\V_{x}=\{x\}\!\times\! \V\subset\Vaff$
(see (\ref{affact})).
If we put $\hat{\hp}_{\hat{\alpha}}
=\{\,{\bf v}\!\in\!\Vaff\,;\hat{\alpha}({\bf v})=0\,\}$, the family
$\{\hat{\hp}_{\hat{\alpha}}\}_{\hat{\alpha}\in \Daff}$ of hyperplanes in
$\Vaff$ is $\Waffh$-invariant. Hence, if we put $\hp_{\hat{\alpha}}
=\hat{\hp}_{\hat{\alpha}}\cap V_{-1}$, the family
$\{\hp_{\hat{\alpha}}\}_{\hat{\alpha}\in \Z\times \Delta \times\{0\}}$
of hyperpalnes in $\V_{-1}$ is also $\Waffh$-invariant.
We denote by $s_{\hat{\alpha}}\in \Waff$ the reflection with respect to
$\hp_{\hat{\alpha}}\neq \emptyset$.
An {\it alc\^ove} is, by definition, a connected component of
$\V_{-1}-\cup_{\hat{\alpha}\in \Daff}\hp_{\hat{\alpha}}$. Then, we have the

\begin{th}
(1) $\Waff$ acts simply transitively on the set of alc\^oves.\\
(2) If $\hp_{0},\hp_{1},\cdots, \hp_{l}$ are walls of an alc\^ove $\alcv$,
for each $i$ there exists a unique affine root $\hat{\alpha}_{i}$ such that
$\hp_{\hat{\alpha}_{i}}=\hp_{i}$ and that $\hat{\alpha}_{i}$ takes positive
values on $\alcv$.\\
(3) The set ${\rm B}(\alcv)
=\{ \hat{\alpha}_{0},\hat{\alpha}_{1},\cdots,\hat{\alpha}_{l}\}$ forms
a base of $\Z\oplus\QQ\oplus\{0\}$.\\
(4) The set $S(\alcv)
=\{ s_{\hat{\alpha}_{0}},s_{\hat{\alpha}_{1}},\cdots,
s_{\hat{\alpha}_{l}}\}$
generates $\Waff$.\\
(5) Any affine root $\hat{\alpha}\in \Daff$ is expressed as
$\hat{\alpha}=\sum_{i=0}^{l}n_{i}\hat{\alpha}_{i}$ where $n_{i}$ are all
non-negative integers or all non-positive integers.
\label{theoremaffWeyl}
\end{th}
A chambre $\Ch$ determines an alc\^ove $\alcv=\{(-1,t)\in\V_{-1}\, ;
t\in \Ch,\tilde{\alpha}(t)<1\, \}$ which gives
${\rm B}(\alcv)
=\{\hat{\alpha}_{0},\cdots , \hat{\alpha}_{l}\}$
where $\hat{\alpha}_{0}=(-1,-\tilde{\alpha},0)$ and
$\hat{\alpha}_i=(0,\alpha_i,0)$ for $i=1,\cdots,l$.
{\it (5)} of the theorem shows that the
set $\Daff$ is decomposed as a disjoint union of the set
$\Delta_{{\rm aff} +}$ of {\it positive affine roots} and the set
$\Delta_{{\rm aff} -}=-\Delta_{{\rm aff} +}$ of {\it negative affine roots}
where an affine root is positive if it takes positive values on the
alc\^ove $\alcv$.

\vsp
$\Waffh$ acts on the set of alc\^oves and we denote by
$\Gamma_{\alcv}$ the isotropy subgroup at $\alcv$. Then we see that
$\Waffh$ decomposes into
semi-direct product of $\Waff$ and $\Gamma_{\alcv}$:
\beq
\Waffh \cong \Waff \semidir \Gamma_{\alcv}\,\,.
\label{Waff'decompo}
\eeq
The subgroup  $\Gamma_{\alcv}$ preserves the decomposition
$\Daff=\Delta_{{\rm aff}+}\cup \Delta_{{\rm aff}-}$ which shows
with the aid of {\it (5)} of the theorem that $\Gamma_{\alcv}$ permutes
the elements
$\hat{\alpha}_{0},\cdots , \hat{\alpha}_{l}$ of ${\rm B}(\alcv)$.
Looking at the transformation rule (\ref{affact*}), we see that
the homogeneous part of $\Gamma_{\alcv}$ permutes the distinct elements
$\alpha_{0}=-\tilde{\alpha},\alpha_{1}, \cdots , \alpha_{l}$ of $\Delta$.
Since the relative disposition of these $l+1$ roots is used
to construct the extended Dynkin diagram,
$\Gamma_{\alcv}$ can be identified with a group
of Dynkin diagram automorphisms.

We explicitly describe the group $\Gamma_{\alcv}$.
For each $j\in \indJ$, the set $S_j=S(\Ch)-\{s_{\alpha_j}\}$ generates
a subgroup $W_j$ of $W$ and determines a length in $W_j$.
Let $w_j$ be the longest element in $W_j$ and we put
$\gamma_j(\theta)=\e^{-i\mu_j\theta}w_jw_0$. Then we can show the
\begin{pn}
 The group $\Gamma_{\alcv}$ is given by
$\Gamma_{\alcv}=\{ \,\gamma_j \,;\, j\in \indJ\, \}$\,.
\label{a.2:propo}
\end{pn}
The embedding $\Pv \hookrightarrow \Waffh$ induces the isomorphism
$\Pv/\Qv\cong \Waffh /\Waff \cong \Gamma_{\alcv}$.
It then follows the latter part of proposition 2.

\vspace{0.4cm}
{\sc A.2 For A General Compact Connected Group}

\vspace{0.25cm}
In general, $\tilH$ is isomorphic to $\R^M{\times}\prod_{n=1}^NG_n$ for some
$M=0,1,2,\cdots$ and some sequence $G_1,\cdots,G_N$ of simple simply
connected groups. The root lattice $\QQ$ is no longer dual to $\Pv$. If $H$
is not semi-simple ($M\ne 0$), $\Pv/\Qv$ is an infinite group.

\vsp
The definition of the Weyl group $W$, hyperplanes, chambres are the same as
in A.1. Theorem 1 holds with the modification that $l$ is replaced by
the rank $l^{ss}$ of the semi-simple part $\prod_nG_n$. (The actual rank is
$\dim \Ch=M+l^{ss}$.) A choice of chambre $\Ch$ determines the decomposition
$\Delta=\Delta_+\cup\Delta_-$ and we put
\beq
\MC=\left\{\,\mu\in\Pv\,;\,\alpha(\mu)=0\,\,{\rm or}\,\,1\,\,\,
\mbox{for any}
\,\,\alpha\in\Delta_+\,\right\}\,.
\eeq
{}From the proposition 2 of A.1, one can see that $\MC\subset\Pv$
is a section of the projection $\Pv\to\Pv/\Qv$.

\vsp
The definition of the affine Weyl groups $\Waff$, $\Waffh$ are the same
as in A.1. However, $\Waffh$ action on $\hat{\V}=\R_r^{M+N}\oplus\V\oplus
\R_c^{M+N}$ needs some modification if $H$ is not semi-simple: It depends on
choice of the physical model which determines the lift of the rotation action
on $LH$ to $\tiL H$. (This could be read by looking at the energy
momentum tensor given in Appendix B.) Hence, we shall consider only
the action on $\Vaff=\hat{\V}/\R_c^{M+N}$ which is the same as in A.1.
Affine root set $\Daff$ is defined as the subset of $\Vaff^*\cong\R_r^{*M+N}
\oplus \V^*$. A system of hyperplanes in
$\{(-1,\cdots,-1)\}{\times}\V\subset \Vaff$ is defined by using
the affine roots and it leads to the definition of an alc\^ove.
Theorem 3 holds under the modification that the number of walls of an
alc\^ove is $l^{ss}+N$ and that ${\rm B}(\alcv)$ forms a base of
$\Z^N\oplus \QQ$.

\vsp
A choice of chambre $\Ch$ determines an alc\^ove $\alcv$ which in turn
determines the decompositions $\Daff=\Delta_{{\rm aff}+}
\cup\Delta_{{\rm aff}-}$ and
$\Waffh\cong\Waff\semidir\Gmalcv$ where $\Gmalcv$ is the isotropy subgroup
of $\Waffh$ at $\alcv$.
For an indexing set $\indJ$ of $\MC=\{\,\mu_j\,;j\in\indJ\}$, we can give
a map $j\in\indJ\mapsto w_j\in W$ so that $\Gmalcv$ is given by
$\Gmalcv=\{\,\gamma_j\,;j\in\indJ\}$ where
$\gamma_j(\theta)=\e^{-i\mu_j\theta}w_jw_0$.
($w_0$ is the longest element of $W$)

\vspace{0.3cm}
In any case, the groups
$$\pi_1(H)\,,\,\,\,\, \Pv/\Qv\,,\,\,\,\, \Waffh/\Waff\,\,\,\,
\mbox{and}\,\,\,\, \Gmalcv$$
are all isomorphic.

\renewcommand{\theequation}{B.\arabic{equation}}\setcounter{equation}{0}
\vsp
\begin{center}
{\bf Appendix B.}
\end{center}

In a two dimensional quantum field theory coupled to
background metric $\met$ and gauge field $A$ for a Lie group $H$,
the energy momentum tensor $T$ and
the current $J$ is defined as the response to variation of $\met$ and $A$:
\beq
\delta Z_{\Sigma}(\met,A;O_1 O_2\cdots)=Z_{\Sigma}(\met,A;
\frac{1}{2\pi i}\int_{\Sigma}\left\{
\frac{i}{2}\hbox{\small $\sqrt{\met}$}d^{2}x\delta\met^{ab}T_{ab}
+J\cd\delta A\right\}O_1 O_2\cdots)\,.
\eeq
The theory is said to be conformally invariant up to anomaly $c$ when
$T_{z\bar z}=-\frac{c}{12}R_{z\bar z}$ on an insertionless region
on which $A$ is flat, where $R_{z\bar z}$ is the curvature of $\met$.
In this appendix, we give expressions of $T_{zz}$ and $J_z$ of
the adjoint ghost system ($c=-2\dim H$) and
give a description of the Sugawara energy
momentum tensor of the level $k$ WZW model with the target $H$
($c=\frac{k}{k+\hvH}\dim H$).

\vsp
We first assume that $H$ is simple and compact. Choose a local complex
coordinate $z$ ($\met_{zz}=0$) and a local holomorphic section $\sigma$
with respect to $A$. To a base $\{e_{\rm a}\}$ of $\h_{\C}$, $\sigma$
associates a local holomorphic frame $\{\sigma_{\rm a}\}$ of
the adjoint bundle and
the dual frame $\{\sigma^{\rm a}\}$ of the coadjoint bundle.
We denote by $\omega$ and $A^{\sigma}$
the Levi-Chivita connection and the connection $A$ represented
via the holomorphic sections $\frac{\partial}{\partial z}$ and $\sigma$
respectively.

\vspace{0.3cm}
{\it Ghost System}

\vsp
We put $c^{\sigma}(z)=\sum_{\rm a}e_{\rm a}\sigma^{\rm a}\cd
c(z)\in \h_{\C}$
and $b_z^{\sigma}(z)
=\sum_{\rm a}e^{\rm a}b_z\cd \sigma_{\rm a}(z)\in \h_{\C}^*$.
Defining the regularized product $:b_z^{\sigma}(z)c^{\sigma}(w):$ by
\beq
b_z^{\sigma}(z)\ot c^{\sigma}(w)
=\frac{\sum_{\rm a}e^{\rm a}\ot e_{\rm a}}{z-w}
+:b_z^{\sigma}(z)\ot c^{\sigma}(w):\,,
\label{def:regbc}
\eeq
we have
\beqa
J_z\cd\sigma X&=&
:b_z^{\sigma}\cd[X,c^{\sigma}]:-2\hvH\tr(A^{\sigma}_z X)\,,
\label{Jgh}\\
\noalign{\vskip0.2cm}
T_{zz}&=&
:\partial_zb^{\sigma}_z\cd c^{\sigma}:
-:b^{\sigma}_z\cd [A_z^{\sigma},c^{\sigma}]:
+\hvH\tr(A_z^{\sigma}A_z^{\sigma})-\frac{c}{12}S_{zz}\,,\label{Tgh}
\eeqa
where $\sigma X=\sum_{\rm a}\sigma_{\rm a}X^{\rm a}$,
$\tr(XY)=\frac{1}{2\hvH}\tr_{\h}(\ad X\ad Y)$ and
$S_{zz}=\partial_z\omega_z-\frac{1}{2}\omega_z^2$.

\vspace{0.3cm}
{\it Group $H$ WZW Model At Level $k$}

\vsp
To the current, we associate an $\h_{\C}^*$-valued
holomorphic differential $J_z^{\sigma}$ defined by
\beq
J_z\cd\sigma X=J^{\sigma}_z\cd X-k\tr(A^{\sigma}_z X)\,.
\eeq
Defining the regularized product $:J^{\sigma}_z(z)J^{\sigma}_z(w):$ by
\beq
J^{\sigma}_z(z)\cd X J^{\sigma}_z(w)\cd Y=\frac{k\tr(XY)}{(z-w)^2}
+\frac{J^{\sigma}_z(w)\cd [X,Y]}{z-w}+
:J^{\sigma}_z(z)\cd XJ^{\sigma}_z(w)\cd Y:\,,
\eeq
the Sugawara energy momentum tensor is expressed as
\beq
T_{zz}=\frac{\eta^{\rm a b}}{2(k+\hvH)}
:J^{\sigma}_z\cd e_{\rm a}J^{\sigma}_z\cd e_{\rm b}:
-J^{\sigma}_z\cd A^{\sigma}_z+\frac{k}{2}\tr(A^{\sigma}_zA^{\sigma}_z)
-\frac{c}{12}S_{zz}\,,
\label{Sug}
\eeq
where $\eta^{\rm ab}\tr(e_{\rm b}e_{\rm c})=\delta^{\rm a}_{\rm c}$.
This leads to differential equations of correlation functions
\cite{KZ,B,E-O}.

\vsp
If $H$ is an arbitrary compact group, $\h$ is decomposed as the sum
$\R^M\oplus \oplus_{n=1}^N\g_n$ of abelian and simple components.
Since one can always find a local holomorphic section $\sigma$ such that
$A^{\sigma}$ is of block diagonal form, generalization of (\ref{Jgh}),
(\ref{Tgh}) and (\ref{Sug}) to this case is obvious under the prescription
that $\hvH=0$ for $H=U(1)$.

\renewcommand{\theequation}{C.\arabic{equation}}\setcounter{equation}{0}
\vsp

\begin{center}
{\bf Appendix C.}
\end{center}

In this appendix, we derive residual gauge fixing term on the Riemann sphere
$\CP$. We assume that the gauge group $H$ is simple and use notations
introduced in Appendix A. Let $\mu\in\overline{\Ch}$ be one of
$\{\,\mu_j\,;j\in \indJ\}$ and let $\Ph_{[\mu]}$ be the holomorphic
$\HC$-bundle with transition function $h_{\infty 0}(z)=z^{-\mu}$.

\vspace{0.4cm}
{\it Description Of Symmetries}.

\vsp
We introduce subsets of the set $\Delta$ of roots:
\beqa
\Delta^{\mu,0}&=&\{\alpha \in \Delta\, ; \alpha(\mu)=0\}\,,\quad
\Delta^{\mu,0}_{\pm}=\Delta^{\mu,0}\cap \Delta_{\pm}\,,\\
\Delta^{\mu,1}&=&\{\alpha\in \Delta\, ; |\alpha(\mu)|=1\}\,,\quad
\Delta^{\mu,1}_{\pm}=\Delta^{\mu,1}\cap \Delta_{\pm}.
\eeqa
A holomorphic automorphism of $\Ph_{[\mu]}$ is given by $\HC$-valued
holomorphic functions $f_0(z)$ and $f_{\infty}(z)$ of $z$ and $z^{-1}$
respectively such that $f_0(z)=z^{a}f_{\infty}(z)z^{-a}$.
An infinitesimal automorphism is thus of the form
\beq
\delta f_0(z)=v+\sum_{\alpha\in\Delta^{\mu,0}}a_{\alpha}e_{\alpha}
+\sum_{\alpha\in \Delta^{\mu,1}_{+}}(b_{\alpha}+zc_{\alpha})e_{\alpha},
\eeq
where $v\in \liet_{\bf c}$ and $e_{\alpha}\in \h_{\C}$ is a root vector
corresponding to $\alpha\in\Delta$. Putting

$\n^{\mu,1}_+=$ the abelian subalgebra of $\h_{\C}$ spanned by
$\{e_{\alpha};\alpha\in \Delta^{\mu,1}_+\}$,

$H^{\mu,0}=$ the subgroup of $H$ of maximal rank with root system
$\Delta^{\mu,0}$,

\noindent we can describe the automorphism group by
\beq
\Aut\,\Ph_{[\mu]}=\left\{f_0(z)=f^0\e^{n_0+zn_{\infty}};f^0\in
H^{\mu,0}_{\!\C}, n_0,n_{\infty}\in \n^{\mu,1}_+ \right\}.
\eeq

\vspace{0.4cm}
{\it Gauge Condition}

\vsp
Note that, for an automorphism $f$ of the form
$f_0(z)=f^0\e^{n_0+zn_{\infty}}$, $f_0(0)=f^0\e^{n_0}$ and
$f_{\infty}(\infty)=f^0\e^{n_{\infty}}$.
We shall put the gauge condition separately at $z=0$ and
at $z=\infty$.
To start with, we take the Iwasawa decomposition of the field
$h$ at $z=0$ and at $z=\infty$ : $h_0(0)=\e^{v_0}\e^{\frac{\phi_0}{2}}g_0$
and similarly for $h_{\infty}(\infty)$ where $g_0\in H$, $\phi_0\in i\liet$
and $v_0$ is spanned by positive root vectors. We also take
the Iwasawa decomposition $f^0=g^0\e^{\frac{\phi^0}{2}}\e^{v^0}$ where
$g^0\in H^{\mu,0}$, $\phi^0\in i\liet$ and $v^0$ is spanned by
root vectors for $\Delta^{\mu,0}_+$.

\vsp
The condition $h_{\infty}(\infty)\in H$ is necessary and enough to fix
$n_{\infty}, v^0$ and $\phi^0$. In order to fix $g^0$, we decompose
$H^{\mu,0}\backslash H$ into pieces $\{U_{\sigma}\}$ so that we can find
a section $s_{\sigma}$ of $H\to H^{\mu,0}\backslash H$ over each piece
$U_{\sigma}$. If we further require $h_{\infty}(\infty)$ to be in some
$s_{\sigma}U_{\sigma}$, then, $g^0$ is fixed.
The rest, $n_0\in \n^{\mu,1}_+$
is fixed by the condition $v^{\alpha}_0=0$ for $\alpha\in \Delta^{\mu,1}_+$
where $v^{\alpha}_0$ is the coefficient of $v_0$.

\vsp
These conditions determine the residual gauge fixing term. The complexity
arising from the gauge fixing of $g^0$ disappears if we integrate $g_0$ over
$H$ with the factor $\frac{1}{\vol H}$. This leads to the following
expression depending only on the combination $hh^*$ which is gauge invariant
in the usual sense:
\beq
\prod_{\alpha >0}\delta^{(2)}(v^{\alpha}_{\infty})c_{\infty}^{\alpha}
\bar c_{\infty}^{\alpha}\prod_{i=1}^l\delta(\phi_{\infty}^i)c_{\infty}^i
\bar c_{\infty}^i\frac{1}{\vol H^{\mu,0}}\!\!\!
\prod_{-\beta\in \Delta^{\mu,0}_-}
\!\!c_{\infty}^{-\beta}\bar c_{\infty}^{-\beta}\!\prod_{\alpha\in
 \Delta^{\mu,1}_+}\!\!
\delta^{(2)}(v_0^{\alpha})c_0^{\alpha}\bar c_0^{\alpha}\,.
\eeq

\renewcommand{\theequation}{D.\arabic{equation}}\setcounter{equation}{0}

\vsp
\begin{center}
{\bf Appendix D.}
\end{center}

This appendix gives an outline of the proof of the transformation rule
(\ref{rulegmax}) of $\gamma_x$. The $\gamma_x$-transform
$(\Ph^{\gamma},f^{\gamma})$ of an $\HC$-bundle $\Ph$ described by
$\sigma(qz)=\sigma(z)h(q;z)$ with a flag $\sigma(1)h_f$
is defined by the relations (\ref{4.4rel-1}) and (\ref{4.4rel-2})
of an admissible section $\sigma^{\gamma}_0$ around $z=1$ and
a section $\sigma'$ over $\C^*-q^{\Z}$.

We shall find an everywhere regular (but multivalued) section
$\sigma^{\gamma}$. We put $\sigma^{\gamma}(z)=\sigma'(z)\tilde{\chi}(z)$
for $z\not\equiv 1$ and require the relation $\sigma^{\gamma}(qz)
=\sigma^{\gamma}(z)h^{\gamma}(q;z)$ to hold. The task is then to find such
$\tilde{\chi}(z)$ that
$$
\left\{\begin{array}{l}
\tilde{\chi}(zq)=h(q;z)^{-1}\tilde{\chi}(z)h^{\gamma}(q;z)\\
\noalign{\vskip0.2cm}
\chi(z)=h_{\gamma}(z-1)^{-1}h_f^{-1}\tilde{\chi}(z)\quad\mbox{is regular as
$z\to 1$}\,.
\end{array}\right.
$$
The latter condition arises by the requirement that $\sigma_0^{\gamma}(z)
=\sigma^{\gamma}(z)\chi(z)^{-1}$ is an admissible section around $z=1$.
In the following, the solution is exhibited as
$(\Ph,f)\to(\Ph^{\gamma}f^{\gamma}):\tilde{\chi}(z)$.
\beq
\begin{array}{rcll}
(\Ph_u^{(0)},1)\!\!&\to&\!\!(\Ph_F^{(1)},y_u) :&\pmatrix{
r_{-u}R_u(z) &
ie^{-2\pi i u}q^{-\frac{1}{4}}r_{-u}R_{u-\frac{\tau}{2}}(z)
\cr
-r_u R_{-u}(z) &
-ie^{-2\pi i u}q^{-\frac{1}{4}}r_u R_{-u-\frac{\tau}{2}}(z) \cr
}\,;\,u\not\sim 0 \\
\noalign{\vskip0.2cm}
(\Ph_{00}^{(0)},\infty)\!\!&\to&\!\!(\Ph_F^{(1)},y_0) :&\pmatrix{
R_0(z)F(z) &\!\! iq^{-\frac{1}{4}}R_{-\frac{\tau}{2}}(z)G(z) \cr
R_0(z) & \!\!iq^{-\frac{1}{4}}R_{-\frac{\tau}{2}}(z) \cr
}\\
\noalign{\vskip0.2cm}
(\Ph_u^{(0)},0)\!\!&\to&\!\!(\Ph_u^{(1)},0) :&\pmatrix{
0& -c(z)^{-1} \cr
c(z) & 0 \cr
}\\
\noalign{\vskip0.2cm}
(\Ph_u^{(0)},\infty)\!\!&\to &\!\!(\Ph_{-u}^{(1)},0) :&\pmatrix{
c(z) & 0 \cr
0 & c(z)^{-1}\cr
}\\
\noalign{\vskip0.2cm}
(\Ph_{00}^{(0)},0)\!\!&\to&\!\!(\Ph_0^{(1)},1) :&\pmatrix{
c(z)H(z) & -c(z)^{-1} \cr
c(z) & 0 \cr
}
\end{array}
\end{equation}
where
\beqa
\noalign{\vskip-0.4cm}
c(z)\!&=&\!\mbox{$\left(\vartheta(\tau,\zeta+\frac{\tau+1}{2})
\right)^{\nibun}$},
\quad\,\,z=\e^{-2\pi i \z}\nonumber\\
R_u(z)\!\!&=&\!\!\vartheta(2\tau, \zeta +2u+\tau)/c(z),
\qquad r_u=c_{\tau}\cdot(z-1)^{\nibun}R_u(z)|_{z=1},\nonumber\\
F(z)\!&=&\!
2z\frac{\partial}{\partial z}\log\vartheta(2\tau,\zeta\!+\!\tau)-1,
\nonumber\\
G(z)\!&=&\!2z\frac{\partial}{\partial z}\log\vartheta(2\tau,\zeta),
\nonumber\\
H(z)\!&=&\!2z\frac{\partial}{\partial z}\log c(z).
\eeqa
in which $c_{\tau}$ is a constant and
$\vartheta$ is the theta function
$\vartheta(\tau,\z)=\sum_{n\in\Z}q^{\frac{1}{2}n^2}z^{-n}$.

\vspace{1cm}
\noindent{\it Acknowledgement}.
I wish to thank Y. Kazama for advice and encouragement
throughout my graduate course.
I thank T. Eguchi, N. Hayashi, A. Kato, N. Kawamoto,
T. Kohno, K. Mohri, K. Yano and T. Yoneya
for valuable discussions and helpful suggestions.
I thank H. Nakajima for explaining the Hecke correspondence and
N. Iwase for instructions on some algebraic topology including
the isomorphism (\ref{isom:theta}).
I thank S. Higuchi for calling attention to the ref. \cite{DijkWitt}.

\end{document}